\documentclass[journal]{IEEEtran}
\usepackage{graphicx}
\usepackage{multirow}
\usepackage[dvipsnames]{xcolor}
\usepackage{color}
\usepackage{xcolor,colortbl}
\usepackage{tikz}
\def\checkmark{\tikz\fill[scale=0.4](0,.35) -- (.25,0) -- (1,.7) -- (.25,.15) -- cycle;}
\usepackage{pifont}
\usepackage[activate]{microtype}
\usepackage[skip=3pt]{caption}
\usepackage{url}
\usepackage{colortbl}
\newcommand{\xmark}{\text{\ding{55}}}
\usepackage{authblk}
\usepackage{booktabs}% http://ctan.org/pkg/booktabs
\newcommand{\tabitem}{~~\llap{\textbullet}~~}
\usepackage{url}
\usepackage[dvipsnames]{xcolor}
\usepackage{soul}

\usepackage[separate-uncertainty = true,multi-part-units = repeat]{siunitx}
\usepackage{array,amsfonts,amsmath}
\usepackage{caption}
\ifCLASSOPTIONcompsoc
  \usepackage[nocompress]{cite}
\else
  \usepackage{cite}
\fi
\ifCLASSINFOpdf
\else
\fi
\usepackage[pscoord]{eso-pic}% The zero point of the coordinate systemis the lower left corner of the page (the default).

\newcommand{\placetextbox}[3]{% \placetextbox{<horizontal pos>}{<vertical pos>}{<stuff>}
  \setbox0=\hbox{#3}% Put <stuff> in a box
  \AddToShipoutPictureFG*{% Add <stuff> to current page foreground
    \put(\LenToUnit{#1\paperwidth},\LenToUnit{#2\paperheight}){\vtop{{\null}\makebox[0pt][c]{#3}}}%
  }%
}%

\usepackage{fancyhdr}
\begin{document}

\placetextbox{0.5}{1}{****\scriptsize{This work is the extended version of paper accepted in IEEE Transactions on Affective Computing 2021. https://ieeexplore.ieee.org/document/9543566}****}
\title{Deep Representation Learning in Speech Processing: Challenges, Recent Advances, and Future Trends}

\author[1,2]{Siddique Latif\thanks{Email: siddique.latif@usq.edu.au}}
\author[1]{Rajib Rana}
\author[2]{Sara Khalifa}
\author[2,3]{Raja Jurdak}
\author [4]{Junaid Qadir}
\author [5,6]{Bj\"{o}rn W.\ Schuller}
%\author [6]{Julien Epps}
%

\affil[1]{University of Southern Queensland, Australia}
\affil[2]{Distributed Sensing Systems Group, Data61, CSIRO Australia}
\affil[3]{Queensland University of Technology (QUT), Brisbane, Australia}
\affil[4]{Information Technology University, Punjab, Pakistan}
\affil[5]{Imperial College London, UK}
\affil[6]{University of Augsburg, Germany}
%\affil[5]{University of New South Wales, Australia}

% \IEEEcompsocthanksitem J. Doe and J. Doe are with Anonymous University.}% <-this % stops an unwanted space
% \thanks{Manuscript received April 19, 2005; revised August 26, 2015.}}

% \markboth{Journal of \LaTeX\ Class Files,~Vol.~14, No.~8, August~2015}%
% {Shell \MakeLowercase{\textit{et al.}}: Bare Demo of IEEEtran.cls for Computer Society Journals}
\maketitle
%\IEEEtitleabstractindextext{%
\begin{abstract}

Research on speech processing has traditionally considered the task of designing hand-engineered acoustic features (feature engineering) as a separate distinct problem from the task of designing efficient machine learning (ML)  models to make prediction and classification decisions. There are two main drawbacks to this approach: firstly, the feature engineering being manual is cumbersome and requires human knowledge; and secondly, the designed features might not be best for the objective at hand. This has motivated the adoption of a recent trend in speech community towards utilisation of representation learning techniques, which can learn an intermediate representation of the input signal automatically that better suits the task at hand and hence lead to improved performance. The significance of representation learning has increased with advances in deep learning (DL), where the representations are more useful and less dependent on human knowledge, making it very conducive for tasks like classification, prediction, etc. The main contribution of this paper is to present an up-to-date and comprehensive survey on different techniques of speech representation learning by bringing together the scattered research across three distinct research areas including Automatic Speech Recognition (ASR), Speaker Recognition (SR), and Speaker Emotion Recognition (SER). %A large body of research has studied representation learning for speech, however, the most cited review performed by Bengio et al. in 2013 was conducted prior to the Adversarial Neural Network era. 
  Recent reviews in speech have been conducted for ASR, SR, and SER, however, none of these has focused on the representation learning from speech---a gap that our survey aims to bridge. We also highlight different challenges and the key characteristics of representation learning models and discuss important recent advancements and point out future trends. Our review can be used by the speech research community as an essential resource to quickly grasp the current progress in representation learning and can also act as a guide for navigating study in this area of research.

\end{abstract}

% \begin{IEEEkeywords}
% Speech emotion recognition, multi task learning 
% \end{IEEEkeywords}}

%}
%\maketitle
\IEEEpeerreviewmaketitle

\section{Introduction}
\label{sec:introduction}

The performance of machine learning (ML) algorithms heavily depends on data representation or features. 
Traditionally most of the actual ML research has focused on feature engineering or the design of pre-processing data transformation pipelines to craft representations that support ML algorithms \cite{bengio2013representation}. Although such feature engineering techniques can help improve the performance of predictive models, the downside is that these techniques are labour-intensive and time-consuming. To broaden the scope of ML algorithms, it is desirable to make learning algorithms less dependent on hand-crafted features. 
% This will lead to real progress towards Artificial Intelligence (AI), where the ML models become less dependent on manually-crafted knowledge-based features and will instead automatically disentangle the underlying hidden explanatory attributes from the real-world data.   
 %In this way, predictive models take advantage of human ingenuity that highlights their weakness.  
 
A key application of ML algorithms has been in analysing and processing speech. Nowadays, speech interfaces have become widely accepted and integrated into various real-life applications and devices. Services like Siri and Google Voice Search have become a part of our daily life and are used by millions of users \cite{herff2016automatic}. Research in speech processing and analysis has always been motivated by a desire to enable machines to participate in verbal human-machine interactions. The research goals of enabling machines to understand human speech, identify speakers, and detect human emotion have attracted researchers' attention for more than sixty years \cite{tran2000fuzzy}. Researchers are now focusing on transforming current speech-based systems into the next generation AI devices that react with humans more friendly and provide personalised responses according to their mental states.
%BS again: I find emotion narrow. They will react to our general state, including cognitive and physical load, health, etc. ... :)
In all these successes, speech representations---in particular, deep learning (DL)-based speech representations---play an important role. Representation learning, broadly speaking, is the technique of learning representations of input data, usually through the transformation of the input data, where the key goal is yielding abstract and useful representations for tasks like classification, prediction, etc. One of the major reasons for the utilisation of representation learning techniques in speech technology is that speech data is fundamentally different from two-dimensional image data. Images can be analysed as a whole or in patches but speech has to be studied sequentially to capture temporal contexts. 
%BS so has video - not sure if 2D vs time makes sense?!
%These properties gave rise to speech-specific DL based representation learning techniques to capture speech attributes. 
%BS is that so? I think mostly, to be frank, a lot has been taken over from image processing :) But ok.

Traditionally, the efficiency of ML algorithms on speech has relied heavily on the quality of hand-crafted features. A good set of features often leads to better performance compared to a poor speech feature set. Therefore, feature engineering, which focuses on creating features from raw speech and has led to lots of research studies, has been an important field of research for a long time. 
%This was not only specific for speech---vision processing and natural language processing (NLP) have demonstrated similar trends \cite{pouyanfar2019survey}. 
DL models, in contrast, can learn feature representation automatically which minimises the dependency on hand-engineered features and thereby give better performance in different speech applications \cite{najafabadi2015deep}. %Researchers have evaluated different DL models for representation learning from speech. 
These deep models can be trained on speech data in different ways such as supervised, unsupervised, semi-supervised, transfer, and reinforcement learning. This survey covers all these feature learning techniques and popular deep learning models in the context of three popular speech applications \cite{huang2014historical}: (1) automatic speech recognition (ASR); (2) speaker recognition (SR); and (3) speech emotion recognition (SER). 

% Representation learning, broadly speaking, is the technique of learning representations of input data, usually through the transformation of the input data, where the key goal is yielding abstract and useful representations for machine learning (ML) tasks such as classification, prediction, etc. In the context of probabilistic models, representation learning aims to learn a representation that captures the probability distribution of the input data. Whereas in deep learning, representation learning is done through the composition of multiple non-linear transformations of the input data.

% Representation learning from speech is a rapidly developing area, therefore, we highlight the fundamental challenges, discuss recent progress made by the speech community and point out future trends.  

\begin{table*}[]
\scriptsize
\centering
\caption{Comparison of our paper with that of the existing surveys.}
\label{summary}
\begin{tabular}{|l|c|c|c|c|c|l|}
\hline
 & \multicolumn{5}{c|}{\textbf{Focus}}   & \multicolumn{1}{c|}{}  \\ \cline{2-6}
 & \multicolumn{1}{l|}{}    & \multicolumn{3}{c|}{\textbf{Speech}}    & \multicolumn{1}{l|}{}& \multicolumn{1}{c|}{}\\ \cline{3-5}
\multirow{-3}{*}{\textbf{Paper}}  & \multicolumn{1}{l|}{\multirow{-2}{*}{\begin{tabular}[c]{@{}l@{}}\textbf{Representation}\\ \textbf{Learning}\end{tabular}}} & \multicolumn{1}{l|}{\textit{ASR}} & \multicolumn{1}{l|}{\textit{SR}} & \multicolumn{1}{l|}{\textit{SER}} & \multicolumn{1}{l|}{\multirow{-2}{*}{\textbf{Deep Learning}}} & \multicolumn{1}{c|}{\multirow{-3}{*}{\textbf{Details}}}    \\ \hline

\begin{tabular}[c]{@{}l@{}}Bengio et al.\  \cite{bengio2013representation}\\2013\end{tabular}& \checkmark{}&\xmark{}  & \xmark{} & \xmark{}& \checkmark{} &\begin{tabular}[c]{@{}l@{}}This paper reviewed the work in the area of unsupervised feature learning \\and deep learning, it also covered advancements in probabilistic models \\and autoencoders. It does not include recent models like VAE and GANs. \end{tabular}     \\ \hline

\begin{tabular}[c]{@{}l@{}}Zhong et al.\\2016\cite{zhong2016overview}\end{tabular} & \checkmark{}&\xmark{}  & \xmark{} & \xmark{}& \checkmark{} & \begin{tabular}[c]{@{}l@{}}In this paper, the history of data representation learning is reviewed from \\traditional to recent DL methods. Challenges for representation learning, \\recent advancement, and future trends are not covered.\end{tabular}     \\ \hline
\begin{tabular}[c]{@{}l@{}}Zhang et al.\ \cite{zhang2018deep}\\2018\end{tabular} &  \checkmark{}  &\checkmark{} & \xmark{}&\xmark{}  & \checkmark{}& \begin{tabular}[c]{@{}l@{}}This paper provides a systematic overview of representative DL approaches \\that are designed for environmentally robust ASR.\end{tabular}     \\ \hline

\begin{tabular}[c]{@{}l@{}}Swain et al.\ \cite{swain2018databases}\\2018\end{tabular} &  \xmark{} & \xmark{} & \xmark{}&\checkmark{}  & \checkmark{}& \begin{tabular}[c]{@{}l@{}}This paper reviewed the literature on various databases, features, and \\classifiers for SER system. \end{tabular}     \\ \hline

\begin{tabular}[c]{@{}l@{}}Nassif et al.\ \cite{nassif2019speech}\\2019\end{tabular} &  \xmark{} & \checkmark{} & \xmark{}&\xmark{} & \checkmark{}& \begin{tabular}[c]{@{}l@{}}This paper presented a systematic review of studies from 2006 to 2018 on\\ DL based speech recognition and highlighted the on the trends of\\ research in ASR.

\end{tabular}     \\ \hline

\rowcolor[HTML]{EFEFEF} 
\bf{Our paper}  &  \checkmark{}&   \checkmark{}     &  \checkmark{}   &  \checkmark{} &  \checkmark{}  & \begin{tabular}[c]{@{}l@{}}Our paper covers different representation learning techniques from \\speech, DL models, discusses different challenges, highlights recent \\advancements and future trends. The main contribution of this paper is to \\bring together scattered research on representation learning of speech \\across three research areas: ASR, SR, and SER.   \end{tabular} \\ \hline
\end{tabular}
\end{table*}

% trim={<left> <lower> <right> <upper>}
\begin{figure*}[!ht]
\centering
%captionsetup{justification=centering}
\includegraphics[trim=0cm 0cm 0cm 0cm,clip=true, width=.9\textwidth]{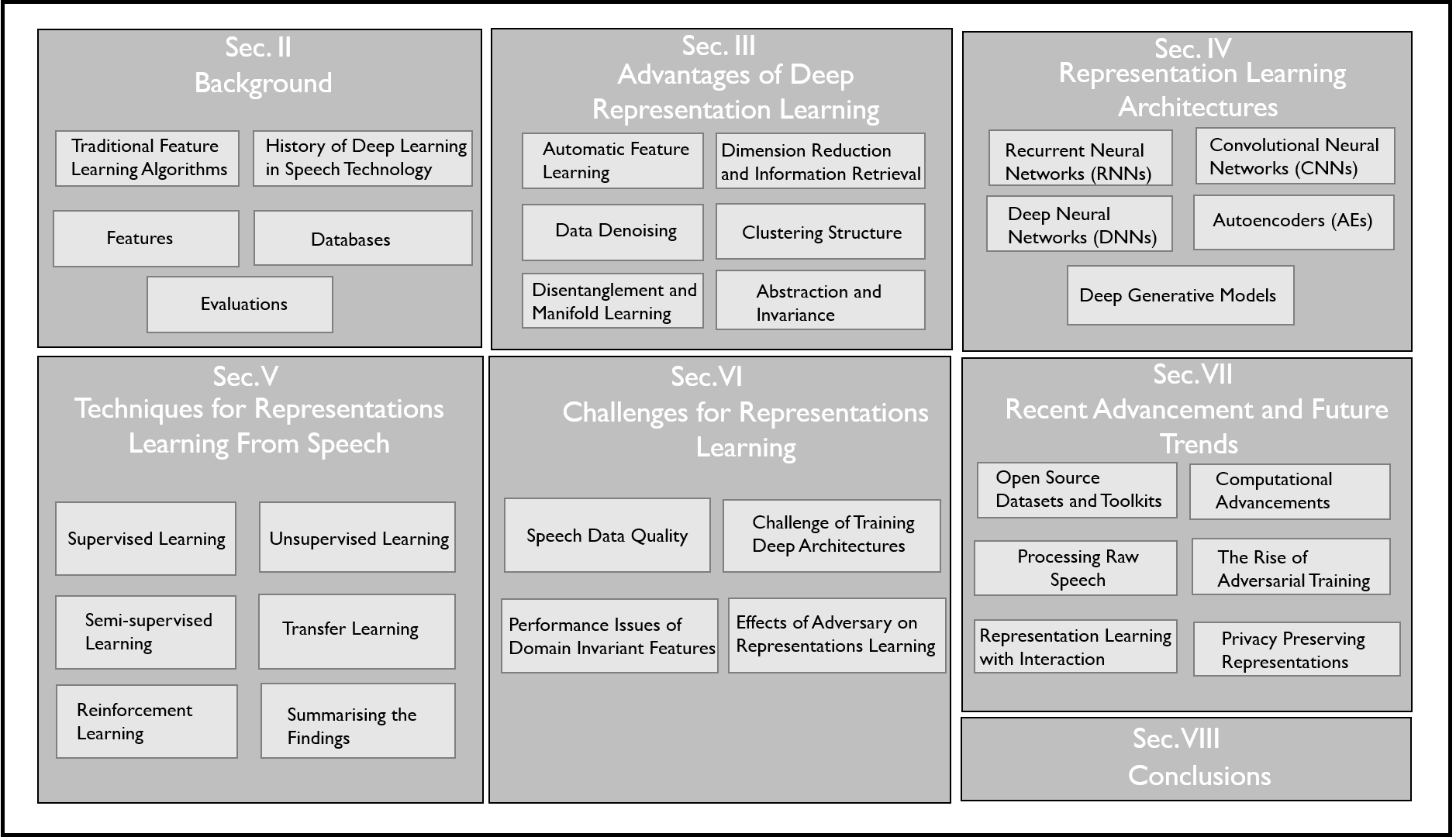}
%\captionsetup{width=0.95\textwidth}
\caption{Organisation of the paper.}
\label{fig:structure}
\end{figure*}

%\subsection{Survey Scope and Novelty}

Despite growing interest in representation learning from speech, existing contributions are scattered across different research areas and a comprehensive survey is missing. To highlight this,  we present the summary of different popular and recently published review papers Table \ref{summary}. The review article published in 2013 by Bengio et al.\  \cite{bengio2013representation} is one of the most cited papers. It is very generic and focuses on appropriate objectives for learning good representations, for computing representations (i.\,e., inference), and the geometrical connections between representation learning, manifold learning, and density estimation. Due to an earlier publication date, this paper had a focus on principal component analysis (PCA), restricted Boltzmann machines (RBMs), autoencoders (AEs) and recently proposed generative models were out of the scope of this paper. The research on representation learning has evolved significantly since then as generative models like variational autoencoders (VAEs) \cite{kingma2013auto}, generative adversarial networks (GANs) \cite{goodfellow2014generative}, etc., have been found to be more suitable for representation learning compared to autoencoders and other classical methods. We cover all these new models in our review. Although, other recent surveys have focused on DL techniques for ASR \cite{nassif2019speech,zhang2018deep}, SR \cite{gomez2019design}, and SER \cite{swain2018databases}, none of these has focused on representation learning from speech. This article bridges this gap by presenting an up-to-date survey of research that focused on representation learning in three active areas: ASR, SR, and SER. Beyond reviewing the literature, we discuss the applications of deep representation learning, present popular DL models and their representation learning abilities, and different representation learning techniques used in the literature. We further highlight the challenges faced by deep representation learning in the  speech and finally conclude this paper by discussing the recent advancement and pointing out future trends. The structure of this article is shown in Figure \ref{fig:structure}.
%BS consider also this one:
%SL: Added Sir

\section{Background}
\label{background}

\subsection{Traditional Feature Learning Algorithms}
%\RR{Siddique, can you look into the comments by the co-authors. This subsection needs attention.}
In the field of data representation learning, the algorithms are generally categorised into two classes: shallow learning algorithms and DL-based models \cite{zhong2019shallow}. Shallow learning algorithms are also considered as traditional methods. They aim to learn transformations of data by extracting useful information. One of the oldest feature learning algorithms, Principal Components Analysis (PCA) \cite{pearson1901liii} has been studied extensively over the last century. 
%BS: is PCA really a feature extraction method? And wasn't energy calculation earlier than that? Perhaps "one of the oldest"?
% Updated 
During this period, various other shallow learning algorithms have been proposed based on various learning techniques and criteria, until the popular deep models in recent years.  Similar to PCA, Linear Discriminant Analysis (LDA) \cite{fisher1936use} is another shallow learning algorithm. 
%BS: I massively doubt that! LDA never found much relevance, as it is supervised. Usually, the only way LDA appears in speech is if someone uses PCA people say "but LDA would be better...". That's a bit like wavelets and FFT. Nobody uses wavelets, but all say "but wavelets would be better"...
%SL: Updated
Both PCA and LDA are linear data transformation techniques, however, LDA is a supervised method that requires class labels to maximise class separability. Other linear feature learning methods includes Canonical Correlation Analysis (CCA) \cite{hardoon2004canonical}, Multi-Dimensional Scaling (MDS) \cite{borg2003modern}, and Independent Component Analysis (ICA)\cite{hyvarinen2000independent}. 
%BS: I unified all to leading capital letters, as it was arbitrarily mixed.
%BS: CCA and ICA are almost irrelavant in the context of speech (only) (from a single source).
The kernel version of some linear feature mapping algorithms are also proposed including kernel PCA (KPCA) \cite{scholkopf1998nonlinear}, and generalised discriminant analysis (GDA) \cite{baudat2000generalized}. They are non-linear versions of PCA and LDA, respectively. %The kernel versions of linear feature learning techniques are considered as first mapping of the data in the observation space to a high dimensional feature space using a nonlinear function and then conducting linear dimensionality reduction in this feature space. 
Another popular technique is Non-negative Matrix Factorisation (NMF) \cite{lee1999learning} that can generate sparse representations of data useful for ML tasks.  

%BS: Should we also mention Non-negative Matrix Factorisation (NMF) somewhere?
% SL: updated Sir

Many methods for nonlinear feature reduction are also proposed to discover the non-linear hidden structure from the high dimensional data \cite{lee2004feature}. They include Locally Linear Embedding (LLE) \cite{roweis2000nonlinear}, Isometric Feature Mapping (Isomap) \cite{tenenbaum2000global}, T-distributed Stochastic Neighbour Embedding (t-SNE) \cite{maaten2008visualizing}, and Neural Networks (NNs) \cite{lecun1998gradient}. In contrast to kernel-based methods, non-linear feature representation algorithms directly learn the mapping functions by preserving the local information of data in the low dimensional space. Traditional representation algorithms have been widely used by researchers of the speech community for transforming the speech representations to more informative features having low dimensional space (e.g., \cite{kocsor2004kernel,takiguchi2007pca}). %In speech analysis, noise removal from audio is very crucial to avoid degrading the system performance. Non-linear feature learning algorithms have been found very helpful to project the main speech components onto low-order features to remove noise \cite{takiguchi2006robust}. 
However, these shallow feature learning algorithms dominate the data representation learning area until the successful training of deep models for representation learning of data by Hinton and Salakhutdinov in 2006 \cite{hinton2006reducing}. This work was quickly followed up with similar ideas by others \cite{bengio2007greedy,poultney2007efficient}, which lead to a large number of deep models suitable for representation learning. We discuss the brief history of the success of DL in speech technology next.

\subsection{Brief History on Deep Learning (DL) in Speech Technology}
For decades, the Gaussian Mixture Model (GMM) and Hidden Markov Model (HMM) based models (GMM-HMM) ruled the speech technology due to their many advantages including their mathematical elegance and capability to model time-varying sequences \cite{gales2008application}. Around 1990, discriminative training was found to produce better results compared to the models trained using maximum likelihood \cite{waibel1989phoneme}. 
%BS REFERENCE!
%SL: updated Sir
Since then researchers started working towards replacing GMM with a feature learning models including neural networks (NNs),
%BS REFERENCE!
restricted Boltzmann machines (RBMs), deep belief networks (DBNs), 
%BS First in SER:
%@inproceedings{stuhlsatz2011deep,
%  title={Deep neural networks for acoustic emotion recognition: raising the benchmarks},
%  author={Stuhlsatz, Andr{\'e} and Meyer, Christine and Eyben, Florian and Zielke, Thomas and Meier, G{\"u}nter and Schuller, Bj{\"o}rn},
%  booktitle={2011 IEEE international conference on acoustics, speech and signal processing (ICASSP)},
%  pages={5688--5691},
%  year={2011},
%  organization={IEEE}
%}
%
%BS first LSTM for SER:
%SL:Updated Sir
%BS early LSTM for ASR:
%SL:updated Sir
and deep neural networks (DNNs) \cite{purwins2019deep}. Hybrid models gained popularity while HMMs continued to be investigated. 
%BS first hybrid LSTM for ASR:
%@inproceedings{wollmer2009tandem,
%  title={A Tandem BLSTM-DBN architecture for keyword spotting with enhanced context modeling},
%  author={W{\"o}llmer, Martin and Eyben, Florian and Graves, Alex and Schuller, Bj{\"o}rn and Rigoll, Gerhard},
%  booktitle={Proc. of NOLISP 2009, ISCA Tutorial and Research Workshop on Non-Linear Speech Processing, Vic, Spain},
%  pages={8--pages},
%  year={2009}
%}

In the meanwhile, researchers also worked towards replacing HMM with other alternatives. In 2012, DNNs were trained on thousands of hours of speech data and they successfully reduced the word error rate (WER) on  ASR task \cite{hinton2012deep}. This is due to their ability to learn a hierarchy of representations from input data. However, soon after recurrent neural networks (RNNs) architectures including long-short term memory (LSTM) and gated recurrent units (GRUs) outperformed DNNs and became state-of-the-art models not only in  ASR \cite{wollmer2010long} but also in  SER \cite{wollmer2008abandoning}. 
%BS See above for my suggestions of first LSTM RNNs in SER and ASR
%SL: Updated Sir
The superior performance of RNN architectures was because of their ability to capture temporal contexts from speech \cite{latif2018phonocardiographic,qayyum2018quran}. 
% Their superior performance made them popular not only for ASR but also for speaker recognition (SR) and speech emotion recognition (SER). 
Later, a cascade of convolutional neural networks (CNNs),
%BS First CNN-LSTM RNN end-to-end for SER:
%SL: Updated Sir
LSTM and fully connected (DNNs) layers were further shown to outperform LSTM-only models by capturing more discriminative attributes from speech \cite{sainath2015convolutional,trigeorgis2016adieu}. The lack of labelled data set the pace for the unsupervised representation learning research. For unsupervised representation from speech, AEs, RBMs, and DBNs were widely used \cite{langkvist2014review}. 
%BS add REFERENCES
% This direction always aspired to achieve better performance with unsupervised learning. Therefore, researchers attempted various models and their combination to achieve better performance in an unsupervised way.

Nowadays, there has been a significant interest in three classes of generative models including VAEs, GANs, and deep auto-regressive models \cite{oord2016pixel,oord2016wavenet}. They have been widely being employed for speech processing---especially VAEs and GANs are becoming very influential models for learning speech representation in an unsupervised way \cite{bollepalli2019generative,hsu2017unsupervised}. In speech analysis tasks, deep models for representation learning can either be applied to speech features or directly on the raw waveform. We present a brief history of speech features in the next section.

\subsection{Speech Features}
In speech processing, feature engineering and designing of models for classification or prediction are often considered as separate problems. Feature engineering is a way of manually designing speech features by taking advantage of human ingenuity. %However, such features might not guarantee the best-performing speech-based systems. Therefore, it becomes imperative to use representation learning after these features. 
For decades, 
%BS corrected:
Mel Frequency Cepstral Coefficients (MFCCs) \cite{furui1986speaker} 
have been used as the principal set of features for speech analysis tasks. Four steps involve for MFCCs extraction: 
%BS corrected:
computation of the Fourier transform, projection of the powers of the spectrum onto the Mel scale, taking the logarithm of the Mel frequencies, and applying Discrete Cosine Transformation (DCT) for compressed representations. 
%BS end corrected.
It is found that the last step removes the information and destroys spatial relations; 
%BS corrected:
therefore, it is usually omitted, which yields the log-Mel spectrum, a popular feature across 
%BS added:
the speech community. 
%BS corrected:
This has been the most popular feature to train DL networks. 

%BS corrected:
The 
Mel-filter bank is inspired by auditory and physiological findings of how humans perceive speech signals \cite{davis1980comparison}. Sometimes, it becomes preferable to use features that capture transpositions as translations. For this, a suitable filter bank is spectrograms that captures how the frequency content of 
%BS corrected:
the 
speech signal changes with time \cite{purwins2000new}. %The 
%BS choose BE or AE:
% neighbouring 
% spectrogram bins of the speech signal are correlated in time and frequency.
In speech research, researchers widely used CNNs for spectrogram 
%BS corrected:
inputs
due to their image like configuration. 
%BS REFERENCE!
% Recently, researchers started using raw speech to avoid hand-designed features, which allows exploiting the improved 
% %BS this is BE again - please unify - I prefer British...
% modelling capability of DL models. However, this has higher computational 
% %BS corrected:
% cost and data requirements.
% For speech analysis tasks, such as 
% %BS: BE has "," ahead of "and" - above, this is used in mixed ways...
% ASR, SR, and SER, 
Log-Mel spectrograms is another speech representation that provides a compact representation and 
%BS corrected:
became the current state of the art because models using these features usually need less data and training to achieve 
%BS added:
similar or better results.

In SER, feature engineering is more dominant and 
%BS corrected:
a minimalistic 
%BS corrected:
sets of features like GeMAPs and eGeMAPs \cite{eyben2015geneva} 
%BS Cite:
%SL:Added citation Sir
are also proposed based on affective physiological changes in voice production and their theoretical significance \cite{eyben2015geneva}. They are also popular being used as benchmark 
%BS corrected:
feature sets. 
However, in speech analysis tasks, some works \cite{neumann2017attentive,jaitly2011learning} 
%BS corrected:
show that the particular choice of features is less important compared to the design of 
%BS corrected:
the model architecture and the amount of training data. The research is continuing in designing such DL models and input features that involve minimum human knowledge.

\subsection{Databases}
Although the success of deep learning is usually attributed to the models' capacity and higher computational power, the most crucial role is played by the availability of large-scale labelled datasets \cite{sun2017revisiting}. In contrast to the vision domain, the speech community started using DNNs with considerably smaller datasets. Some popular conventional corpora used for ASR and SR includes TIMIT \cite{fisher1986darpa}, Switchboard  \cite{godfrey1992switchboard}, WSJ \cite{paul1992design}, AMI \cite{hain2007ami}. Similarly, EMO-DB \cite{burkhardt2005database}, FAU-AIBO \cite{schuller2009interspeech}, RECOLA \cite{ringeval2013introducing}, and GEMEP \cite{banziger2012introducing} are some popular classical datasets. Recently, larger datasets are being created and released to the research community to engage the industry as well as the researchers. We summarise some of these recent and publicly available datasets in Table \ref{table:corpora} that are widely used in the speech community.

\begin{table*}[]
\centering
\caption{Speech corpora and their details.}
\tiny
\begin{tabular}{|c|c|c|c|l|l|}
\hline
\textbf{Application}  & \textbf{Corpus} & \textbf{Language} &  \textbf{Mode} & \textbf{Size}  & \textbf{Details}  \\ \hline

\multirow{6}{*}{\begin{tabular}[c]{@{}c@{}}\textbf{Speech and} \\ \textbf{Speaker} \\ \textbf{Recognition}\end{tabular}} 
& \begin{tabular}[c]{@{}c@{}}LibriSpeech \cite{panayotov2015librispeech}\end{tabular} 
& \begin{tabular}[c]{@{}c@{}}English\end{tabular}
& \begin{tabular}[c]{@{}c@{}}Audio\end{tabular}
& \begin{tabular}[l]{@{}l@{}}1\,000 hours of speech\\ of 2\,484 speakers\end{tabular} 
& \begin{tabular}[l]{@{}l@{}}Designed for speech recognition and also used for speaker identification \\and verification.\end{tabular}  \\ \cline{2-6} 

% & \begin{tabular}[c]{@{}c@{}}VoxCeleb1 \& 2 \cite{panayotov2015librispeech}\end{tabular} 
% & \begin{tabular}[c]{@{}c@{}}Multiple\end{tabular}
% & \begin{tabular}[c]{@{}c@{}}Audio-Visual\end{tabular}
% & \begin{tabular}[l]{@{}l@{}}153\,516 utterances of \\ 1\,251 celebrities\end{tabular} 
% & \begin{tabular}[l]{@{}l@{}}This data is extracted from videos uploaded to YouTube and \\designed  for speaker identification and verification.\end{tabular}   \\ \cline{2-6}

& \begin{tabular}[c]{@{}c@{}}VoxCeleb2 \cite{chung2018voxceleb2}\end{tabular} 
& \begin{tabular}[c]{@{}c@{}}Multiple\end{tabular}
& \begin{tabular}[c]{@{}c@{}}Audio-Visual\end{tabular}
& \begin{tabular}[l]{@{}l@{}}1\,128\,246 utterances \\of 6\,112 celebrities\end{tabular} 
& \begin{tabular}[l]{@{}l@{}}This data is extracted from videos uploaded to YouTube and designed  for\\ speaker identification and verification.\end{tabular} \\ \cline{2-6}

& \begin{tabular}[c]{@{}c@{}}TED-LIUM \cite{rousseau2012ted}\end{tabular} 
& \begin{tabular}[c]{@{}c@{}}English \end{tabular}
& \begin{tabular}[c]{@{}c@{}}Audio\end{tabular}
& \begin{tabular}[l]{@{}l@{}}118 hours of speech \\ of 698 speakers\end{tabular} 
& \begin{tabular}[l]{@{}l@{}}This corpus is extracted from 818 TED Talks for ASR.\end{tabular}  \\ \cline{2-6} 

& \begin{tabular}[c]{@{}c@{}}THCHS-30 \cite{wang2015thchs}\end{tabular} 
& \begin{tabular}[c]{@{}c@{}}Chinese\end{tabular}
& \begin{tabular}[c]{@{}c@{}}Audio\end{tabular}
& \begin{tabular}[l]{@{}l@{}}30 hours of speech \\ from 30 speakers\end{tabular} 
& \begin{tabular}[l]{@{}l@{}}This corpus is recorded for Chinese speech recognition. \end{tabular}  \\ \cline{2-6} 

& \begin{tabular}[c]{@{}c@{}}AISHELL-1 \cite{bu2017aishell}\end{tabular} 
& \begin{tabular}[c]{@{}c@{}}Mandarin\end{tabular}
& \begin{tabular}[c]{@{}c@{}}Audio\end{tabular}
& \begin{tabular}[l]{@{}l@{}} 170 hours of speech \\ from 400 speakers\end{tabular} 
& \begin{tabular}[l]{@{}l@{}}An open source Mandarin ASR corpus. \end{tabular}   \\ \cline{2-6}

& \begin{tabular}[c]{@{}c@{}}Tuda-De \cite{milde2018open}\end{tabular} 
& \begin{tabular}[c]{@{}c@{}}German\end{tabular}
& \begin{tabular}[c]{@{}c@{}}Audio\end{tabular}
& \begin{tabular}[l]{@{}l@{}} 127 hour of speech \\ from 147 speakers\end{tabular} 
& \begin{tabular}[l]{@{}l@{}}A corpus of German utterances was publicly released for distant \\speech recognition. \end{tabular}   \\\hline

% & \begin{tabular}[c]{@{}c@{}}Thchs-30\cite{wang2015thchs}\end{tabular} 
% & \begin{tabular}[c]{@{}c@{}}Chinese\end{tabular}
% & \begin{tabular}[c]{@{}c@{}}Audio\end{tabular}
% & \begin{tabular}[l]{@{}l@{}} 35 hours of \\ speech data \\from 50 speakers\end{tabular} 
% & \begin{tabular}[l]{@{}l@{}}An open source Mandarin ASR corpus. \end{tabular}  \\\hline
%BS The SER corpora are enitrely outdated!!!! THERE HAS TO BE FAU AEC, RECOLA, SEWA, GEMEPS, USoM - these have all been corpora used in challenges and the info on the corpora is completely missing info on the type of emotion (acted, spontaneous, elicited) and labelling (classes, continuous (time, value)). 

\multirow{6}{*}{\begin{tabular}[c]{@{}c@{}}\textbf{Speech} \\ \textbf{Emotion}\\\textbf{Recogntion}\end{tabular}} 
& \begin{tabular}[c]{@{}c@{}}EMO-DB \cite{burkhardt2005database}\end{tabular} 
& \begin{tabular}[c]{@{}c@{}}German\end{tabular}
& \begin{tabular}[c]{@{}c@{}}Audio\end{tabular}
& \begin{tabular}[l]{@{}l@{}} 10 actors and \\494 utterances\end{tabular} 
& \begin{tabular}[l]{@{}l@{}}An acted corpus on 10 German sentences which which usually\\ used in everyday communication. \end{tabular}  \\ \cline{2-6} 
%BS EMO-DB is NOT "stimulated" but merely acted

% & \begin{tabular}[c]{@{}c@{}}SAVEE \cite{jackson2014surrey}\end{tabular} 
% & \begin{tabular}[c]{@{}c@{}}English\end{tabular}
% & \begin{tabular}[c]{@{}c@{}}Audio-Visual\end{tabular}
% & \begin{tabular}[l]{@{}l@{}}4 male actors and \\480 utterances\end{tabular} 
% & \begin{tabular}[l]{@{}l@{}}A stimulated corpus recorded in seven emotions released for SER\end{tabular} \\ \cline{2-6} 

& \begin{tabular}[c]{@{}c@{}}SEMAINE \cite{mckeown2012semaine}\end{tabular} 
& \begin{tabular}[c]{@{}c@{}}English\end{tabular}
& \begin{tabular}[c]{@{}c@{}}Audio-Visual\end{tabular}
& \begin{tabular}[l]{@{}l@{}}150 participants and \\959 conversations\end{tabular} 
& \begin{tabular}[l]{@{}l@{}}An induced corpus recorded to build sensitive artificial listener agents\\ that can engage a person in a sustained and emotionally coloured conversation. \end{tabular}  \\ \cline{2-6}

& \begin{tabular}[c]{@{}c@{}}IEMOCAP \cite{busso2008iemocap}\end{tabular} 
& \begin{tabular}[c]{@{}c@{}}English\end{tabular}
& \begin{tabular}[c]{@{}c@{}}Audio-Visual\end{tabular}
& \begin{tabular}[l]{@{}l@{}}12  hours of speech\\ from 10 speakers\end{tabular} 
& \begin{tabular}[l]{@{}l@{}}To collect this data, an interactive setting served to elicit authentic emotions \\and create a larger emotional corpus to study multimodal interactions.\end{tabular}  \\ \cline{2-6}

& \begin{tabular}[c]{@{}c@{}}MSP-IMPROV\cite{busso2017msp}\end{tabular} 
& \begin{tabular}[c]{@{}c@{}}English \end{tabular}
& \begin{tabular}[c]{@{}c@{}}Audio-Visual\end{tabular}
& \begin{tabular}[l]{@{}l@{}}9 hours of audiovisual\\ data of 12 actors\end{tabular} 
& \begin{tabular}[l]{@{}l@{}}This corpus is recorded from dyadic interactions of actors to \\study emotions.\end{tabular}  \\ \hline
\end{tabular}
\label{table:corpora}
\vspace{-2mm}
\end{table*}
%BS I find the text pretty verbous above in the table - could be shortened to type of application or so

\subsection{Evaluations}
Evaluation measures vary across speech tasks. The performance of ASR systems is usually measured using word error rates (WER), which is the fraction of the sum of insertion, deletion, and substitution divided by the total number of words in the reference transcription. In speaker verification systems, two types of errors---namely, \textit{false rejection} (fr), where a valid identity is rejected, and \textit{false acceptance} (fa), where a fake identity is accepted---are used. These two errors are measured experimentally on test data. Based on these two errors, a detection error trade-offs (DETs) curve is drawn to evaluate the performance of the system. DET is plotted using 
the probability of false acceptance $(P_{fa})$ as a function of the probability of false rejection $(P_{fr})$. Another popular evaluation measure is the equal error rate (EER) which corresponds to the operating point where $P_{fa}$ = $P_{fr}$. 
%BS added:
Similarly, the area under curve (AUC) of the receiver operating curve (ROC) is often found. 
%BS end addition. 
The details on other evaluation measures for the speaker verification task can be found in \cite{bimbot2004tutorial}. Both speaker identification and emotion recognition use classification accuracy as a metric. 
%BS added:
However, as data is often imbalanced across the classes in naturalistic emotion corpora, the accuracy is usually used as so-called unweighted accuracy (UA) or unweighted average recall (UAR), which represents the average recall across classes, unweighted by the number of instances by classes. This has been introduced by the first challenge in the field---the Interspeech 2009 Emotion Challenge \cite{Schuller11-RRE} and has since been picked up by other challenges across the field. Also, SER systems use regression to predict emotional attributes such as continuous arousal and valence or dominance.

\section{Applications of Deep Representation Learning}
\label{usecases}

\begin{figure}[!ht]
\centering
%captionsetup{justification=centering}
\includegraphics[width=0.25\textwidth]{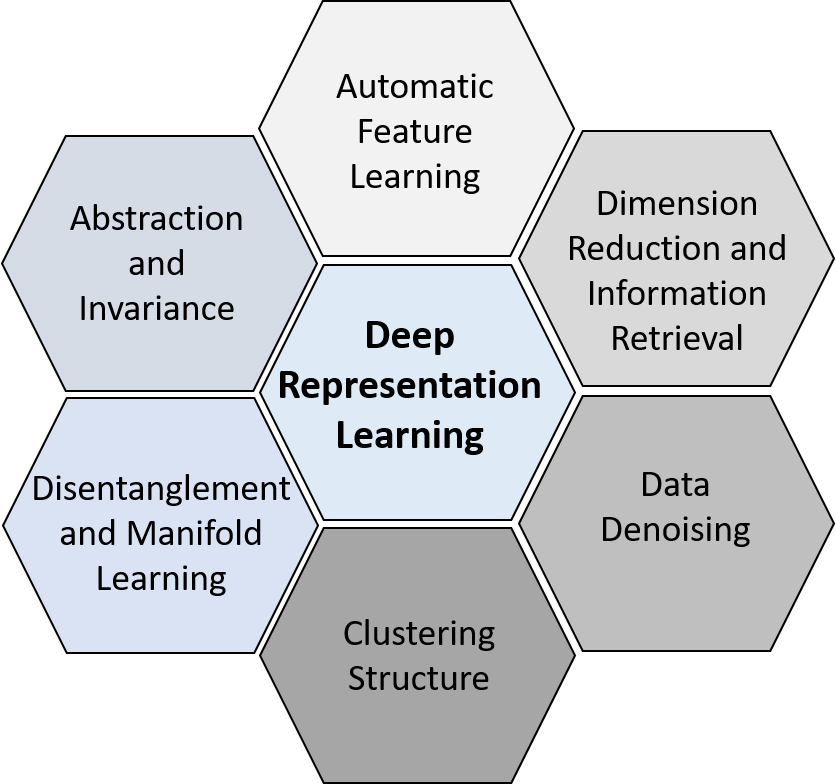}
%\captionsetup{width=0.95\textwidth}
\caption{Applications of deep representation learning.}
\label{fig:advantanges}
\vspace{-3mm}
\end{figure}

Learning representations is a fundamental problem in AI and it aims to capture useful information or attributes of data, where deep representation learning involves DL models for this task. Various applications of deep representation learning have been summarised in Figure~\ref{fig:advantanges}.

\subsection{Automatic Feature Learning}
 Automatic feature learning is the process of constructing explanatory variables or features that can be used for classification or prediction problems. Feature learning algorithms can be supervised or unsupervised \cite{goodfellow2016deep}. Deep learning (DL) models are composed of multiple hidden layers and each layer provides a kind of representation of the given data \cite{bengio2009learning}. It has been found that automatically 
 %BS learned --> learnt 
 learnt feature representations are
 %BS added
 -- given enough training data -- usually 
 more efficient and repeatable than hand-crafted or manually designed features which allow building better faster predictive models \cite{zhong2016overview}. Most importantly, 
 %BS corrected this wrong formulation - what you meant is imho not feature extraction, but:
 automatically learnt feature representation is in most cases 
 more flexible and powerful and can be applied to any data science problem in the fields of vision processing \cite{jing2019self}, text processing \cite{liang2017text}, and speech processing \cite{noda2015audio}. 

\subsection{Dimension Reduction and Information Retrieval}

Broadly speaking, dimensionality reduction methods are commonly used for two purposes: (1) to eliminate data redundancy and irrelevancy
%BS added
for higher efficiency and often increased performance
, 
%BS added
and 
(2) to make the data more understandable and interpretable by reducing the number of input variables \cite{zhong2016overview}. In some applications, it is very difficult to analyse the high dimensional data with a limited number of training samples \cite{usman2017using}. Therefore, dimension reduction becomes imperative to retrieve important variables or information relevant to the specified problems. It has been validated that the use of more interpretable features in a lower dimension can provide competitive performance or even better performance when used for designing predictive models  \cite{ latif2017variational}. 

Information retrieval is a process of finding information based on 
%BS the --> a
a user query by examining a collection of data \cite{mitra2018introduction}. The queried material can be text, 
%BS turned to plural 
documents, images, 
or audio, and users can express their queries in the form of a text, voice, or image \cite{mitra2017neural,kim2018one}. Finding a suitable representation of 
%BS:
a 
query to perform retrieval is a challenging task and DL-based representation learning techniques are playing 
%BS:
an 
important role in this field. The major advantages of using 
%BS turned to singular 
representation 
learning models for information retrieval is that they can learn features automatically with little or no prior knowledge \cite{sordoni2016learning}.

\subsection{Data Denoising}
Despite the success of deep models in different fields, these models remain brittle to the noise \cite{kurakin2016adversarial}. To deal with noisy conditions, 
%BS deleted researchers - also engineers and others do :)
one 
often performs data augmentation by adding artificially-noised examples to the training set \cite{liu2018towards}. However, data augmentation may not help always, because the distribution of noise is not always known. In contrast, representation learning methods can be effectively utilised to learn noise-robust features learning and they often provide better results compared to data augmentation \cite{latif2018adversarial}. 
%BS added:
In addition, the speech can be denoised such as by DL-based speech enhancement systems \cite{Liu19-NIT}.
%BS please add citation:

\subsection{Clustering Structure}
Clustering is one of the most traditional and frequently used data representation methods. It aims to categorise similar classes of data samples into one cluster using similarity measures (e.\,g., Euclidean distance). A large number of data clustering techniques have been proposed \cite{xu2005survey}. Classical clustering methods usually have poor performance on high-dimensional data, and suffer from high computational complexity on large-scale datasets \cite{min2018survey}. In contrast, DL-based clustering methods can process large and high dimensional data (e.\,g., images, text, speech) with a reasonable time complexity and they have emerged as effective tools for clustering structures \cite{min2018survey}. %They map the raw data into a low dimensional space by producing more suitable representations that make the most similar structure closer together. 

\subsection{Disentanglement and Manifold Learning}
Disentangled representation is a method that disentangles or represents each feature into narrowly defined variables and encodes them as separate dimensions \cite{bengio2013representation}. Disentangled representation learning differs from other feature extraction or dimensionality reduction techniques as it explicitly aims to learn such representations 
%BS corrected following word order and axis --> axes
that aligns axes with the generative factors of the input data \cite{dicarlo2007untangling,higgins2018towards}. Practically, data is generated from independent factors of variation. Disentangled representation learning aims to capture these factors by different independent variables in the representation. In this way, latent variables are interpretable, generalisable, and robust against adversarial attacks \cite{alemi2016deep}. 

Manifold learning aims to describe data as low-dimensional manifolds embedded in high-dimensional spaces \cite{hinton2003stochastic}. Manifold learning can retain a meaningful structure in very low dimensions compared to 
%BS deleted "the"
linear dimension reduction methods \cite{errity2006investigation}. Manifold learning algorithms attempt to describe the high dimensional data as a non-linear function of fewer underlying parameters by 
%BS deleted "its" 
preserving the intrinsic geometry \cite{cayton2005algorithms,golchin2014overview}. Such parameters have a widespread application in pattern recognition, speech analysis, and computer vision \cite{ma2011manifold}.

\subsection{Abstraction and Invariance}
The architecture of DNNs is inspired by the hierarchical structure of the brain \cite{dayan2001theoretical}. It is anticipated that deep architectures might capture abstract representations \cite{simpson2015abstract}. Learning abstractions is equivalent to discovering a universal model that can be across all tasks to facilitate generalisation and knowledge transfer. More abstract features are generally invariant to the local changes and are non-linear functions of the raw input \cite{cutler2008abstract}. Abstraction representations also capture high-level continuous-valued attributes that are only sensitive to some very specific types of changes in the input signal. Learning such sorts of invariant features 
%BS corrected:
has 
more predictive power which has always been required 
%BS corrected:
by the 
artificial intelligence (AI) community \cite{deng2018abstraction}.  

\section{Representation Learning Architectures}
\label{Architectures}
In 2006, DL-based 
%BS corrected 
automatic feature discovery 
was initiated by Hinton and his colleagues \cite{hinton2006reducing} and followed up by other researchers \cite{bengio2007greedy,poultney2007efficient}. %The central idea was to learn a hierarchy of feature representations in different layers of a deep model. 
This has led to a breakthrough in representation learning research and many novel DL models have been proposed. In this section, we will discuss these models and highlight the mechanics of representation learning using them.

\subsection{Deep Neural Networks (DNNs)}
Historically, the idea of deep neural networks (DNNs) is an extension of ideas emerging from research on artificial neural networks (ANNs) \cite{deng2014tutorial}. Feed Forward Neural Networks (FNNs) or Multilayer Perceptrons (MLPs) \cite{svozil1997introduction} with multiple hidden layers are indeed a good example of deep architectures. DNNs consist of multiple layers, including 
%BS added 
an 
input layer, hidden layers, and an output layer, of processing units called  ``neurons''. These neurons in each layer are densely connected with the neurons of the 
%BS corrected:
adjacent layers. 
%Each connection has some weight that is updated during training. 
The goal of DNNs is to approximate some function $f$. For instance, a DNN classifier maps 
%BS:
an input $x$ to a category label $y$ by using a mapping function $y=f(x;\theta)$ and learns the value of the parameters $\theta$ that result in the best function approximation. Each layer of a DNN performs representation learning based on the input provided to it. For example, in case of a classifier, all hidden layers except the last layer (\textbf{softmax}) learn a representation of input data to make classification task easier. A well trained DNN network learns a hierarchy of distributed representations \cite{bengio2009learning}. Increasing the depth of DNNs promotes reusing of learnt 
representations and 
%BS: 
enables 
the learning of 
%BS: 
a 
deep hierarchy of representations at different levels of abstraction.  Higher levels of abstraction are generally associated with invariance to local changes of the input \cite{bengio2013representation}. These representations 
%BS deleted "are"
proved very helpful in designing different speech-based systems.

\subsection{Convolutional Neural Networks (CNNs)}
Convolutional neural networks (CNNs) \cite{krizhevsky2012imagenet} are 
%BS 
a 
specialised kind of deep architecture for processing of data having a grid-like topology. Examples include image data that have 2D grid pixels and time-series data (i.\,e., 1D grid) having samples at regular intervals of time to create 
%BS: 
a 
grid-like structure.  CNNs are 
%BS: 
a variant of the standard FNNs. They introduce convolutional and pooling layers into the structure of DNNs, which take into account the spatial representations of the data and make the network more efficient by introducing sparse interactions, parameter sharing, and equivariant representations. The convolution operation in the convolution layer is the fundamental building block of CNNs. It consists of several learnable kernels that are convolved with the input to compute the output feature map.  This operation is defined as: 
\begin{equation}
    (h_{k})_{ij}= \big(W_{k}\otimes q\big) + b_{k},
\end{equation}
where $(h_k)_{ij}$ represents the $(i, j)^{th}$ element for the $k^{th}$ output feature map, $q$ is the input feature maps, $W_k$ and $b_k$
represent the $k^{th}$ filter and bias, respectively. The symbol $\otimes$ denotes the 2D convolution operation. After the convolution operation, 
%BS 
a 
pooling operation is applied, which 
%BS 
facilitates nonlinear downsampling of the feature map and 
%BS 
makes the representations invariant
to small translations in the input \cite{goodfellow2016deep}. Finally, it is common to use DNN layers to accumulate the outputs from the previous layers to yield a stochastic likelihood representation for classification or regression.

In contrast to DNNs, the training process of CNNs is easy due to fewer parameters \cite{lecun1995convolutional}. CNNs are found very powerful in extracting low-level representations at the initial layers and high-level features (textures and semantics) in the higher layers \cite{latif2019direct}. The convolution layer in CNNs acts as data-driven filterbank that is able to capture representations from speech \cite{palaz2015analysis} that are more generalised \cite{palaz2015convolutional}, discriminative \cite{latif2019direct}, and contextual  \cite{aldeneh2017using}. %CNNs are commonly used for 2D (image data), 1D (e.\,g., audio data), and 3D (e.\,g., video data) and have been extensively leveraged by the speech community. 

\subsection{Recurrent Neural Networks (RNNs)}
Recurrent neural networks (RNNs) \cite{sutskever2014sequence} are an extension of FNNs by introducing recurrent connections within layers. They use 
%BS 
the previous state of the model as additional input at each time step which creates a memory in its hidden state having information from all previous inputs. This makes RNNs to have stronger representational memory compared to hidden Markov models (HMMs), whose discrete hidden states bound their memory \cite{cho2014learning}.  Given an input sequence $x ({t})=(x_{1},.....,x_{T})$ at the current time step $t$, they calculates the hidden state $h_t$ using 
%BS
the 
previous hidden state $h_{t-1}$ and outputs a vector sequence $y=(y_{1},.....,y_{T})$. The standard equations for RNNs are given below:
\begin{equation}
h_{t}= \textit{H}(W_{xh}x_{t}+W_{hh}h_{t-1}+b_{h})
\end{equation}
\begin{equation}
y_{t}= (W_{xh}x_{t}+b_{y}), 
\end{equation}
where $W$ terms are the weight matrices (i.\,e., $W_{xh}$ is 
%BS 
a 
weight matrix of an input-hidden layer), $b$ is the bias vector, and $\textit{H}$ denotes the hidden layer function. Simple RNNs usually fail to model the long-term temporal contingencies due to the vanishing gradient problem. To deal with this problem, multiple specialised RNN architectures 
%BS
exist 
including long short-term memory (LSTM) \cite{hochreiter1997long} and gated recurrent units (GRUs) \cite{cho2014learning} with gating mechanism to add and forget the information selectively. Bidirectional RNNs \cite{schuster1997bidirectional} were proposed to model future context by passing the input sequence through two separate recurrent hidden layers. These separate recurrent layers are connected to the same output to access the temporal context in both directions to model both past and future.

RNNs introduce recurrent connections to allow parameters to be shared across time which makes them very powerful in learning temporal dynamics from sequential data (e.\,g., audio, video). Due to these abilities, RNNs especially LSTMs have had an enormous impact in speech community and they are incorporated in state-of-the-art ASR systems \cite{sainath2019two}.  
%BS IMPORTANT: We should add at least a mention on CTC here (Connectionist Temporal Classification) by Alex Graves.

\subsection{Autoencoders (AEs)}
The idea of 
%BS 
an 
autoencoding  network \cite{hinton1994autoencoders} is to learn a mapping from high-dimensional data to a lower-dimensional feature space such that the input observations can be approximately reconstructed from the lower-dimensional representation. The function $f_{\theta}$ is called the encoder that 
%BS 
maps the input vector $x$ into feature/representation vector $h=f_{\theta}(x)$. The decoder network is responsible to map 
%BS
a 
feature vector $h$ to 
%BS corrected: 
reconstruct the 
input vector $\hat{x}=g_{\theta}(h)$. The decoder network parameterises the decoder function $g_{\theta}$. Overall, the parameters are optimised by minimising the following cost function:
\begin{equation}
\label{AE}
    \mathcal{L}(x,g_{\theta}(f_{\theta}(x)))=\lVert{x-\hat{x}}\rVert_{2}^{2}.
\end{equation}

The set of parameters $\theta$ of the encoder and decoder networks are simultaneously learnt by attempting to incur a minimal reconstruction error. If the input data have correlated structures; then, the autoencoders (AEs) can learn some of these correlations \cite{usman2017using}. To capture useful representations $h$, the cost function of Equation \ref{AE} is usually optimised with an additional constraint to prevent the AE from learning the useless identity function having zero reconstruction error. This is achieved through various ways in the different forms of AEs, as discussed below in more detail.

% , The constraint can be imposed on either on the
% architecture of the autoencoder by limiting the feature $h$ dimension or by adding a regularisation term in the objective function. We will such autoencoders in next. 
\subsubsection{Undercomplete Autoencoders (AEs)}
One way of learning useful feature representations $h$ is to regularise the autoencoder by imposing constraints to have a low dimensional feature size. In this way, the AE is forced to learn the salient features/representations of data from high dimensional space to 
%BS from now on, I will simply do "a"/"the" additions / corrections - same for comma - it is too many to write "BS" each time :)
a low dimensional feature space. If an autoencoder uses a linear activation function with the mean squared error criterion; then, the resultant
architecture will become equivalent to the PCA algorithm, and its hidden units will learn the principal components of input data \cite{bengio2013representation}. However, an autoencoder with non-linear activation functions can learn a more useful feature representation compared to PCA \cite{hinton2006reducing}. 

\subsubsection{Sparse Autoencoders (AEs)}
An AE network can also discover a useful feature representation of data, even when the size of the feature representations is larger than the input vector $x$ \cite{usman2017using}. This is done by using the idea of sparsity regularisation \cite{poultney2007efficient} by imposing an additional constraint on the hidden units. Sparsity can be achieved either by penalising hidden unit biases \cite{poultney2007efficient} or the outputs' hidden unit, however, it hurts numerical optimisation. Therefore, imposing sparsity directly on the outputs of hidden units is very popular and has several variants. One way to 
%BS:
realise a sparse AEs is to incorporate an additional term in the loss function to penalise the KL divergence between average activation of the hidden unit and the desired sparsity ($\rho$) \cite{ng2011sparse}. Let us consider $a_{j}$ as the  activation of a hidden unit $j$ for a given input $x_{i}$
%BS changed throught the doc: ", then " --> "; then, "
; then, the average activation $\hat\rho$ over the training set is given by: 
\begin{equation}
   \hat\rho= \frac{1}{n}\sum_{i=1}^{n}\big[ a_{j} (x_{i})\big],
\end{equation}
where $n$ is the number of training samples. Then, the cost function of a sparse autoencoder will become: 
\begin{equation}
   \mathcal{L}(x,g_{\theta}(f_{\theta}(x)))+\lambda \sum_{i=1}^{n}\text{KL},  (\hat\rho\Vert \rho)
\end{equation}
where $\mathcal{L}(x,g_{\theta}(f_{\theta}(x)))$ is the cost function of the standard autoencoder. Another way to penalise a hidden unit is to use $l_{1}$ as  penalty by 
%BS phrase was bad. Added: 
which 
the following objective becomes: 
\begin{equation}
   \mathcal{L}(x,g_{\theta}(f_{\theta}(x)))+\lambda \lVert z \rVert_{1}. 
\end{equation}
Sparseness plays a key role in learning a more meaningful representation of input data \cite{ng2011sparse}. It has been found that sparse AEs are simple to train and can learn better representation compared to denoising autoencoders (DAE) and RBMs \cite{makhzani2013k}. In particular, sparse encoders can learn useful information and attributes from speech, which can facilitate better classification performance \cite{deng2013sparse}.  

\subsubsection{Denoising Autoencoders (DAEs)}

Denoising autoencoders (DAEs) are considered as a stochastic version of the basic AE. They are trained to reconstruct a clean input from its corrupted version \cite{vincent2008extracting}. The objective function of a  DAE is given by:
\begin{equation}
        \mathcal{L}(x,g_{\theta}(f_{\theta}(\tilde{x}))), 
\end{equation}
where $\tilde{x}$ is the corrupted version of $x$, which is done 
%BS 
via 
stochastic mapping $\tilde{x}\sim q_{D}(\tilde{x}|x)$. During training, DAEs are still minimising the same reconstruction loss between a clean $x$ and its reconstruction from $h$. The difference is that $h$ is learnt by applying a  deterministic mapping $f_{\theta}$ to a corrupted input $\tilde{x}$. It thus learns higher level feature representations that are robust to the corruption process. The features learnt by a DAE are reported qualitatively better for tasks like classification and also better than RBM features \cite{bengio2013representation}. 

\subsubsection{Contractive Autoencoders (CAEs)}
Contractive autoencoders (CAEs) proposed by Rifai et al.\ \cite{rifai2011contractive} with the motivation to learn robust representations 
%BS:
are similar to a DAEs. CAEs are forced to learn useful representations that are robust to infinitesimal input variations. This is achieved by adding an analytic contractive penalty to Equation \ref{AE}. The penalty term is the Frobenius norm of the Jacobian matrix of the hidden layer with respect to the input $x$. The loss function for a CAE is given by:
\begin{equation}
   \mathcal{L}(x,g_{\theta}(f_{\theta}(x)))+\lambda \sum_{i=1}^{m}\lVert\Delta_{x}z_{i}\rVert^2,
\end{equation}
where $m$ is the number of hidden units, and $z_i$ is the activation of hidden unit $i$.

\subsection{Deep Generative Models}
Generative models are powerful in learning the distribution of any kind of data (audio, images, or video) and aim to generate new data points. Here, we discuss four generative models due to their popularity in the speech community.

\subsubsection{Boltzmann Machines and Deep Belief Networks}
Deep Belief Networks (DBNs) \cite{hinton2006fast} are a powerful probabilistic generative model that consists of multiple layers of stochastic latent variables, where each layer is a Restricted Boltzmann Machine (RBM) \cite{ackley1985learning}. Boltzmann Machines (BM) are a bipartite graph in which visible units are connected to hidden units using undirected connections with weights. A BM is restricted in the sense that there are no hidden-hidden and visible-visible connections. %When RBMs are used as classifiers, they are trained on the joint distribution of input data and their labels. These labels are assigned to the new input which has the highest probability under the model. 
A RBM is an energy-based model whose joint probability distribution between visible layer ($v$) and hidden layer ($h$) is given by: 
\begin{equation}
  P(v,h)= \frac{1}{Z}exp(-E(v,h)).
\end{equation}
$Z$ is the normalising constant also known as the partition function, and $E(v, h)$ is an energy
function defined by the following equation: 
\begin{equation}
    E(v,h)=  -\sum_{i=1}^{D}\sum_{j=1}^{k}W_{ij}v_{i}h_{j}-\sum_{i=1}^{D}b_{i}v_{i}-\sum_{j=1}^{k}a_{j}h_{j}, 
\end{equation}
where $v_{i}$ and $h_{i}$ are the binary states of hidden and visible units. $W_{ij}$ are the weights between visible and hidden nodes. $b_{i}$ and $a_{j}$ represent the bias terms for visible and hidden units respectively. 
% The conditional probabilities for the visible and hidden units are given by the following equations: 

% \begin{equation}
%     P(v_{i}=1|h)= g\big(b_{i}^{v}+ \sum_{j}h_{j}W_{ij}\big)
% \end{equation}

% \begin{equation}
%     P(h_{j}=1|v)= g\big(b_{j}^{h}+ \sum_{i}v_{i}W_{ij}\big)
% \end{equation}
% where $g$ is the sigmoid function:
 
%  \begin{equation}
%      g(x)= \frac{1}{1+e^{-x}}
%  \end{equation}
%BS: changed everywhere learned --> learnt, et al. --> et al.\ , i.e. --> i.\,e., and same for e.g.

During the training phase, an RBM uses Markov Chain Monte Carlo (MCMC)-based algorithms \cite{hinton2006fast} to maximise the log-likelihood of the training data. %Because the RBM have intractable partition functions, therefore, techniques like  Gibbs sampling, Contrastive Divergence (CD), Persistent Contrastive Divergence (PCD) are used for training. 
% The stack of generatively pre-trained RBMs creates a powerful DBN model that can be fine-tuned for performance improvement. Training of DBNs 
% %BS:
% is 
% accomplished using a layer by layer training of RBMs from bottom to up. %The second layer (RBM) is trained to model the distribution defined by sampling (Gibbs, CD, PCD) the hidden units of the first RBM when data is given to the first RBM. 
% The value of weights between visible and hidden units is updated so that the energy between the two layers is at minimum. This procedure may be repeated indefinitely, to have as many layers of the DBN architecture
% %BS
% as needed
% . 
Training based on MCMC computes the gradient of the log-likelihood, which poses a significant learning problem \cite{bengio2014deep}, Moreover, DBNs are trained using layer-wise training that is also computationally inefficient. In recent years, generative models like GANs and VAEs have been proposed that can be trained via direct back-propagation and avoid the difficulties of MCMC based training. We discuss GANs and VAEs in more detail next.  

\begin{table*}[]
\scriptsize
\centering
\caption{Summary of some popular representation learning models.}
\begin{tabular}{|l|l|c|c|c|c|c|l|}
\hline
\multirow{2}{*}{\textbf{Model}} & \multirow{2}{*}{\textbf{Characteristics}}  & \multicolumn{5}{c|}{\textbf{Applications in Speech Processing}} & \multirow{2}{*}{\textbf{References}} \\ \cline{3-7}
 &    & \multicolumn{1}{c|}{\begin{tabular}[c]{@{}c@{}}\textit{Automatic} \\ \textit{Feature Learning}\end{tabular}} & \begin{tabular}[c]{@{}l@{}}\textit{Dimension} \\ \textit{Reduction}\end{tabular} & \begin{tabular}[c]{@{}l@{}}\textit{Clustering}\\ \textit{Structures}\end{tabular} & \textit{Disentanglement} & \multicolumn{1}{c|}{\begin{tabular}[c]{@{}c@{}}Data \\ \textit{Denoising}\end{tabular}} &                             \\ \hline
 
 DNNs& \begin{tabular}[c]{@{}l@{}}Good for learning a hierarchy of representations. They can learn \\invariant and discriminative representations. Features learnt\\ by DNNs are more generalised compared to traditional methods.\end{tabular} & \checkmark{}  &  \xmark &  \xmark   & \checkmark{} &\xmark & \cite{yu2013feature}  \\ \hline
 
CNNs & \begin{tabular}[c]{@{}l@{}}Originated from image recognition and were also extended for NLP \\and speech. They can learn a high-level abstraction from speech.\end{tabular} 
&  \checkmark{}   &    \xmark     &          \xmark    &\checkmark{}     &     \xmark &\begin{tabular}[c]{@{}l@{}} \cite{krizhevsky2012imagenet}\\\cite{lecun1995convolutional}\end{tabular}  \\ \hline

RNNs & \begin{tabular}[c]{@{}l@{}}Good for sequential modelling. They can learn temporal structures \\from speech and outperformed DNNs\end{tabular} 
&  \checkmark{}   &    \xmark     &          \xmark    &\checkmark{}     &     \xmark &\begin{tabular}[c]{@{}l@{}} \cite{cho2014learning}\\\cite{li2015constructing}\end{tabular}  \\ \hline

AEs & \begin{tabular}[c]{@{}l@{}}Powerful unsupervised representation learning models that\\ encode the data in sparse and compress representations.\end{tabular} 
&  \checkmark{}   &   \checkmark{}    &         \checkmark{}    &\checkmark{}    &     \checkmark{} &\begin{tabular}[c]{@{}l@{}}\cite{bengio2013representation}\\\cite{vincent2008extracting} \end{tabular}  \\ \hline

VAEs & \begin{tabular}[c]{@{}l@{}}Stochastic variational inference and learning model. Popular\\ in learning disentangled representations from speech.\end{tabular} 
&  \checkmark{}   &   \checkmark{}    & \checkmark{}    &\checkmark{}     &     \checkmark{} &\begin{tabular}[c]{@{}l@{}} \cite{kingma2013auto} \end{tabular}  \\ \hline

GANs& \begin{tabular}[c]{@{}l@{}}Game-theoretical framework and very powerful for data generation \\and robust to overfitting. They can learn disentangled representation\\ that are very suitable for speech analysis.\end{tabular} 
&  \checkmark{}   &   \checkmark{}     &       \checkmark{}     &\checkmark{}     &     \checkmark{} &\begin{tabular}[c]{@{}l@{}} \cite{goodfellow2014generative}\\\cite{radford2015unsupervised}  \end{tabular}  \\ \hline
\end{tabular}
\vspace{-2mm}
 \label{table:architectures}
\end{table*}
\subsubsection{Generative Adversarial Networks (GANs)}
%BS changed to networks. Also, decapitalised ...
Generative adversarial networks (GANs) \cite{goodfellow2014generative} use adversarial training to directly shape the output distribution of the network via back-propagation %and avoid the challenges that come with MCMC based training approaches. 
They include two neural networks---a generator, $G$, and a discriminator, $D$, which play a min-max adversarial game defined by the following optimisation program: 
\begin{equation}
    \underset{G}{\text{min}} \  \underset{D}{\text{max}} \quad \mathrm{E}_x[\log(D(x))] + \mathrm{E}_y[\log(1 - D(G(y)))].
\end{equation}

The generator, $G$, maps the latent vectors, $z$, drawn from some known prior, $p_{z}$ (e.\,g., Gaussian), to fake data points, $G(z)$. The discriminator, $D$, is tasked with differentiating between generated samples (fake), $G(z)$, and real data samples, $x$, (drawn from data distribution, $p_{\text{data}}$). The generator network, $G(z)$, is trained to maximally confuse the discriminator into believing that samples it generates come from the data distribution. This makes the GANs very powerful. They have become very popular 
%BS: added, as all here could be simply an overview for deep learning -- it lacks largely the connection to speech...
and are being exploited in various ways by speech community either for speech synhtesising or to augment the training material by generated feature observations or speech itself.
%BS end.
Researchers proposed various other architecture on the idea of GAN. These models include conditional GAN \cite{radford2015unsupervised}, BiGAN \cite{donahue2016adversarial}, InfoGAN \cite{chen2016infogan}, etc. These days, GAN based architectures are widely being used for representation learning not only from images but also from speech and related fields.

\subsubsection{Variational Autoencoders}
Variational Autoencoders (VAEs) are probabilistic models that use a stochastic encoder for modelling the posterior distribution $q(z|x)$, and a generative network (decoder) that models the conditional log-likelihood $ \text{log} p(x|z)$. Both of  these networks 
%BS added:
are 
%BS I am now adding such minor things w/o comment, as it is too many...
jointly trained to maximise the following variational lower bound on the data loglikelihood:

\begin{equation}
    \text{log} p(x)> \mathbb{E}_{q(z|x)} \text{log} p(x|z)- \text{KL}(q(z|x)||p(z)).
\end{equation}

The first term is the standard reconstruction term of an AE and the second term is the KL divergence between the prior $p(z)$ and the
posterior distribution $q(z|x)$.  The second term acts as a regularisation term and without it, the model is simply a standard autoencoder. VAEs are becoming very popular in learning representation from speech. Recently, various variants of VAEs are proposed in the literature, which include $\beta$-VAE \cite{higgins2017beta}, InfoVAE \cite{zhao2017infovae}, PixelVAE \cite{gulrajani2016pixelvae}, and many more \cite{tschannen2018recent}. All these VAEs are very powerful in learning disentangled and hierarchical representations and are also popular in clustering multi-category structures of data \cite{tschannen2018recent}.

\subsubsection{Autoregressive Networks (ANs)}
Autoregressive networks (ANs) are directed probabilistic models with no latent random variables. They model the joint distribution of high-dimensional data as a product of conditional distributions using the following probabilistic chain-rule:
\begin{equation}
    p(x)=\prod_{i=1}^{n^{2}}p(x_{i}|x_{<t},\theta), 
\end{equation}
where $x_{t}$ is the $t^{th}$ variable of $x$ and $\theta$ are the parameters of the AN model. The conditional probability distributions in ANs are usually modelled with a neural network that receives $x<t$ as input and outputs a distribution over possible $x_{t}$. Some of the popular ANs includes PixelRNN \cite{oord2016pixel}, PixelCNN \cite{van2016conditional}, and WaveNet \cite{oord2016wavenet}.  ANs are powerful density estimators and they capture details over global data without learning a hierarchical latent representation unlike latent variable models such as GANs, VAEs, etc. In speech technology, WaveNet is very popular and has powerful acoustic modelling capabilities. They are used for speech synthesis \cite{oord2016wavenet}, denoising \cite{rethage2018wavenet},  and also in unsupervised representation learning setting in conjunction with VAEs \cite{chorowski2019unsupervised}. 

%\subsection{Summary}

In this section, we have discussed DL models that use representation learning of speech. In Table \ref{table:architectures} we highlight the key characteristics of DL models in terms of their representation learning abilities.  All these models can be trained in different ways to learn useful representation from speech, which we have reviewed in the next section.

%%%%%%%%%%%%%%%%%%%%%%%%%%%%%%%%%%%%%%%%%%%%%%%%%%%%%%%%%%%%%%%%%%%%%%%%%%%%%%%%%%%%%

%%%%%%%%%%%%%%%%%%%%%%%%%%%%%%%%%%%%%%%%%%%%%%%%%%%%%%%%%%%%%%%%%%%%%%%%%%%%%%%%%%%%%

% \section{Speech Corpora and Features}
% \label{dataandFeatures}

%%%%%%%%%%%%%%%%%%%%%%%%%%%%%%%%%%%%%%%%%%%%%%%%%%%%%%%%%%%%%%%%%%%%%%%%%
%%%%%%%%%%%%%%%%%%%%%%%%%%%%%%%%%%%%%%%%%%%%%%%%%%%%%%%%%%%%%%%%%%%%%%%%%

\section{Techniques for Representation Learning From Speech}
\label{Learning}
\begin{figure}[t!]
\centering
%captionsetup{justification=centering}
\includegraphics[width=0.49\textwidth]{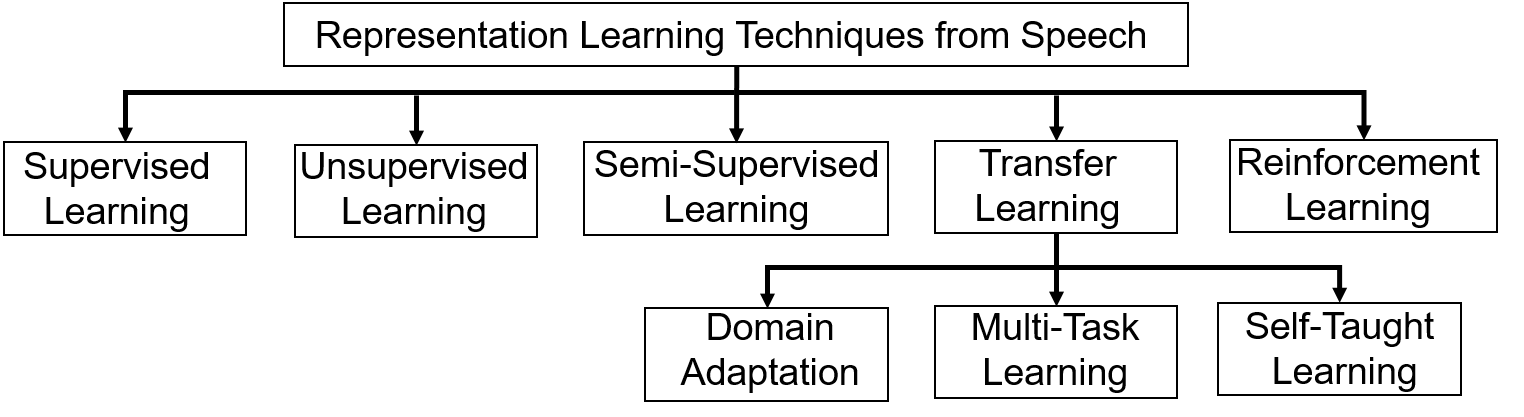}
%\captionsetup{width=0.95\textwidth}
\caption{Representation Learning Techniques.}
\label{fig:RL_taxonomy}
\end{figure}

Deep models can be used in different ways to automatically discover suitable representations for the task at hand, and this section covers these techniques for learning features from speech for ASR, SR, and SER. Figure \ref{fig:RL_taxonomy} shows the different learning techniques that can be used to capture information from data. These techniques have different important attributes that we highlight in Table \ref{mthods}.   

\begin{table*}[t!]
\centering
\scriptsize
\caption{Comparison of different types of representation learning.}
\label{mthods}
\begin{tabular}{|l|l|l|l|l|l|}
\hline
\textbf{Types/Aspect} & \begin{tabular}[c]{@{}l@{}}\textbf{Supervised Learning}\end{tabular} & \begin{tabular}[c]{@{}l@{}}\textbf{Unsupervised Learning}\end{tabular} & \begin{tabular}[c]{@{}l@{}}\textbf{Semi-Supervised Learning}\end{tabular} & \begin{tabular}[c]{@{}l@{}}\textbf{Transfer Learning}\end{tabular} & \begin{tabular}[c]{@{}l@{}}\textbf{Reinforcement Learning}\end{tabular} \\ \hline

\textit{Attributes}& 
\begin{tabular}[c]{@{}l@{}}\tabitem Learn explicitly\\ \tabitem Data with labels\\ \tabitem Direct feedback is given \\\tabitem Predict outcome/future \\\tabitem No exploration \end{tabular} 

& \begin{tabular}[c]{@{}l@{}}\tabitem Learn patterns and\\ structure\\ \tabitem Data without labels\\ \tabitem No direct feedback \\\tabitem No prediction \\\tabitem No exploration\end{tabular} 

& \begin{tabular}[c]{@{}l@{}}\tabitem Blend on both supervised \\and unsupervised\\ \tabitem Data with and without labels\\ \tabitem Direct feedback is given \\\tabitem Predict outcome/future \\\tabitem No exploration \end{tabular} 

& \begin{tabular}[c]{@{}l@{}}\tabitem Transfer knowledge from\\ one supervised task to other\\ \tabitem Labelled data for \\different task\\ \tabitem Direct feedback is given \\\tabitem Predict outcome/future\\ \tabitem No exploration \end{tabular} 
& \begin{tabular}[c]{@{}l@{}}\tabitem Reward based learning\\ \tabitem Policy making with \\feedback \\\tabitem Predict outcome/future \\\tabitem Adaptable to changes \\through exploration  \end{tabular}  \\ \hline

\textit{Uses}&
\begin{tabular}[c]{@{}l@{}}\tabitem Classification\\ \tabitem Regression\\ \end{tabular} 
& \begin{tabular}[c]{@{}l@{}}\tabitem Clustering\\ \tabitem Association\end{tabular}
& \begin{tabular}[c]{@{}l@{}}\tabitem Classification\\ \tabitem Clustering\end{tabular}
& \begin{tabular}[c]{@{}l@{}}\tabitem Classification\\ \tabitem Regression\end{tabular}
& \begin{tabular}[c]{@{}l@{}}\tabitem Classification\\ \tabitem Control\end{tabular}  \\ \hline
\end{tabular}
\end{table*}

\subsection{Supervised Learning}
Deep learning (DL) models can learn representations from data in both unsupervised and supervised manners. In the supervised case,
features representations are learnt on datasets by considering label information. In the speech domain, supervised representation learning methods are widely employed for feature learning from speech. RBMs and DBNs are found very capable in learning features from speech for different tasks including ASR \cite{jaitly2011learning,lee2009unsupervised}, speaker recognition \cite{ali2018speaker,yaman2012bottleneck}, and SER \cite{cairong2016novel}. Particularly, DBNs trained in a greedy layer-wise training \cite{hinton2006fast} can learn usef
ul features from the speech signal \cite{dahl2010phone}. 

Convolutional neural networks (CNNs) \cite{krizhevsky2012imagenet} are another popular supervised model that is widely used for feature extraction from speech. They have shown very promising results for speech and speaker recognition tasks by learning more generalised features from raw speech compared to ANNs and other feature-based approaches \cite{palaz2015analysis,palaz2015convolutional,muckenhirn2018towards}. After the success of CNNs in ASR, researchers also attempted to explore them for SER \cite{trigeorgis2016adieu,tzirakis2018end,sarma2018emotion,latif2019direct}, where they used CNNs 
%BS: 
in combination with LSTM networks for modelling long term dependencies in an emotional speech. Overall, it has been found that LSTMs 
%BS:
(or GRUs) 
can help CNNs in learning more useful features from speech  \cite{sainath2015learning,sainath2015convolutional}.    

Despite the promising results, the success of supervised learning is limited by the requisite of transcriptions or labels for speech-related tasks. It cannot exploit  the 
%BS:
plethora of 
freely available unlabelled datasets. It is also important to note that the labelling of these datasets is very expensive in terms of time and resources. To tackle these issues, unsupervised learning comes into play to learn representations from unlabelled data. We are discussing the potentials of unsupervised learning in the next section. 
\subsection{Unsupervised Learning}
Unsupervised learning facilitates the analysis of input data without corresponding labels and aims to learn the underlying inherent structure or distribution of data. In the real world,  data (speech, image, text) have extremely rich structures and algorithms trained in an unsupervised way to create understandings of data rather than learning for particular tasks. Unsupervised representation learning from large unlabelled datasets is an active area of research. In the context of speech analysis, it can exploit the practically available unlimited amount of unlabelled corpora to learn good intermediate feature representations, which can then be used to improve the performance of a variety of supervised tasks such as speech emotion recognition \cite{neumann2019improving}.

Regarding unsupervised representation learning, researchers mostly utilised variants of autoencoders (AEs)  to learn suitable features from speech data. AEs can learn high-level semantic content (e.\,g., phoneme identities) that are invariant to confounding low-level details (pitch contour or background noise) in speech \cite{chorowski2019unsupervised}. 

In ASR and SR, most of the studies utilised VAEs for unsupervised representation learning from speech \cite{chorowski2019unsupervised,hsu2019disentangling}. VAEs can jointly learn a generative model and an inference model, which allows them to capture latent variables from observed data. In \cite{hsu2017unsupervised}, the authors used FHVAE to capture interpretable and disentangled representations from speech without any supervision. They evaluated the model on two speech corpora and demonstrated that FHVAE can satisfactorily extract linguistic contents from speech and outperform an i-vector baseline speaker verification task while reducing WER for ASR. 
%BS IMPORTANT: a reference for the last claim perhaps?
Other autoencoding architectures like DAEs are also found very promising in finding speech representations in an unsupervised way. Most importantly, they can produce robust representation for noisy speech recognition \cite{feng2014speech,weninger2014deep,zhao2015music}. 

Similarly, classical models like RBMs have proved to be very successful for learning representation from speech. For instance, Jaitly and Hinton used RBMs for phoneme recognition in \cite{jaitly2011learning}, and showed that RBMs can learn more discriminative features that achieved better performance compared to MFCCs. %When RBMs are stacked on top of each other, this creates a new powerful model, called a Deep Belief Net (DBN)
%BS added:
%(see also above)
Interestingly, RBMs can also learn filterbanks from raw speech. In \cite{sailor2016unsupervised}, Sailor and Patil used a convolutional RBM (ConvoRBM)  to learn auditory-like sub-band filters from the raw speech signal. The authors showed that unsupervised deep auditory features learnt by ConvoRBM can outperform using Mel filterbank features for ASR. Similarly, DBNs trained on features such as MFCCs \cite{mohamed2009deep} or Mel scale filter banks \cite{dahl2010phone} create high-level feature representations. 

Similar to ASR and SR, models including AEs, DAEs, and VAEs are mostly used for unsupervised representation learning. In \cite{ghosh2016representation}, Ghosh et al.\ used stacked AEs for learning emotional representations from speech. They found that stacked AE can learn highly discriminative features from the speech that are suitable for the emotion classification task. Other studies \cite{ghosh2015learning,huang2015speech} also used AEs for capturing emotional representation from speech and found they are very powerful in learning discriminative features. DAEs are exploited in \cite{xia2014modeling,xia2013using} to show the suitability of DAEs for SER. In \cite{latif2017variational}, the authors used VAEs for learning latent representations of speech emotions. They showed that VAEs can learn better emotional representations suitable for classification in contrast to standard AEs. 

%BS added:
As outlined above, 
recently, adversarial learning (AL) is becoming very popular in learning unsupervised representation form speech. It involves more than one network and enables the learning in an adversarial way, which enables to learn more discriminative \cite{chang2017learning} and robust \cite{yu2017adversarial} features. Especially GANs \cite{donahue2018synthesizing}, adversarial autoencoders (AAEs)  \cite{sahu2018adversarial} and other AL \cite{ravanelli2018learning} based models are becoming popular in modelling speech not only in ASR but also SR and SER.  

Despite all these successes, the performance of a representation learnt in an unsupervised way is generally harder to compare with supervised methods. Semi-supervised representation learning techniques can solve this issue by simultaneously utilising both labelled and unlabelled data. We discuss semi-supervised representation learning methods in the next section. 

\subsection{Semi-supervised Learning}
The success in DL has predominately been enabled by key factors like advanced algorithms, processing hardware, 
%BS added: 
open sharing of codes and papers, 
and most importantly the availability of large-scale labelled datasets
%BS added:
and pre-trained networks on these
, e.\,g., ImageNet. However, a large labelled database or pre-trained network for every problem like speech emotion recognition is not always available \cite{latif2018transfer,latif2019unsupervised,latif2018cross}. It is very difficult, expensive, and time-consuming to annotate such data as it requires expert human efforts \cite{rana2019multi}. Semi-supervised learning solves this problem by utilising the large unlabelled data, together with the labelled data to build better classifiers. It reduces human efforts and provides higher accuracy, therefore, semi-supervised models are of great interest both in theory and practice \cite{zhu2005semi}.

Semi-Supervised learning is very popular in SER and researchers tried various models to learn emotional representation from speech.  Huang et al.\ \cite{Huang} used CNN in semi-supervised for capturing affect-salient representations and reported superior performance compared to well-known hand-engineered features. Ladder network-based semi-supervised methods are very popular in SER and used in \cite{huang2018speech,parthasarathy2019semi,parthasarathy2018ladder}. A ladder network is an unsupervised DAE that is trained along with a supervised classification or regression task. It can learn more generalised representations suitable for SER compared to the standard methods. Deng et al.\  \cite{deng2018semisupervised} proposed a semi-supervised model by combining an AE and a classifier. They considered unlabelled samples from unlabelled data as an extra garbage class in the classification problem. Features learnt by a semi-supervised AE performed better compared to an unsupervised AE. In \cite{rana2019multi}, the authors trained an AAE by utilising the additional unlabelled emotional data to improve SER performance. They showed that additional data help to learn more generalised representations that perform better compared to supervised and unsupervised methods.

In ASR, semi-supervised learning is mainly used to circumvent the lack of sufficient training data by creating features fronts ends \cite{thomas2013deep}, by using multilingual acoustic representations \cite{cui2015multilingual}, and by extracting an intermediate representation from large unpaired datasets \cite{karita2018semi} to improve the performance of the system. In SR, DNNs were used to learn representations for both the target speaker and interference for speech separation in a semi-supervised way \cite{tu2014speech}. Recently, a GAN based model is exploited for a speaker diarisation system with superior results using semi-supervised training \cite{pal2019study}.

\subsection{Transfer Learning}
Transfer learning (TL) involves methods that utilise any knowledge resources (i.e., data, model, labels, etc.) to increase model learning and generalisation for the target task \cite{bengio2012deep}. The idea behind TL is ``Learning to Learn'', which specifies that learning from scratch (\textit{tabula rasa learning}) is often limited, and experience should be used for deeper understanding \cite{thrun2012learning}. TL encompasses different approaches including multitask learning (MTL), model adaptation, knowledge transfer, covariance shift, etc. In the speech processing field, representation learning gained much interest in these approaches of TL. In this section, we cover three popular TL techniques that are being used in today's speech technology including domain adaptation, multi-task learning, and self-taught learning. 
 \subsubsection{Domain Adaptation}
Deep domain adaptation is a sub-field of TL and it has emerged to address the problem of unavailability of labelled data. It aims to eliminate the training-testing mismatch. Speech is a typical 
%BS:
example of 
heterogeneous data and a mismatch always exists between the probability distributions of source and target domain data, which can degrade the performance of the system \cite{sun2017unsupervised}. To build more robust systems for speech-related applications in real-life, domain adaptation techniques are usually applied in the training pipeline of deep models to learn representations that explicitly minimise the difference between the source and target domains. 

Researchers have attempted different methods of domain adaptation using representation learning to achieve robustness under noisy conditions in ASR systems. In \cite{sun2017unsupervised}, the authors used DNN based unsupervised representation method to eliminate the difference between the training and the testing data. They evaluate the model with clean training data and noisy test data and found that relative error reduction is achieved due to the elimination of mismatch between the distribution of train and test data. Another study \cite{swietojanski2016learning} explored the unsupervised domain adaptation of the acoustic model by learning hidden unit contributions. The authors evaluated the proposed adaptation method on four different datasets and achieved improved results compared to unadapted methods.  Hsu et al.\ \cite{hsu2017unsupervised} used a VAE based domain adaptation method to learn a latent representation of speech and create additional labelled training data (source domain) having a distribution similar to the testing data (target domain). The authors were able to reduce the absolute word error rate (WER) by 35\,\% in contrast to a non-adapted baseline. Similarly, in \cite{hsu2018extracting} domain invariant representations are extracted using Factorised Hierarchical Variational Autoencoder (FHVAE) for robust ASR. Also, some studies \cite{tang2018study,hsu2018unsupervised} explored unsupervised representation learning-based domain adaptation for distant conversational speech recognition. They found that a representations-learning-based approach outperformed unadapted models and other baselines. For unsupervised speaker adaptation, Fan et al.\  \cite{fan2016unsupervised} used multi-speaker DNNs to takes advantage of shared hidden representation and achieved improved results. 

Many researchers exploited DNN models for learning transferable representations in multi-lingual ASR \cite{qin2018towards,huang2013cross}. Cross-lingual transfer learning is important for the practical application, and it has been found that learnt features can be transferred to improve the performance of both resource-limited and resource-rich languages \cite{thomas2013deep}. The representations learnt in this way are referred to as bottleneck features and these can be used to train models for languages even without any transcriptions \cite{knill2014language}. Recently, adversarial learning of representations for domain adaptation methods is becoming very popular. Researchers trained different adversarial models and were able to improve the robustness against noise \cite{denisov2018unsupervised}, adaptation of acoustic models for accented speech \cite{sun2018domain,shinohara2016adversarial}, gender variabilities \cite{hosseini2018multi}, and speaker and environment variabilities \cite{meng2018adversarial,meng2018speaker,meng2019adversarial,tripathi2018adversarial}. These studies showed that representation learning using the adversarially trained models can improve the ASR performance on unseen domains. 

In SR, Shon et al.\ used DAE to minimise the mismatch between the training and testing domain by utilising out-of-domain information  \cite{shon2017autoencoder}. Interestingly, domain adversarial training is utilised by Wang et al.\  \cite{wang2018unsupervised} to learn speaker-discriminative representations. Authors empirically showed that the adversarial training help to solve dataset mismatch problem and outperform other unsupervised domain adaptation methods. Similarly, a GAN is recently utilised by Bhattacharya et al.\ \cite{bhattacharya2019generative} to learn speaker embeddings for a domain robust end-to-end speaker verification system. They achieved significantly better results over the baseline. 

In SER, domain adaptation methods are also very popular to enable the system to learn representations that can be used to perform emotion identification across different corpora and different languages. Deng et al.\ \cite{deng2014autoencoder} used AE with shared hidden layers to learn common representations for different emotional datasets. These authors were able to minimise the mismatch between different datasets and able to increase the performance. In another study \cite{deng2017universum}, the authors used a Universum AE for cross-corpus SER. They were able to learn more generalised representations using the Universum AE, which achieves promising results compared to standard AEs.  Some studies exploited GANs for SER.  For instance, Wang et al.\ \cite{wang2018unsupervised} used adversarial training to capture common representations for both the source and target language data. Zhou et al.\ \cite{zhou2018transferable} used a class-wise domain adaptation method using adversarial training to address cross-corpus mismatch issues and showed that adversarial training is useful when the model is to be trained on target language with minimal labels.  Gideon et al.\ \cite{gideon2019barking} used an adversarial discriminative domain generalisation method for cross-corpus emotion recognition and achieved better results. Similarly, \cite{latif2019unsupervised} utilised GANs in an unsupervised way to learn language invariant, and evaluated over four different language datasets. They were able to significantly improve the SER across different language using language invariant features. 

\subsubsection{Multi-Task Learning}
Multi-task learning (MTL) has led to successes in different applications of ML, from NLP \cite{collobert2008unified} and speech recognition \cite{deng2013new} to computer vision \cite{girshick2015fast}. MTL aims to optimise more than one loss function in contrast to single-task learning and uses auxiliary tasks to improve 
%BS:
on the main task of interest 
\cite{caruana1998learning}. %Multi-task representation learning can improve the performance of the main task by capturing underlying relevant common factors from the data \cite{bouguelia2017multi,bengio2013representation}. 
Representations learned in MTL scenario become more generalised, which are very important in the field of speech processing, since speech contains multi-dimensional information (message, speaker, gender, or emotion) that can be used as auxiliary tasks. As a result, MTL increases performance without requiring external speech data. 

In ASR, researchers have used MTL with different auxiliary tasks including gender \cite{stadermann2005multi}, speaker adaptation \cite{huang2015rapid,price2014speaker}, speech enhancement \cite{lu2004multitask,chen2015speech}, etc. Results in these studies have shown that learning shared representations for different tasks act as complementary information about the acoustic environment and gave a lower word error rate (WER). Similar to ASR, researchers also explored MTL in SER with significantly improved results \cite{rana2019multi,han2019emobed}. For SER, studies used emotional attributes (e.\,g., arousal and valance) as auxiliary tasks \cite{kim2017towards,parthasarathy2017jointly,xia2017multi,Lotfian2018} as a way to improve the performance of the system. Other auxiliary tasks that researchers considered in SER are speaker and gender recognition \cite{tao2018advanced,rana2019multi,zhang2017cross} to improve the accuracy of the system compared to single-task learning. 

MTL is an effective approach to learn shared representation  that leads to no major increase of the computational power, while it improves the recognition accuracy of a system and also decreases the chance of overfitting \cite{ranapironkov2016multi,rana2019multi}. However, MTL implies the preparation of labels for considered auxiliary tasks. Another problem that hinders MTL is dealing with temporality differences among tasks. For instance, the modelling of speaker recognition requires different temporal information than phenom recognition does  \cite{pironkov2016multi}. Therefore, it is viable to use memory-based deep neural networks like the recurrent networks---ideally with LSTM or GRU cells---to deal with this issue. 

\subsubsection{Self-Taught Learning}
Self-taught learning \cite{raina2007self} is a new paradigm in ML, which combines semi-supervised and TL. It utilises both labelled and unlabelled data, however, unlabelled data do not need to belong to the same class labels or generative distribution as the labelled data. %The unlabelled data are used as a source from which the knowledge is transferred to the tasks performed on labelled target data.
Such a loose restriction on unlabelled data in self-taught learning significantly simplifies learning from a huge volume of unlabelled data. This fact differentiates it from semi-supervised learning. 

We found very few studies on audio based applications using self-taught learning. In  \cite{ons2013self}, the authors used self-taught learning and developed an assistive vocal interface for users with a speech impairment. The designed interface is maximally adapted using self-taught learning to the end-users and can be used for any language, dialect, grammar, and vocabulary. In another study \cite{feng2018autoencoder}, the authors proposed an AE-based sample selection method using self-taught learning. They selected highly relevant samples from unlabelled data and combined with training data. The proposed model was evaluated on four benchmark datasets covering computer vision, NLP, and speech recognition with results showing that the proposed framework can decrease the negative transfer while improving the knowledge transfer performance in different scenarios.

% This framework includes two modules including a source sample re-weighting module and a classifier. In the first module, each unlabelled source sample is assigned a weight based on its relevance to labelled target samples. In the second module, source samples having larger relevance weights are combined with the training set of target data for training a classifier. They evaluated the proposed model on four benchmark datasets covering computer vision, NLP, and speech recognition with results showing that the proposed framework can decrease the negative transfer while improving the knowledge transfer performance in different scenarios. 

\subsection{Reinforcement Learning}
Reinforcement Learning (RL) follows the principle of behaviourist psychology where an agent learns to take actions in an environment and tries to maximise the accumulated reward over its lifetime. In a RL problem, the agent and its environment can be modelled being in a state $s$ $\in$ $S$ and the agent can perform actions $a$ $\in$ $A$, each of which may be members of either discrete or continuous sets and can be multi-dimensional. A state $s$ contains all related information about the current situation to predict future states. The goal of RL is to find a mapping from states to actions, called policy $\pi$, that picks actions $a$ in given states $s$ by maximising the cumulative expected reward. The policy $\pi$ can be deterministic or probabilistic. %The RL agent is tasked to discover the relations between states, actions, and rewards. Hence, exploration is required which can either be directly embedded in the policy or performed separately and only as part of the learning process. 
RL approaches are typically based on the Markov decision process (MDP) consisting of the set of states $S$, the set of actions $A$, the rewards $R$, and transition probabilities $T$ that capture the dynamics of a system. RL has been repeatedly successful in solving various problems \cite{kober2013reinforcement}. Most importantly, deep RL that combines deep learning with RL principles. Methods 
%BS:
such as deep Q-learning 
have significantly advanced the field \cite{arulkumaran2017deep}. 

Few studies used RL-based approaches to learn representations. For instance, in \cite{gelada2019deepmdp}, the authors introduced DeepMDP, a parameterised latent space model that is trained by minimising two tractable latent space losses including prediction of rewards and prediction of the distribution over the next latent states. They showed that the optimisation of these two objectives guarantees the quality of the embedding function as a representation of the state space. They also show that utilising DeepMDP as an auxiliary task in the Atari 2600 domain leads to large performance improvements. 
Zhang et al.\ \cite{zhang2018learning} used RL for learning optimised structured representation learning from text. They found that an RL model can learn task-friendly representations by identifying task-relevant structures without any explicit structure annotations, which yields competitive performance. %Different other studies \cite{van2017hybrid,jaderberg2016reinforcement} also exploited RL-based approaches for representation learning and achieve considerable success.    

Recently, RL is also gaining interest in the speech community and researchers have proposed multiple approaches to model different speech problems. Some of the popular RL-based solutions include dialog modelling and optimisation  \cite{cuayahuitl2006reinforcement,li2016deep}, speech recognition \cite{lee1990corrective}, and emotion recognition \cite{lakomkin2018emorl}. However, the problem of representation learning of speech signals is not explored using RL.

%BS: IMPORTANT: What about Active Learning and Cooperative Learning?

\subsection{Active Learning and Cooperative Learning}
Active Learning aims at achieving improved accuracy with fewer training samples by selecting data from which it learns. This idea of cleverly picking training samples rather than random selection gives better predictive models with less human effort for labelling data \cite{zhang2017active}. An active learner selects samples from a large pool of unlabelled data and subsequently asks queries to an oracle (e.g., human annotator) for labelling. 
% Most of the active learning methods select either representative or informative samples for the query.  Selecting the most representative samples for the overall input patterns of unlabelled data usually gives better results when there are no or few initial labelled samples \cite{wang2015querying}. In this case, the representation mechanism used to represent data while performing active learning has a significant role on the performance \cite{zhang2017active}. 
In speech processing, accurate labelling of speech utterances is extremely important and time-consuming. It has larger abundantly available unlabelled data. In this situation, active learning can help by allowing the learning model to select the samples from it learns, which leads to better performance with less training. Studies (e.g., \cite{hakkani2004unsupervised,riccardi2005active}) utilised classical ML-based active learning for ASR with the aim to minimise the effort required in transcribing and labelling data. However, it has been showed in \cite{huang2016active,zhang2015cooperative} that utilisation of deep models for active learning in speech processing can improve the performance and significantly reduce the requirement of labelled data. 

%Recently, models like GANs are found very promising for active learning in image domain \cite{zhu2017generative}, however, such powerful representation learning models need to be investigated for speech processing. 

%\subsection{Cooperative Learning}
Cooperative learning  \cite{zhang2015cooperative,dong2003human} combines active and semi-supervised learning to best exploit available data. It is an efficient way of sharing the labelling work between human and machine which leads to reduce the time and cost of human annotation \cite{schuller2015speech}. In cooperative learning, predicted samples with insufficient confidence value are subjected to human annotations and other with with high confidence values are labelled by machines. The models trained via cooperative learning perform better compared to active or semi-supervised learning \cite{zhang2015cooperative}. In speech processing, a few studies utilised ML-based cooperative learning and showed its potential to significantly reduce data annotation efforts. For instance, \cite{wagner2018applying}, authors applied cooperative learning speed up the process of annotation of large multi-modal corpora. Similarly, the proposed model in \cite{zhang2015cooperative} achieved the same performance with 75\% fewer labelled instances compared to the model trained on the whole training data. These finding shows the potential of cooperative learning in speech processing, however, DL-based representation learning methods need to be investigated in this setting.

\subsection{Summarising the Findings}
A summary of various representation learning techniques has been presented in Table \ref{review}. We segregated the studies based on the learning techniques used to train the representation learning models. Studies on supervised learning methods typically use models to learn discriminative and noise-robust representations. Supervised training of models like CNNs, LSTM/GRU RNNs, and CNN-LSTM/GRU-RNNs are widely exploited for learning of representations from raw speech. %The reason for their popularity is that CNN layers act as data-driven filterbanks that can model spectral envelope of raw speech and LSTM/GRU-RNNs can model contextual information. Therefore, most of the studies on raw speech either used CNN, LSTM/GRU-RNNs, or their combination for three speech applications covered in this study. 
%BS: LSTM --> LSTM/GRU RNN (note that LSTM is only the cell type...)

Unsupervised learning is to learn patterns in the data. We covered unsupervised representation learning for three speech applications. Autoencoding networks are widely used in unsupervised feature learning from speech. Most importantly,  DAEs are very popular due to their denoising abilities. They can learn high-level representations from speech that are robust to noise corruption. Some studies also exploited AEs and RBM based architectures for unsupervised feature learning due to their non-linear dimension reduction and long-range of features extraction abilities. Recently, VAEs are becoming very popular in learning speech representation due to their generative nature and distribution learning abilities. They can learn salient and robust features from speech that are very essential for speech applications including ASR, SR, and SER. 

Semi-supervised representation learning models are widely used in SER because speech corpora have smaller sizes compared to ASR and SR. These studies tried to exploit additional data to improve the performance of SER. The popular models include AE-based models \cite{deng2018semisupervised,rana2019multi} and other discriminative architectures \cite{zhang2018leveraging,Huang}. %Semi-supervised GANs are also exploited in \cite{chang2017learning} for SER to learn better quality of speech representations.
In ASR, semi-supervised learning is mostly exploited for learning noise-robust representations \cite{tu2015speech,narayanan2014investigation} and for feature extraction \cite{thomas2013deep,cui2015multilingual,liu2014graph}.  %Architectures like DNNs, AEs, and VAEs are widely used for semi-supervised representation learning in ASR. 

Transfer learning methods---especially domain adaptation and MTL---are very popular in ASR, SR, and SER. The domain adaptations methods in ASR, SER, and SR  are mainly used to achieve adaptation by learning such representations that are robust against noise, speaker, and language and corpus difference. %In SER, domain adaptation methods also used to learn such representations that can be used for emotion identification in cross-corpus and cross-languages. Similarly, domain adaptive representation learning models are also used in SR to minimise the mismatch between training and testing domains. 
In all these speech applications,  adversarially learnt representations are found to better solve the issue of domain mismatch. MTL methods are most popular in SER, where researchers tried to utilise additional information available (e.\,g., speaker, or gender) in speech to learn more generalised representations that help to improve the performance. %Among these studies, networks like DNNs and LSTM/GRU-RNNs are most popularl, however, also models based on adversarial training have been used.  
%BS the above paragraph was particularly full of English mistakes :)

% Overall, from this literature search, we have observed that representation learning is crucial and it has been widely studied in the speech community. We present our literature research in a condensed form in Table \ref{review} by highlighting different representation learning techniques and popular models used by researchers in three speech applications. Among these models, LSTM/GRU-RNN-based architectures are the mostly applied models due to their temporal context capturing abilities. VAEs are widely used in learning speech representations in an unsupervised manner. Interestingly, in all three speech applications, recent papers are mostly focused on generative models and adversarial representation learning is gaining significant traction not only in an unsupervised manner, but also as semi-supervised methods, in domain adaptation, and MTL scenarios.    

% \usepackage{multirow}
\begin{table*}[ht!]
\centering
\caption{Review of representation learning techniques used in different studies. }
\begin{tabular}{|l|c|l|l|}
\hline
Learning Type         & \multicolumn{1}{l|}{Applications} & Aim & Models        \\ \hline
\multirow{6}{*}{Supervised}                                                         & ASR                           & \multirow{3}{*}{\begin{tabular}[c]{@{}l@{}}To learn discriminative and robust\\ representation\end{tabular}}
& \begin{tabular}[c]{@{}l@{}}DBNs (\cite{mohamed2010investigation,mohamed2011deep,dahl2010phone,gholamipoor2014feature,huang2013audio,hu2016dbn,yu2010roles}), DNNs (\cite{yu2013feature,yu2011improved,wu2015deep,yu2012exploiting,dighe2016exploiting}),\\ CNNs \cite{abdel2014convolutional,sainath2015convolutional,mitra2015time}, \\GANs (\cite{serdyuk2016invariant})\end{tabular} \\ \cline{2-2} \cline{4-4}
& SR &   & \begin{tabular}[c]{@{}l@{}}DBNs (\cite{campbell2014using,ali2018speaker,ghahabi2014vector,huang2013audio,vasilakakis2013speaker,yamada2013improvement}), DNNs \cite{heigold2016end,heo2017joint,huang2015investigation,snyder2018x,icsik2015s}, \\CNNs (\cite{chung2018voxceleb2,zhang2016end,zhang2017end,mclaren2014application,lukic2016speaker,ranjan2017improved}), LSTM (\cite{wang2017does,shon2018frame,marchi2018generalised,mclaren2015advances})\\ CNN-RNNs (\cite{li2017deep}), GANs \cite{michelsanti2017conditional}\end{tabular} \\ \cline{2-2} \cline{4-4} 
 & SER&                            
  & \begin{tabular}[c]{@{}l@{}}DNNs (\cite{lorenzo2018investigating,wang2017learning,stolar2014optimized}), RNNs (\cite{lee2015high,huang2016attention}),\\ CNNs (\cite{kim2017learning,zheng2015experimental,mao2014learning,li2018attention}), \\CNN-RNNs (\cite{zhao2018exploring,luo2018investigation,lim2016speech})\end{tabular} \\ \cline{2-4} 
   & ASR& \multirow{3}{*}{To learn representation from raw speech}
   & \begin{tabular}[c]{@{}l@{}}CNNs (\cite{palaz2015analysis,palaz2015convolutional,palaz2013estimating,palaz2015analysis,kabil2018learning,golik2015convolutional,dai2017very}), \\CNN-LSTM (\cite{sainath2015learning,liu2018deep,zeghidour2018learning,zazo2016feature})\end{tabular} \\ \cline{2-2} \cline{4-4} 
   & SR &
 & \begin{tabular}[c]{@{}l@{}}CNNs (\cite{ravanelli2018speaker,jung2018avoiding,bengiospeaker}), CNN-LSTM (\cite{jung2018complete})\end{tabular} \\ \cline{2-2} \cline{4-4} 
  & SER &
  & \begin{tabular}[c]{@{}l@{}}CNN-LSTM (\cite{trigeorgis2016adieu,tzirakis2018end,latif2019direct}),\\ DNN-LSTM \cite{sarma2018emotion,sarma2019improving}\end{tabular} \\ \hline

\multirow{6}{*}{Unsupervised} 
& ASR
& \multirow{3}{*}{\begin{tabular}[c]{@{}l@{}}To learn speech feature and and noise\\ robust representation.\end{tabular}}
& \begin{tabular}[c]{@{}l@{}}DBNs (\cite{lee2009unsupervised}), DNNs (\cite{lu2013combining}) CNNs (\cite{hau2011exploring}),\\ LSTM (\cite{chung2019unsupervised}) AEs (\cite{chung2016audio}), VAEs (\cite{hsu2018unsupervised})\\DAEs (\cite{feng2014speech,ishii2013reverberant,weninger2014deep,xia2013speech,zhao2015music})\end{tabular} \\ \cline{2-2} \cline{4-4}

  & SR  & 
  & \begin{tabular}[c]{@{}l@{}}DBNs (\cite{lee2009unsupervised}), DAEs (\cite{zhang2015deep}), \\VAEs (\cite{van2017neural,hsu2017unsupervised,hsu2018hierarchical,hsu2019disentangling}), AL (\cite{ravanelli2018learning}), GAN (\cite{pal2019speaker})\end{tabular} \\ \cline{2-2} \cline{4-4} 
    & SER   &
    & \begin{tabular}[c]{@{}l@{}}AEs (\cite{ghosh2016representation,ghosh2015learning,cibau2013speech,huang2015speech}), DAEs \cite{xia2014modeling,xia2013using}, \\VAEs \cite{latif2017variational,eskimez2018unsupervised}, AAEs (\cite{sahu2018adversarial}), GANs (\cite{sahu2019modeling})\end{tabular} \\ \cline{2-4}
    
 & ASR& \multirow{1}{*}{To learn feature from raw speech.}
 & \begin{tabular}[c]{@{}l@{}}RBMs (\cite{sailor2016unsupervised}), VAEs \cite{chorowski2019unsupervised}, GANs (\cite{donahue2018synthesizing})\end{tabular}  \\ \hline

 \multirow{3}{*}{\begin{tabular}[c]{@{}l@{}}Semi-Supervised\\ Learning\end{tabular}} 
 & ASR  & \multirow{3}{*}{\begin{tabular}[c]{@{}l@{}}To learn speech feature representations \\ in semi-supervised way.\end{tabular}}
 
 & \begin{tabular}[c]{@{}l@{}}DNNs (\cite{thomas2013deep,cui2015multilingual}), AEs (\cite{karita2018semi,chung2019semi}), GANs \cite{zhao2018wasserstein}\end{tabular} \\ \cline{2-2} \cline{4-4} 
 & SR   &
 & \begin{tabular}[c]{@{}l@{}}DNNs (\cite{tu2014speech}), GANs \cite{pal2019study}\end{tabular} \\ \cline{2-2} \cline{4-4} 
     & SER   &   
& \begin{tabular}[c]{@{}l@{}}DNNs(\cite{zhang2018leveraging}), CNNs (\cite{Huang}), AEs (\cite{deng2018semisupervised}), \\AAEs (\cite{rana2019multi})\end{tabular} \\ \hline
 
\multirow{3}{*}{\begin{tabular}[c]{@{}l@{}}Domain\\ Adaptation\end{tabular}}        
& ASR  & \multirow{3}{*}{\begin{tabular}[c]{@{}l@{}}To learn representation to minimise the\\ acoustic mismatch between the training\\ and testing conditions.\end{tabular}} 

& \begin{tabular}[c]{@{}l@{}}DNNs \cite{huang2016unified,swietojanski2012unsupervised}, AEs (\cite{tang2018study}), VAEs (\cite{hsu2017unsupervised}) \end{tabular} \\ \cline{2-2} \cline{4-4} 
   & SR & 
   & \begin{tabular}[c]{@{}l@{}}AL (\cite{wang2018unsupervised}), DAE (\cite{shon2017autoencoder}), GANs (\cite{bhattacharya2019generative})\end{tabular} \\ \cline{2-2} \cline{4-4} 
   & SER     &  
   & \begin{tabular}[c]{@{}l@{}}DBNs \cite{latif2018crosscorpus}, CNNs (\cite{neumann2018cross}), AL (\cite{abdelwahab2018domain,tu2019towards,zhou2018transferable}),\\ AEs \cite{deng2014introducing,deng2017universum,deng2014autoencoder,deng2013sparse,deng2017recognizing}\end{tabular} \\ \hline

\multirow{3}{*}{\begin{tabular}[c]{@{}l@{}}Multi-Task\\ Learning\end{tabular}}      
& ASR   & \multirow{3}{*}{\begin{tabular}[c]{@{}l@{}}To learn common representations\\ using multi-objective training.\end{tabular}} 

& \begin{tabular}[c]{@{}l@{}} DNNs \cite{xu2017multi,seltzer2013multi}, RNNs (\cite{tan2016speaker,li2015modeling,tang2017collaborative}), AL (\cite{shinohara2016adversarial})\end{tabular} \\ \cline{2-2} \cline{4-4} 
  & SR &
  & \begin{tabular}[c]{@{}l@{}}DNNs (\cite{qian2016noise,chen2015multi}), CNNs (\cite{yadav2018learning}), RNNs (\cite{tang2016multi})\end{tabular} \\ \cline{2-2} \cline{4-4} 
   & SER &      
   & \begin{tabular}[c]{@{}l@{}}DBNs (\cite{xia2015multi}), DNNs (\cite{zhang2017multi,parthasarathy2017jointly,ma2018speech,Lotfian2018}), \\CNN-LSTM \cite{zhang2019attention}, LSTM (\cite{eyben2012multitask,le2017discretized,tao2018advanced,kim2017towards}), \\AL (\cite{li2019speaker}), GANs (\cite{chang2017learning})\end{tabular} \\ \hline

\end{tabular}
\label{review}
\end{table*}

\section{Challenges For Representation Learning}
\label{Challanges}
\begin{figure}[!ht]
\centering
%captionsetup{justification=centering}
\includegraphics[width=0.4\textwidth]{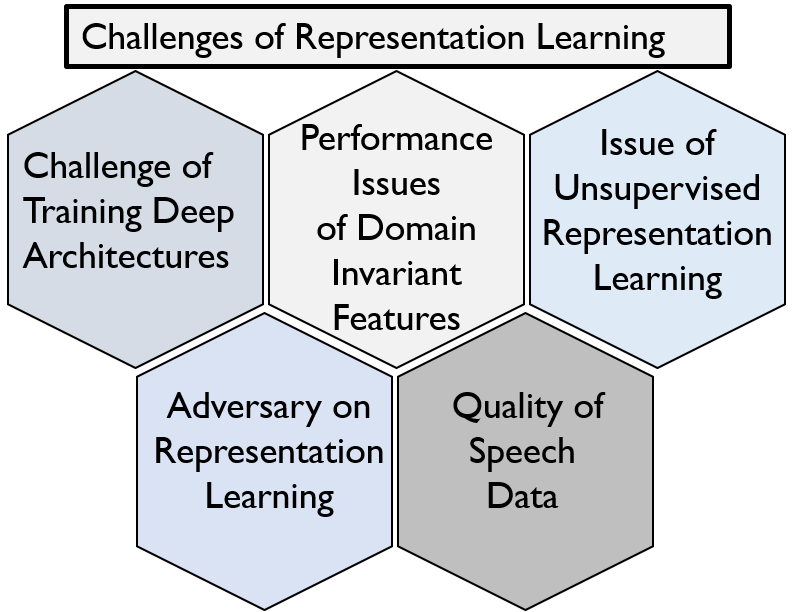}
%\captionsetup{width=0.95textwidth}
\caption{Challenges of representation learning.}
\label{fig:challenges}
\vspace{-3mm}
\end{figure}

In this section, we discuss the challenges faced by representation learning. The summary of these challenges is presented in Figure \ref{fig:challenges}.
\subsection{Challenge of Training Deep Architectures}
Theoretical and empirical evidence show the deep models 
%BS:
usually 
have superior performance over classical machine learning techniques. It is also empirically validated that deep learning models require much more data to learn certain attributes efficiently \cite{srivastava2015training}. For instance, adding top-5 layers in the network on the 1000-class ImageNet dataset
%BS
trained network 
has increased the accuracy from $\sim$84\,\% \cite{krizhevsky2012imagenet} to $\sim$95\,\%  \cite{szegedy2015going}. However, training deep  learning models is not straightforward; it becomes considerably more difficult to optimise a deeper network \cite{saxe2013exact,he2015delving}. For deeper models, network parameters become very large and tuning of different hyper-parameter is also very difficult. Due to the availability of modern graphics processing units (GPUs) and recent advancement in optimisation \cite{jia2017improving} and training strategies \cite{salimans2016weight,ioffe2015batch}, the training of DNNs considerably accelerated; however, it is still an open research problem. 

Training of representation learning models is a more tricky and difficult task. Learning high-level abstraction means more non-linearity and learning representations associated with input manifolds becoming even more complex if the model might need to unfold and distort complicated input manifolds. Learning such representation which involves disentangling and unfolding of complex manifolds requires more intense and difficult training \cite{bengio2013representation}. Natural speech has very complex manifolds \cite{li2017unsupervised}) and inherently contains information about the message, gender, age, health status, personality, friendliness, mood, and emotion. All of this information is entangled together \cite{gong2018towards},  and the disentanglement of these attributes in some latent space is a very difficult task that requires extensive training. Most importantly,  the training of unsupervised representation learning models is much more difficult in contrast to supervised ones. As highlighted in \cite{bengio2013representation}, in supervised learning,  there is a clear objective to optimise. For instance, the classifiers are trained to learn such representations or features that minimise the misclassifications error. Representation learning models do not have such training objectives like classification or regression problems do.

%BS:
As outlined, GANs are a novel approach for generative modelling, they aim to learn the distribution of real data points. In recent years, they have been widely utilised for representation learning in different fields including speech.  However, they also proved difficult to train and face different failure modes, mainly vanishing gradients issues,  convergence problems, and mode collapse issues. Different remedies are proposed to tackle these issues. For instance, modified minimax loss \cite{goodfellow2014generative} can help to deal with vanishing gradients, the Wasserstein loss \cite{arjovsky2017wasserstein}
%BS:
and training of ensembles alleviate mode collapse, and noise addition to the discriminator inputs \cite{arjovsky2017towards} or penalising discriminator weights \cite{roth2017stabilizing} act as regularisation to improve a GAN's convergence. These are some earlier attempts to solve these issues; however, there is still room to improve the GANs training problems.

\subsection{Performance Issues of Domain Invariant Features}
To achieve generalisation in DL models, we need a large amount of data with similar training and testing examples. However, the performance of DL models drops significantly if test samples deviate from the distribution of the training data.  %This is one of the major problems in representation learning from speech. %This problem is also common to in vision \cite{li2018learning} and NLP \cite{xu2018emo2vec}. 
Learning speech representations that are invariant to variabilities in speakers, language, etc., are very difficult to capture. The performance of representations learnt from one corpus do not work well to another corpus having different recording conditions. This issue is common to all three applications of speech covered in this paper. In the past few years, researchers have achieved competitive performance by learning speaker invariant representations \cite{asakawa2007automatic,lu2004multitask}. However, language invariant representation is still very challenging. %Although emotions are considered language invariant, the performance of SER systems degrades when they are tested across different language emotional corpora  \cite{latif2019unsupervised}.  
The main reason is that we have speech corpora covering only a few languages in contrast to the number of spoken languages in the world. There are more than 5\,000 spoken languages in the world, but only 389 languages account for 94\,\% of the world's population\footnote{\url{https://www.ethnologue.com/statistics}}. We do not have speech corpora even for 389 languages to enable across language speech processing research. %Therefore, research in language and speech representation learning is facing the problem of data scarcity. 
This variation, imbalance, diversity, and dynamics in speech and language corpora are causing hurdles to designing generalised representation learning algorithms. %This variation, imbalance, diversity, and dynamics in speech and language corpora are causing hurdles to designing generalised representation learning algorithms. %Recent studies are focusing on representation learning for zero resource languages \cite{latif2019unsupervised,hermann2018multilingual}, but it will take some time to come across with solutions to address this issue. 

%Recent studies are focusing on representation learning for zero resource languages \cite{latif2019unsupervised,hermann2018multilingual}, but it will take some time to come across with solutions to address this issue. 

\subsection{Adversary on Representation Learning}
DL has undoubtedly offered tremendous improvements in the performance of state-of-the-art speech representation learning systems. However, recent works on adversarial examples pose enormous challenges for robust representation learning from speech by showing the susceptibility of DNNs to adversarial examples having imperceptible perturbations \cite{latif2018adversarial}. Some popular adversarial attacks include the fast gradient sign method (FGSM) \cite{goodfellow2014explaining}, Jacobian-based saliency map attack (JSMA) \cite{papernot2016limitations}, and DeepFool \cite{moosavi2016deepfool}; they compute the perturbation noise based on the gradient of targeted output. Such attacks are also evaluated against speech-based systems. For instance, Carlini and Wagner \cite{carlini2018audio} evaluated an iterative optimisation-based attack against DeepSpeech \cite{hannun2014deep} (a state-of-the-art ASR model) with 100\,\% success rate. Some other attempts also proposed different adversarial attacks \cite{alzantot2018did,schonherr2018adversarial} and against speech-based systems. The success of adversarial attacks against DL models shows that the representations learnt by them are not good \cite{akhtar2018threat}. Therefore, research is ongoing to tackle the challenge of adversarial 
%BS
attacks 
by exploring what DL models capture from the input data and how adversarial examples can be defined as a combination of previously learnt representations without any knowledge of adversaries.

\subsection{Quality of Speech Data}
 Representation learning models aim to identify potentially useful and ultimately understandable patterns. This demands not just more data, but more comprehensive and diverse data. Therefore, for learning a good representation, data must be correct and properly labelled, and unbiased. The quality of speech data can be poor due to various reasons. For example, different background noises and music can corrupt the speech data. Similarly, the noise of microphones or recording devices can also pollute the speech signal. Although studies use ‘noise injection’ techniques to avoid overfitting, this works for moderately high signal-to-noise ratios~\cite{yin2015noisy}. 
% With the rise of data collection through smartphone This has been idenfied as an impMost of the classical DL models are designed for high-quality data and they do not perform well on low-quality data \cite{zhang2018survey}.
This has been an active research topic and in the past few years, different DL models have been proposed that can learn representations from noisy data. For instance, DAEs \cite{vincent2008extracting} can learn a representation of data with noise, imputation AE \cite{bu2014incomplete} can learn a representation from incomplete data, and non-local AE \cite{wang2016non} can learn reliable features from corrupted data. Such techniques are also very popular in the speech community for noise invariant representation learning and we 
%BS: I changed a lot of past tense to present tense - usually, past tense should only go into the conclusion...
highlight this in Table \ref{review}. However, 
% the performance of models on noisy speech is always degraded \cite{latif2018adversarial}. 
there is still a need for such DL models that can deal with the quality of data not only for speech but also for other domains. 

%\section{Recent Advancement and Future Trends}
\section{Recent Advancements and Future Trends}
\label{Trends}

\subsection{Open Source Datasets and Toolkits}
There are a large number of speech databases available for speech analysis research. Some of the popular benchmark datasets including TIMIT \cite{fisher1986darpa}, WSJ \cite{paul1992design}, AMI \cite{hain2007ami}, and many other databases are not freely available. They are usually purchased from commercial organisations like LDC\footnote{\url{https://www.ldc.upenn.edu/}}, ELRA\footnote{\url{http://catalog.elra.info/}}, and Speech Ocean\footnote{\url{http://www.speechocean.com/}}.  The licence fees of these datasets are affordable for most of the research institutes; however, their fee is expensive (e.\,g., the WSJ corpus license costs 2500 USD) for young researchers who want to start their research on speech, particularly for researchers in developing countries. Recently, a free data movement is started in the speech community and different good quality datasets are made free for the public to invoke more research in this field. VoxForge\footnote{\url{http://www.voxforge.org/}} and OpenSLR\footnote{\url{http://www.openslr.org/}} are two popular platforms that contain freely available speech and speaker recognition datasets. Most of the SER corpora are developed by research institutes and they are freely available for research proposes.

\begin{table}[t!]
%\scriptsize
\centering
\caption{Some popular tools for speech feature extraction and model implementations.}
\begin{tabular}{|l|l|l|}
\hline\hline
\multicolumn{3}{|l|}{\textbf{\textit{Feature Extraction Tools}}}    \\ 
\hline\hline
\textit{Tools}  & \multicolumn{2}{l|}{\textit{Description}}   \\ \hline
\begin{tabular}[c]{@{}l@{}}Librosa \cite{mcfee2015librosa}\end{tabular} 
& \multicolumn{2}{l|}{\begin{tabular}[c]{@{}l@{}}A Python based toolkit for music\\ and audio analysis\end{tabular}}    \\ \hline

\begin{tabular}[c]{@{}l@{}}pyAudioAnalysis \\\cite{giannakopoulos2015pyaudioanalysis}\end{tabular}   & \multicolumn{2}{l|}{\begin{tabular}[c]{@{}l@{}}A Python library facilities a wide range of\\ feature extraction and also classification of\\ audio signals, supervised and unsupervised\\ segmentation and content visualisation.\end{tabular}}     \\ \hline

\begin{tabular}[c]{@{}l@{}}openSMILE \cite{eyben2010opensmile}\end{tabular} 
& \multicolumn{2}{l|}{\begin{tabular}[c]{@{}l@{}} It enables to extract a large number of \\audio feature in real time. It is written in C++.\end{tabular}}   \\ \hline

\hline
\multicolumn{3}{|l|}{\textbf{\textit{Speech Recognition Toolkits}}} \\ \hline\hline
\textit{Toolkit}  & \begin{tabular}[c]{@{}l@{}}\textit{Programming}\\ \textit{Language}\end{tabular} & \begin{tabular}[c]{@{}l@{}}\textit{Trained Models}\end{tabular} \\ \hline
\begin{tabular}[c]{@{}l@{}} CMU Sphinx \cite{lamere2003cmu}\end{tabular}     
& \begin{tabular}[c]{@{}l@{}}Jave, C, Python,\\ and others\end{tabular}
& \begin{tabular}[c]{@{}l@{}}English plus 10 \\other languages\end{tabular} \\ \hline

\begin{tabular}[c]{@{}l@{}}Kaldi \cite{povey2011kaldi}\end{tabular}     
& \begin{tabular}[c]{@{}l@{}}C++, Python\end{tabular}
& \begin{tabular}[c]{@{}l@{}}English\end{tabular} \\ \hline

\begin{tabular}[c]{@{}l@{}}Julius \cite{lee2001julius} \end{tabular}     
& \begin{tabular}[c]{@{}l@{}}C, Python\end{tabular}
& \begin{tabular}[c]{@{}l@{}}Japanese\end{tabular} \\ \hline

\begin{tabular}[c]{@{}l@{}}ESPnet \cite{watanabe2018espnet}\end{tabular}     
& \begin{tabular}[c]{@{}l@{}}Python\end{tabular}
& \begin{tabular}[c]{@{}l@{}}English, Japanese,\\Mandarin  \end{tabular} \\ \hline

\begin{tabular}[c]{@{}l@{}}HTK \cite{woodland1994large}\end{tabular}     
& \begin{tabular}[c]{@{}l@{}}C, Python\end{tabular}
& \begin{tabular}[c]{@{}l@{}}None\end{tabular} \\ \hline

\hline
\multicolumn{3}{|l|}{\textbf{\textit{Speaker Identification}}}     \\ \hline\hline
\begin{tabular}[c]{@{}l@{}}ALIZE \cite{larcher2013alize}\end{tabular}     
& \begin{tabular}[c]{@{}l@{}} C++\end{tabular}
& \begin{tabular}[c]{@{}l@{}}English\end{tabular} \\ \hline

\hline
\multicolumn{3}{|l|}{\textbf{\textit{Speech Emotion Recognition}}} \\ \hline\hline

\begin{tabular}[c]{@{}l@{}}OpenEAR \cite{eyben2009openear}\end{tabular}     
& \begin{tabular}[c]{@{}l@{}}C++\end{tabular}
& \begin{tabular}[c]{@{}l@{}}English, German\end{tabular} \\ \hline
\hline
\end{tabular}
\label{Table:tools}
\end{table}

Another important progress made by researchers of the speech community is the development of open-source toolkits for speech processing and analysis. These tools help the researchers not only for feature extraction, but also for the development of models. %These tools are built-in different languages which provide the flexibility of selection from different platforms. 
The details of such tools is presented in a tabular form in Table \ref{Table:tools}. It can be noted that ASR
%BS added:
---as the largest field of activity---
has more open source toolkits compared to SR and SER. %Models for SR and SER can also be implemented using Kaldi \cite{povey2011kaldi}. 
The development of such toolkits and speech corpus is providing great benefits to the speech research community and will continue to be needed to speed up the research progress on speech. 

\subsection{Computational Advancements}
In contrast to classical ML models, DL has a significantly larger number of parameters and involves huge amounts of matrix multiplications with many other operations. Traditional central processing units (CPUs) support such processing, therefore, advanced parallel computing is necessary for the development of deep networks. This is achieved by utilisation of graphics processing units (GPUs), %with the compute unified device architecture (CUDA) \cite{nickolls2008scalable} and the CUDA deep neural network library (cuDNN) \cite{chetlur2014cudnn}. 
which contain thousands of cores that can perform exceptionally fast matrix multiplications. %This leads to the fast training of DNNs. 
In contrast to CPUs and GPUs, advanced Tensor Processing Units (TPUs) developed by Google offer 15-30 $\times$ higher processing speeds and 30-80 $\times$ higher performance-per-watt \cite{jouppi2017datacenter}. A recent paper \cite{arute2019quantum} on quantum supremacy using programmable superconducting processor shows amazing results by performing computation a Hilbert space of dimension (253 $\approx$ 9 $\times$ 1015) far beyond the reach of the fastest supercomputers available today. It was the first computation on a quantum processor. This will lead to more progress and the computational power will continue to grow at a double-exponential rate. This will disrupt the area of representation learning from a vast amount of unlabelled data by unlocking new computational capabilities of quantum processors. 

% Researchers will be able to use representation learning models on a large scale of data to learn hidden patterns that can be used to improve the performance of various predictive tasks. 

\subsection{Processing Raw Speech}
In the past few years, the trend of using hand-engineered acoustic features is progressively changing and DL is gaining popularity as a viable alternative to learn from raw speech directly. %Researchers get promising results using CNNs to learn low-level speech representations from raw waveforms, which allow a network to better capture important speech characteristics such as pitch and formants \cite{ravanelli2018speaker}. However, the proper design of the feature extraction layer is crucial to achieving this goal \cite{latif2019direct}. 
This has removed the feature extraction module from the pipeline of the ASR, SR, and SER systems. %Feature extraction methods involve different steps based on standard signal processing techniques that might be computationally expensive, also the performance is not guaranteed. Therefore, the processing of raw speech using DL models is becoming very popular in the speech community. %Most of the studies (e.\,g., \cite{palaz2015analysis, sainath2015learning, bengiospeaker, trigeorgis2016adieu}) on raw speech utilised supervised representation learning technique. Some of them also exploited unsupervised models to learn representations from speech \cite{sailor2016unsupervised,chorowski2019unsupervised}. 
Recently, important progress is also made by Donahue et al.\ \cite{donahue2018synthesizing} in audio generation. They proposed WaveGAN for the unsupervised synthesis of raw-waveform audio and showed that their model can learn to produce intelligible words when trained on a small vocabulary speech dataset, and can also synthesise music audio and bird vocalisations. Other recent works \cite{engel2019gansynth,kumar2019melgan} also explored audio synthesis using GANs; however, such work is at the initial stage and will likely open new prospects of future research as it transpired with the use of GANs in the domain of vision (e.\,g., with DeepFakes \cite{chesney2018deep}). 

%BS a lot here is merely repetition from above... I added some "as outlined" and similar...
\subsection{The Rise of Adversarial Training}
The idea of adversarial training was proposed in 2014 \cite{goodfellow2014generative}. It leads to widespread research in various ML domains including speech representation learning.  %Adversarial models, especially GANs, are becoming a state-of-the-art model for representation learning not only from images but also from speech \cite{han2019adversarial}.  
Speech-based systems---principally, ASR, SR, and SER systems---need to be robust under environmental acoustic variabilities arising from environmental, speaker, and recording conditions. This is very crucial for industrial applications of these systems. 

GANs are being used as a viable tool for robust speech representation learning \cite{serdyuk2016invariant} and also speech enhancement \cite{michelsanti2017conditional} to tackle the noise issues. A popular variant of GAN, cycle-consistent Generative Adversarial Networks (CycleGAN) \cite{zhu2017unpaired} is being used for domain adaptation for low-resource scenarios (where a limited amount of target data is available for adaptation) \cite{nidadavolu2019low}. These results using CycleGANs on speech are very promising for domain adaptation. This will also lead to designing such systems that can learn domain invariant representation learning, especially for zero-resource languages to enable speech-based cross-culture applications. 

Another interesting utilisation of GANs is learning from synthetic data. Researchers succeeded in the synthesis of speech signals also by GANs  \cite{donahue2018synthesizing}. Synthetic data can be utilised for such applications where large label data is not available. In SER, larger labelled data is not available. Learning representation from synthetic data can help to improve the performance of system and researchers have explored the use of synthetic data for SER \cite{sahu2018adversarial,bao2019cyclegan}. %In \cite{sahu2018adversarial},  AAEs are used to generate synthetic samples and they showed that representations learnt from synthetic data can be used to augment SER systems and help to improve the performance. Another recent study \cite{bao2019cyclegan} uses CycleGANs for style transfer of representations extracted from a large unlabelled speech corpus into target emotions representations and enables to improve the performance of the system. Similarly, in \cite{chatziagapi2019data} used conditional GAN to generate synthetic data for the minority class to minimise the data imbalance issue. They were able to significantly improve the performance of the system. 
This shows the feasibility of learning from synthetic data and  will lead to interesting research to solve the problems in the speech domain where data scarcity is a major problem.

\subsection{Representation Learning with Interaction}
Good representation disentangles the underlying explanatory factors of variation. However, it is an open research question that what kind of training framework can potentially learn disentangled representations from input data. Most of the research work on representation learning used static settings without involving the interaction with the environment. Reinforcement learning (RL) facilitates the idea of learning while interacting with the environment. If RL is used to disentangle factors of variation by interacting with the environment, a good representation can be learnt. This will lead to faster convergence, in contrast, to blindly attempting to solve given problems. Such an idea has recently been validated by Thomas et al. \cite{thomas2018disentangling}, where the authors used RL to disentangle the independently controllable factors of variation by using a specific objective function. The authors empirically showed that the agent can disentangle these aspects of the environment without any extrinsic reward. This is an important finding that will act as the key to further research in this direction.

\subsection{Privacy Preserving Representations}

When people use speech-based services such as voice authentication or speech recognition, they grant complete access to their recordings. These services can extract user's information such as gender, ethnicity, and emotional state and can be used for undesired purposes. %Similarly, the users' recordings can also be edited or used to create a fake speech that the user never spoke or the voiceprints can be used to fool voice-authentication systems. 
Various other privacy-related issues arise while using speech-based services \cite{pathak2013privacy}. It is desirable in speech processing applications that there are suitable provisions for ensuring that there is no unauthorised and undisclosed eavesdropping and violation of privacy. Privacy preserved representation learning is a relatively unexplored research topic. Recently, researchers have started to utilise privacy-preserving representation learning models to protect speaker identity \cite{srivastava2019privacy}, gender identity \cite{jaiswal2019privacy}. %Researchers are now focusing on designing such privacy preserved representation learning methods that can hide the information of users to enable the secure cloud-based applications of speech technology. 
To preserve users' privacy, federated learning \cite{shokri2015privacy} is another alternative setting where the training of a shared global model is performed using multiple participating computing devices. This happens under the coordination of a central server, however, the training data remains decentralised.

%BS you could add on federated learning?

\section{Conclusions}
\label{Conclusions}

In this article, we have focused on providing a comprehensive review of representation learning for speech signals using deep learning approaches in three principal speech processing areas: automatic speech recognition (ASR), speaker recognition (SR), and speech emotion recognition (SER). In all of these three areas, the use of representation learning is very promising, and there is an ongoing research on this topic in which different models and methods are being explored to disentangle speech attributes suitable for these tasks. The literature review performed in this work shows that LSTM/GRU-RNNs in combination with CNNs are suitable for capturing speech attributes. Most of the studies have used LSTM  models in a supervised way. In unsupervised representation learning, DAEs and VAEs are widely deployed architectures in the speech community, with GAN-based models also attaining prominence for speech enhancement and feature learning. Apart from providing a detailed review, we have also highlighted the challenges faced by researchers working with representation learning techniques and avenues for future work. It is hoped that this article will become a definitive guide to researchers and practitioners interested to work either in speech signal or deep representation learning in general. 
%BS added:
We are curious whether in the longer run, representation learning will be the standard paradigm in speech processing. If so, we are currently witnessing the change of a paradigm moving away from signal processing and expert-crafted features into a highly data-driven era---with all its advantages, challenges, and risks.

% \bibliographystyle{IEEEtran}
% \bibliography{Reference}

\begin{thebibliography}{100}
\providecommand{\url}[1]{#1}
\csname url@samestyle\endcsname
\providecommand{\newblock}{\relax}
\providecommand{\bibinfo}[2]{#2}
\providecommand{\BIBentrySTDinterwordspacing}{\spaceskip=0pt\relax}
\providecommand{\BIBentryALTinterwordstretchfactor}{4}
\providecommand{\BIBentryALTinterwordspacing}{\spaceskip=\fontdimen2\font plus
\BIBentryALTinterwordstretchfactor\fontdimen3\font minus
  \fontdimen4\font\relax}
\providecommand{\BIBforeignlanguage}[2]{{%
\expandafter\ifx\csname l@#1\endcsname\relax
\typeout{** WARNING: IEEEtran.bst: No hyphenation pattern has been}%
\typeout{** loaded for the language `#1'. Using the pattern for}%
\typeout{** the default language instead.}%
\else
\language=\csname l@#1\endcsname
\fi
#2}}
\providecommand{\BIBdecl}{\relax}
\BIBdecl

\bibitem{bengio2013representation}
Y.~Bengio, A.~Courville, and P.~Vincent, ``Representation learning: A review
  and new perspectives,'' \emph{IEEE transactions on pattern analysis and
  machine intelligence}, vol.~35, no.~8, pp. 1798--1828, 2013.

\bibitem{herff2016automatic}
C.~Herff and T.~Schultz, ``Automatic speech recognition from neural signals: a
  focused review,'' \emph{Frontiers in neuroscience}, vol.~10, p. 429, 2016.

\bibitem{tran2000fuzzy}
D.~T. Tran, ``Fuzzy approaches to speech and speaker recognition,'' Ph.D.
  dissertation, university of Canberra, 2000.

\bibitem{najafabadi2015deep}
M.~M. Najafabadi, F.~Villanustre, T.~M. Khoshgoftaar, N.~Seliya, R.~Wald, and
  E.~Muharemagic, ``Deep learning applications and challenges in big data
  analytics,'' \emph{Journal of Big Data}, vol.~2, no.~1, p.~1, 2015.

\bibitem{huang2014historical}
X.~Huang, J.~Baker, and R.~Reddy, ``A historical perspective of speech
  recognition,'' \emph{Commun. ACM}, vol.~57, no.~1, pp. 94--103, 2014.

\bibitem{zhong2016overview}
G.~Zhong, L.-N. Wang, X.~Ling, and J.~Dong, ``An overview on data
  representation learning: From traditional feature learning to recent deep
  learning,'' \emph{The Journal of Finance and Data Science}, vol.~2, no.~4,
  pp. 265--278, 2016.

\bibitem{zhang2018deep}
Z.~Zhang, J.~Geiger, J.~Pohjalainen, A.~E.-D. Mousa, W.~Jin, and B.~Schuller,
  ``Deep learning for environmentally robust speech recognition: An overview of
  recent developments,'' \emph{ACM Transactions on Intelligent Systems and
  Technology (TIST)}, vol.~9, no.~5, p.~49, 2018.

\bibitem{swain2018databases}
M.~Swain, A.~Routray, and P.~Kabisatpathy, ``Databases, features and
  classifiers for speech emotion recognition: a review,'' \emph{International
  Journal of Speech Technology}, vol.~21, no.~1, pp. 93--120, 2018.

\bibitem{nassif2019speech}
A.~B. Nassif, I.~Shahin, I.~Attili, M.~Azzeh, and K.~Shaalan, ``Speech
  recognition using deep neural networks: A systematic review,'' \emph{IEEE
  Access}, vol.~7, pp. 19\,143--19\,165, 2019.

\bibitem{kingma2013auto}
D.~P. Kingma and M.~Welling, ``Auto-encoding variational bayes,'' \emph{arXiv
  preprint arXiv:1312.6114}, 2013.

\bibitem{goodfellow2014generative}
I.~Goodfellow, J.~Pouget-Abadie, M.~Mirza, B.~Xu, D.~Warde-Farley, S.~Ozair,
  A.~Courville, and Y.~Bengio, ``Generative adversarial nets,'' in
  \emph{Advances in neural information processing systems}, 2014, pp.
  2672--2680.

\bibitem{gomez2019design}
J.~G{\'o}mez-Garc{\'\i}a, L.~Moro-Vel{\'a}zquez, and J.~I. Godino-Llorente,
  ``On the design of automatic voice condition analysis systems. part ii:
  Review of speaker recognition techniques and study on the effects of
  different variability factors,'' \emph{Biomedical Signal Processing and
  Control}, vol.~48, pp. 128--143, 2019.

\bibitem{zhong2019shallow}
G.~Zhong, X.~Ling, and L.-N. Wang, ``From shallow feature learning to deep
  learning: Benefits from the width and depth of deep architectures,''
  \emph{Wiley Interdisciplinary Reviews: Data Mining and Knowledge Discovery},
  vol.~9, no.~1, p. e1255, 2019.

\bibitem{pearson1901liii}
K.~Pearson, ``Liii. on lines and planes of closest fit to systems of points in
  space,'' \emph{The London, Edinburgh, and Dublin Philosophical Magazine and
  Journal of Science}, vol.~2, no.~11, pp. 559--572, 1901.

\bibitem{fisher1936use}
R.~A. Fisher, ``The use of multiple measurements in taxonomic problems,''
  \emph{Annals of eugenics}, vol.~7, no.~2, pp. 179--188, 1936.

\bibitem{hardoon2004canonical}
D.~R. Hardoon, S.~Szedmak, and J.~Shawe-Taylor, ``Canonical correlation
  analysis: An overview with application to learning methods,'' \emph{Neural
  computation}, vol.~16, no.~12, pp. 2639--2664, 2004.

\bibitem{borg2003modern}
I.~Borg and P.~Groenen, ``Modern multidimensional scaling: Theory and
  applications,'' \emph{Journal of Educational Measurement}, vol.~40, no.~3,
  pp. 277--280, 2003.

\bibitem{hyvarinen2000independent}
A.~Hyv{\"a}rinen and E.~Oja, ``Independent component analysis: algorithms and
  applications,'' \emph{Neural networks}, vol.~13, no. 4-5, pp. 411--430, 2000.

\bibitem{scholkopf1998nonlinear}
B.~Sch{\"o}lkopf, A.~Smola, and K.-R. M{\"u}ller, ``Nonlinear component
  analysis as a kernel eigenvalue problem,'' \emph{Neural computation},
  vol.~10, no.~5, pp. 1299--1319, 1998.

\bibitem{baudat2000generalized}
G.~Baudat and F.~Anouar, ``Generalized discriminant analysis using a kernel
  approach,'' \emph{Neural computation}, vol.~12, no.~10, pp. 2385--2404, 2000.

\bibitem{lee1999learning}
D.~D. Lee and H.~S. Seung, ``Learning the parts of objects by non-negative
  matrix factorization,'' \emph{Nature}, vol. 401, no. 6755, p. 788, 1999.

\bibitem{lee2004feature}
F.~Lee, R.~Scherer, R.~Leeb, A.~Schl{\"o}gl, H.~Bischof, and G.~Pfurtscheller,
  \emph{Feature mapping using PCA, locally linear embedding and isometric
  feature mapping for EEG-based brain computer interface}.\hskip 1em plus 0.5em
  minus 0.4em\relax Citeseer, 2004.

\bibitem{roweis2000nonlinear}
S.~T. Roweis and L.~K. Saul, ``Nonlinear dimensionality reduction by locally
  linear embedding,'' \emph{science}, vol. 290, no. 5500, pp. 2323--2326, 2000.

\bibitem{tenenbaum2000global}
J.~B. Tenenbaum, V.~De~Silva, and J.~C. Langford, ``A global geometric
  framework for nonlinear dimensionality reduction,'' \emph{science}, vol. 290,
  no. 5500, pp. 2319--2323, 2000.

\bibitem{maaten2008visualizing}
L.~v.~d. Maaten and G.~Hinton, ``Visualizing data using t-sne,'' \emph{Journal
  of machine learning research}, vol.~9, no. Nov, pp. 2579--2605, 2008.

\bibitem{lecun1998gradient}
Y.~LeCun, L.~Bottou, Y.~Bengio, P.~Haffner \emph{et~al.}, ``Gradient-based
  learning applied to document recognition,'' \emph{Proceedings of the IEEE},
  vol.~86, no.~11, pp. 2278--2324, 1998.

\bibitem{kocsor2004kernel}
A.~Kocsor and L.~T{\'o}th, ``Kernel-based feature extraction with a speech
  technology application,'' \emph{IEEE Transactions on Signal Processing},
  vol.~52, no.~8, pp. 2250--2263, 2004.

\bibitem{takiguchi2007pca}
T.~Takiguchi and Y.~Ariki, ``Pca-based speech enhancement for distorted speech
  recognition.'' \emph{Journal of multimedia}, vol.~2, no.~5, 2007.

\bibitem{hinton2006reducing}
G.~E. Hinton and R.~R. Salakhutdinov, ``Reducing the dimensionality of data
  with neural networks,'' \emph{science}, vol. 313, no. 5786, pp. 504--507,
  2006.

\bibitem{bengio2007greedy}
Y.~Bengio, P.~Lamblin, D.~Popovici, and H.~Larochelle, ``Greedy layer-wise
  training of deep networks,'' in \emph{Advances in neural information
  processing systems}, 2007, pp. 153--160.

\bibitem{poultney2007efficient}
C.~Poultney, S.~Chopra, Y.~L. Cun \emph{et~al.}, ``Efficient learning of sparse
  representations with an energy-based model,'' in \emph{Advances in neural
  information processing systems}, 2007, pp. 1137--1144.

\bibitem{gales2008application}
M.~Gales, S.~Young \emph{et~al.}, ``The application of hidden markov models in
  speech recognition,'' \emph{Foundations and Trends{\textregistered} in Signal
  Processing}, vol.~1, no.~3, pp. 195--304, 2008.

\bibitem{waibel1989phoneme}
A.~Waibel, T.~Hanazawa, G.~Hinton, K.~Shikano, and K.~J. Lang, ``Phoneme
  recognition using time-delay neural networks,'' \emph{IEEE transactions on
  acoustics, speech, and signal processing}, vol.~37, no.~3, pp. 328--339,
  1989.

\bibitem{purwins2019deep}
H.~Purwins, B.~Li, T.~Virtanen, J.~Schl{\"u}ter, S.-Y. Chang, and T.~Sainath,
  ``Deep learning for audio signal processing,'' \emph{IEEE Journal of Selected
  Topics in Signal Processing}, vol.~13, no.~2, pp. 206--219, 2019.

\bibitem{hinton2012deep}
G.~Hinton, L.~Deng, D.~Yu, G.~E. Dahl, A.-r. Mohamed, N.~Jaitly, A.~Senior,
  V.~Vanhoucke, P.~Nguyen, T.~N. Sainath \emph{et~al.}, ``Deep neural networks
  for acoustic modeling in speech recognition: The shared views of four
  research groups,'' \emph{IEEE Signal processing magazine}, vol.~29, no.~6,
  pp. 82--97, 2012.

\bibitem{wollmer2010long}
M.~W{\"o}llmer, Y.~Sun, F.~Eyben, and B.~Schuller, ``Long short-term memory
  networks for noise robust speech recognition,'' in \emph{Proc. INTERSPEECH
  2010, Makuhari, Japan}, 2010, pp. 2966--2969.

\bibitem{wollmer2008abandoning}
M.~W{\"o}llmer, F.~Eyben, S.~Reiter, B.~Schuller, C.~Cox, E.~Douglas-Cowie, and
  R.~Cowie, ``Abandoning emotion classes-towards continuous emotion recognition
  with modelling of long-range dependencies,'' in \emph{Proc. 9th Interspeech
  2008 incorp. 12th Australasian Int. Conf. on Speech Science and Technology
  SST 2008, Brisbane, Australia}, 2008, pp. 597--600.

\bibitem{latif2018phonocardiographic}
S.~Latif, M.~Usman, R.~Rana, and J.~Qadir, ``Phonocardiographic sensing using
  deep learning for abnormal heartbeat detection,'' \emph{IEEE Sensors
  Journal}, vol.~18, no.~22, pp. 9393--9400, 2018.

\bibitem{qayyum2018quran}
A.~Qayyum, S.~Latif, and J.~Qadir, ``Quran reciter identification: A deep
  learning approach,'' in \emph{2018 7th International Conference on Computer
  and Communication Engineering (ICCCE)}.\hskip 1em plus 0.5em minus
  0.4em\relax IEEE, 2018, pp. 492--497.

\bibitem{sainath2015convolutional}
T.~N. Sainath, O.~Vinyals, A.~Senior, and H.~Sak, ``Convolutional, long
  short-term memory, fully connected deep neural networks,'' in \emph{2015 IEEE
  International Conference on Acoustics, Speech and Signal Processing
  (ICASSP)}.\hskip 1em plus 0.5em minus 0.4em\relax IEEE, 2015, pp. 4580--4584.

\bibitem{trigeorgis2016adieu}
G.~Trigeorgis, F.~Ringeval, R.~Brueckner, E.~Marchi, M.~A. Nicolaou,
  B.~Schuller, and S.~Zafeiriou, ``Adieu features? end-to-end speech emotion
  recognition using a deep convolutional recurrent network,'' in \emph{2016
  IEEE international conference on acoustics, speech and signal processing
  (ICASSP)}.\hskip 1em plus 0.5em minus 0.4em\relax IEEE, 2016, pp. 5200--5204.

\bibitem{langkvist2014review}
M.~L{\"a}ngkvist, L.~Karlsson, and A.~Loutfi, ``A review of unsupervised
  feature learning and deep learning for time-series modeling,'' \emph{Pattern
  Recognition Letters}, vol.~42, pp. 11--24, 2014.

\bibitem{oord2016pixel}
A.~v.~d. Oord, N.~Kalchbrenner, and K.~Kavukcuoglu, ``Pixel recurrent neural
  networks,'' \emph{arXiv preprint arXiv:1601.06759}, 2016.

\bibitem{oord2016wavenet}
A.~v.~d. Oord, S.~Dieleman, H.~Zen, K.~Simonyan, O.~Vinyals, A.~Graves,
  N.~Kalchbrenner, A.~Senior, and K.~Kavukcuoglu, ``Wavenet: A generative model
  for raw audio,'' \emph{arXiv preprint arXiv:1609.03499}, 2016.

\bibitem{bollepalli2019generative}
B.~Bollepalli, L.~Juvela, and P.~Alku, ``Generative adversarial network-based
  glottal waveform model for statistical parametric speech synthesis,''
  \emph{arXiv preprint arXiv:1903.05955}, 2019.

\bibitem{hsu2017unsupervised}
W.-N. Hsu, Y.~Zhang, and J.~Glass, ``Unsupervised learning of disentangled and
  interpretable representations from sequential data,'' in \emph{Advances in
  neural information processing systems}, 2017, pp. 1878--1889.

\bibitem{furui1986speaker}
S.~Furui, ``Speaker-independent isolated word recognition based on emphasized
  spectral dynamics,'' in \emph{ICASSP'86. IEEE International Conference on
  Acoustics, Speech, and Signal Processing}, vol.~11.\hskip 1em plus 0.5em
  minus 0.4em\relax IEEE, 1986, pp. 1991--1994.

\bibitem{davis1980comparison}
S.~Davis and P.~Mermelstein, ``Comparison of parametric representations for
  monosyllabic word recognition in continuously spoken sentences,'' \emph{IEEE
  transactions on acoustics, speech, and signal processing}, vol.~28, no.~4,
  pp. 357--366, 1980.

\bibitem{purwins2000new}
H.~Purwins, B.~Blankertz, and K.~Obermayer, ``A new method for tracking
  modulations in tonal music in audio data format,'' in \emph{Proceedings of
  the IEEE-INNS-ENNS International Joint Conference on Neural Networks. IJCNN
  2000. Neural Computing: New Challenges and Perspectives for the New
  Millennium}, vol.~6.\hskip 1em plus 0.5em minus 0.4em\relax IEEE, 2000, pp.
  270--275.

\bibitem{eyben2015geneva}
F.~Eyben, K.~R. Scherer, B.~W. Schuller, J.~Sundberg, E.~Andr{\'e}, C.~Busso,
  L.~Y. Devillers, J.~Epps, P.~Laukka, S.~S. Narayanan \emph{et~al.}, ``The
  geneva minimalistic acoustic parameter set (gemaps) for voice research and
  affective computing,'' \emph{IEEE Transactions on Affective Computing},
  vol.~7, no.~2, pp. 190--202, 2015.

\bibitem{neumann2017attentive}
M.~Neumann and N.~T. Vu, ``Attentive convolutional neural network based speech
  emotion recognition: A study on the impact of input features, signal length,
  and acted speech,'' \emph{arXiv preprint arXiv:1706.00612}, 2017.

\bibitem{jaitly2011learning}
N.~Jaitly and G.~Hinton, ``Learning a better representation of speech
  soundwaves using restricted boltzmann machines,'' in \emph{2011 IEEE
  International Conference on Acoustics, Speech and Signal Processing
  (ICASSP)}.\hskip 1em plus 0.5em minus 0.4em\relax IEEE, 2011, pp. 5884--5887.

\bibitem{sun2017revisiting}
C.~Sun, A.~Shrivastava, S.~Singh, and A.~Gupta, ``Revisiting unreasonable
  effectiveness of data in deep learning era,'' in \emph{Proceedings of the
  IEEE international conference on computer vision}, 2017, pp. 843--852.

\bibitem{fisher1986darpa}
M.~Fisher~William, ``The darpa speech recognition research database:
  Specifications and status/william m. fisher, george r. doddington, kathleen
  m. goudie-marshall,'' in \emph{Proceedings of DARPA Workshop on Speech
  Recognition}, 1986, pp. 93--99.

\bibitem{godfrey1992switchboard}
J.~J. Godfrey, E.~C. Holliman, and J.~McDaniel, ``Switchboard: Telephone speech
  corpus for research and development,'' in \emph{[Proceedings] ICASSP-92: 1992
  IEEE International Conference on Acoustics, Speech, and Signal Processing},
  vol.~1.\hskip 1em plus 0.5em minus 0.4em\relax IEEE, 1992, pp. 517--520.

\bibitem{paul1992design}
D.~B. Paul and J.~M. Baker, ``The design for the wall street journal-based csr
  corpus,'' in \emph{Proceedings of the workshop on Speech and Natural
  Language}.\hskip 1em plus 0.5em minus 0.4em\relax Association for
  Computational Linguistics, 1992, pp. 357--362.

\bibitem{hain2007ami}
T.~Hain, L.~Burget, J.~Dines, G.~Garau, V.~Wan, M.~Karafi, J.~Vepa, and
  M.~Lincoln, ``The ami system for the transcription of speech in meetings,''
  in \emph{2007 IEEE International Conference on Acoustics, Speech and Signal
  Processing-ICASSP'07}, vol.~4.\hskip 1em plus 0.5em minus 0.4em\relax IEEE,
  2007, pp. IV--357.

\bibitem{burkhardt2005database}
F.~Burkhardt, A.~Paeschke, M.~Rolfes, W.~F. Sendlmeier, and B.~Weiss, ``A
  database of german emotional speech,'' in \emph{Ninth European Conference on
  Speech Communication and Technology}, 2005.

\bibitem{schuller2009interspeech}
B.~Schuller, S.~Steidl, and A.~Batliner, ``The interspeech 2009 emotion
  challenge,'' in \emph{Tenth Annual Conference of the International Speech
  Communication Association}, 2009.

\bibitem{ringeval2013introducing}
F.~Ringeval, A.~Sonderegger, J.~Sauer, and D.~Lalanne, ``Introducing the recola
  multimodal corpus of remote collaborative and affective interactions,'' in
  \emph{2013 10th IEEE international conference and workshops on automatic face
  and gesture recognition (FG)}.\hskip 1em plus 0.5em minus 0.4em\relax IEEE,
  2013, pp. 1--8.

\bibitem{banziger2012introducing}
T.~B{\"a}nziger, M.~Mortillaro, and K.~R. Scherer, ``Introducing the geneva
  multimodal expression corpus for experimental research on emotion
  perception.'' \emph{Emotion}, vol.~12, no.~5, p. 1161, 2012.

\bibitem{panayotov2015librispeech}
V.~Panayotov, G.~Chen, D.~Povey, and S.~Khudanpur, ``Librispeech: an asr corpus
  based on public domain audio books,'' in \emph{2015 IEEE International
  Conference on Acoustics, Speech and Signal Processing (ICASSP)}.\hskip 1em
  plus 0.5em minus 0.4em\relax IEEE, 2015, pp. 5206--5210.

\bibitem{chung2018voxceleb2}
J.~S. Chung, A.~Nagrani, and A.~Zisserman, ``Voxceleb2: Deep speaker
  recognition,'' \emph{arXiv preprint arXiv:1806.05622}, 2018.

\bibitem{rousseau2012ted}
A.~Rousseau, P.~Del{\'e}glise, and Y.~Esteve, ``Ted-lium: an automatic speech
  recognition dedicated corpus.'' in \emph{LREC}, 2012, pp. 125--129.

\bibitem{wang2015thchs}
D.~Wang and X.~Zhang, ``Thchs-30: A free chinese speech corpus,'' \emph{arXiv
  preprint arXiv:1512.01882}, 2015.

\bibitem{bu2017aishell}
H.~Bu, J.~Du, X.~Na, B.~Wu, and H.~Zheng, ``Aishell-1: An open-source mandarin
  speech corpus and a speech recognition baseline,'' in \emph{2017 20th
  Conference of the Oriental Chapter of the International Coordinating
  Committee on Speech Databases and Speech I/O Systems and Assessment
  (O-COCOSDA)}.\hskip 1em plus 0.5em minus 0.4em\relax IEEE, 2017, pp. 1--5.

\bibitem{milde2018open}
B.~Milde and A.~K{\"o}hn, ``Open source automatic speech recognition for
  german,'' in \emph{Speech Communication; 13th ITG-Symposium}.\hskip 1em plus
  0.5em minus 0.4em\relax VDE, 2018, pp. 1--5.

\bibitem{mckeown2012semaine}
G.~McKeown, M.~Valstar, R.~Cowie, M.~Pantic, and M.~Schroder, ``The semaine
  database: Annotated multimodal records of emotionally colored conversations
  between a person and a limited agent,'' \emph{IEEE Transactions on Affective
  Computing}, vol.~3, no.~1, pp. 5--17, 2012.

\bibitem{busso2008iemocap}
C.~Busso, M.~Bulut, C.-C. Lee, A.~Kazemzadeh, E.~Mower, S.~Kim, J.~N. Chang,
  S.~Lee, and S.~S. Narayanan, ``Iemocap: Interactive emotional dyadic motion
  capture database,'' \emph{Language resources and evaluation}, vol.~42, no.~4,
  p. 335, 2008.

\bibitem{busso2017msp}
C.~Busso, S.~Parthasarathy, A.~Burmania, M.~AbdelWahab, N.~Sadoughi, and E.~M.
  Provost, ``Msp-improv: An acted corpus of dyadic interactions to study
  emotion perception,'' \emph{IEEE Transactions on Affective Computing}, no.~1,
  pp. 67--80, 2017.

\bibitem{bimbot2004tutorial}
F.~Bimbot, J.-F. Bonastre, C.~Fredouille, G.~Gravier, I.~Magrin-Chagnolleau,
  S.~Meignier, T.~Merlin, J.~Ortega-Garc{\'\i}a, D.~Petrovska-Delacr{\'e}taz,
  and D.~A. Reynolds, ``A tutorial on text-independent speaker verification,''
  \emph{EURASIP Journal on Advances in Signal Processing}, vol. 2004, no.~4, p.
  101962, 2004.

\bibitem{Schuller11-RRE}
B.~Schuller, A.~Batliner, S.~Steidl, and D.~Seppi, ``{Recognising Realistic
  Emotions and Affect in Speech: State of the Art and Lessons Learnt from the
  First Challenge},'' \emph{Speech Communication}, vol.~53, no. 9/10, pp.
  1062--1087, November/December 2011.

\bibitem{goodfellow2016deep}
I.~Goodfellow, Y.~Bengio, and A.~Courville, \emph{Deep learning}.\hskip 1em
  plus 0.5em minus 0.4em\relax MIT press, 2016.

\bibitem{bengio2009learning}
Y.~Bengio \emph{et~al.}, ``Learning deep architectures for ai,''
  \emph{Foundations and trends{\textregistered} in Machine Learning}, vol.~2,
  no.~1, pp. 1--127, 2009.

\bibitem{jing2019self}
L.~Jing and Y.~Tian, ``Self-supervised visual feature learning with deep neural
  networks: A survey,'' \emph{arXiv preprint arXiv:1902.06162}, 2019.

\bibitem{liang2017text}
H.~Liang, X.~Sun, Y.~Sun, and Y.~Gao, ``Text feature extraction based on deep
  learning: a review,'' \emph{EURASIP journal on wireless communications and
  networking}, vol. 2017, no.~1, pp. 1--12, 2017.

\bibitem{noda2015audio}
K.~Noda, Y.~Yamaguchi, K.~Nakadai, H.~G. Okuno, and T.~Ogata, ``Audio-visual
  speech recognition using deep learning,'' \emph{Applied Intelligence},
  vol.~42, no.~4, pp. 722--737, 2015.

\bibitem{usman2017using}
M.~Usman, S.~Latif, and J.~Qadir, ``Using deep autoencoders for facial
  expression recognition,'' in \emph{2017 13th International Conference on
  Emerging Technologies (ICET)}.\hskip 1em plus 0.5em minus 0.4em\relax IEEE,
  2017, pp. 1--6.

\bibitem{latif2017variational}
S.~Latif, R.~Rana, J.~Qadir, and J.~Epps, ``Variational autoencoders for
  learning latent representations of speech emotion: A preliminary study,''
  \emph{arXiv preprint arXiv:1712.08708}, 2017.

\bibitem{mitra2018introduction}
B.~Mitra, N.~Craswell \emph{et~al.}, ``An introduction to neural information
  retrieval,'' \emph{Foundations and Trends{\textregistered} in Information
  Retrieval}, vol.~13, no.~1, pp. 1--126, 2018.

\bibitem{mitra2017neural}
B.~Mitra and N.~Craswell, ``Neural models for information retrieval,''
  \emph{arXiv preprint arXiv:1705.01509}, 2017.

\bibitem{kim2018one}
J.~Kim, J.~Urbano, C.~C. Liem, and A.~Hanjalic, ``One deep music representation
  to rule them all? a comparative analysis of different representation learning
  strategies,'' \emph{Neural Computing and Applications}, pp. 1--27, 2018.

\bibitem{sordoni2016learning}
A.~Sordoni, ``Learning representations for information retrieval,'' 2016.

\bibitem{kurakin2016adversarial}
A.~Kurakin, I.~Goodfellow, and S.~Bengio, ``Adversarial machine learning at
  scale,'' \emph{arXiv preprint arXiv:1611.01236}, 2016.

\bibitem{liu2018towards}
X.~Liu, M.~Cheng, H.~Zhang, and C.-J. Hsieh, ``Towards robust neural networks
  via random self-ensemble,'' in \emph{Proceedings of the European Conference
  on Computer Vision (ECCV)}, 2018, pp. 369--385.

\bibitem{latif2018adversarial}
S.~Latif, R.~Rana, and J.~Qadir, ``Adversarial machine learning and speech
  emotion recognition: Utilizing generative adversarial networks for
  robustness,'' \emph{arXiv preprint arXiv:1811.11402}, 2018.

\bibitem{Liu19-NIT}
S.~Liu, G.~Keren, and B.~W. Schuller, ``{N-HANS: Introducing the Augsburg
  Neuro-Holistic Audio-eNhancement System},'' \emph{arxiv.org}, no. 1911.07062,
  November 2019, 5 pages.

\bibitem{xu2005survey}
R.~Xu and D.~C. Wunsch, ``Survey of clustering algorithms,'' 2005.

\bibitem{min2018survey}
E.~Min, X.~Guo, Q.~Liu, G.~Zhang, J.~Cui, and J.~Long, ``A survey of clustering
  with deep learning: From the perspective of network architecture,''
  \emph{IEEE Access}, vol.~6, pp. 39\,501--39\,514, 2018.

\bibitem{dicarlo2007untangling}
J.~J. DiCarlo and D.~D. Cox, ``Untangling invariant object recognition,''
  \emph{Trends in cognitive sciences}, vol.~11, no.~8, pp. 333--341, 2007.

\bibitem{higgins2018towards}
I.~Higgins, D.~Amos, D.~Pfau, S.~Racaniere, L.~Matthey, D.~Rezende, and
  A.~Lerchner, ``Towards a definition of disentangled representations,''
  \emph{arXiv preprint arXiv:1812.02230}, 2018.

\bibitem{alemi2016deep}
A.~A. Alemi, I.~Fischer, J.~V. Dillon, and K.~Murphy, ``Deep variational
  information bottleneck,'' \emph{arXiv preprint arXiv:1612.00410}, 2016.

\bibitem{hinton2003stochastic}
G.~E. Hinton and S.~T. Roweis, ``Stochastic neighbor embedding,'' in
  \emph{Advances in neural information processing systems}, 2003, pp. 857--864.

\bibitem{errity2006investigation}
A.~Errity and J.~McKenna, ``An investigation of manifold learning for speech
  analysis,'' in \emph{Ninth International Conference on Spoken Language
  Processing}, 2006.

\bibitem{cayton2005algorithms}
L.~Cayton, ``Algorithms for manifold learning,'' \emph{Univ. of California at
  San Diego Tech. Rep}, vol.~12, no. 1-17, p.~1, 2005.

\bibitem{golchin2014overview}
E.~Golchin and K.~Maghooli, ``Overview of manifold learning and its application
  in medical data set,'' \emph{International journal of biomedical engineering
  and science (IJBES)}, vol.~1, no.~2, pp. 23--33, 2014.

\bibitem{ma2011manifold}
Y.~Ma and Y.~Fu, \emph{Manifold learning theory and applications}.\hskip 1em
  plus 0.5em minus 0.4em\relax CRC press, 2011.

\bibitem{dayan2001theoretical}
P.~Dayan, L.~F. Abbott, and L.~Abbott, ``Theoretical neuroscience:
  computational and mathematical modeling of neural systems,'' 2001.

\bibitem{simpson2015abstract}
A.~J. Simpson, ``Abstract learning via demodulation in a deep neural network,''
  \emph{arXiv preprint arXiv:1502.04042}, 2015.

\bibitem{cutler2008abstract}
A.~Cutler, ``The abstract representations in speech processing,'' \emph{The
  Quarterly Journal of Experimental Psychology}, vol.~61, no.~11, pp.
  1601--1619, 2008.

\bibitem{deng2018abstraction}
F.~Deng, J.~Ren, and F.~Chen, ``Abstraction learning,'' \emph{arXiv preprint
  arXiv:1809.03956}, 2018.

\bibitem{deng2014tutorial}
L.~Deng, ``A tutorial survey of architectures, algorithms, and applications for
  deep learning,'' \emph{APSIPA Transactions on Signal and Information
  Processing}, vol.~3, 2014.

\bibitem{svozil1997introduction}
D.~Svozil, V.~Kvasnicka, and J.~Pospichal, ``Introduction to multi-layer
  feed-forward neural networks,'' \emph{Chemometrics and intelligent laboratory
  systems}, vol.~39, no.~1, pp. 43--62, 1997.

\bibitem{krizhevsky2012imagenet}
A.~Krizhevsky, I.~Sutskever, and G.~E. Hinton, ``Imagenet classification with
  deep convolutional neural networks,'' in \emph{Advances in neural information
  processing systems}, 2012, pp. 1097--1105.

\bibitem{lecun1995convolutional}
Y.~LeCun, Y.~Bengio \emph{et~al.}, ``Convolutional networks for images, speech,
  and time series,'' \emph{The handbook of brain theory and neural networks},
  vol. 3361, no.~10, p. 1995, 1995.

\bibitem{latif2019direct}
S.~Latif, R.~Rana, S.~Khalifa, R.~Jurdak, and J.~Epps, ``Direct modelling of
  speech emotion from raw speech,'' \emph{arXiv preprint arXiv:1904.03833},
  2019.

\bibitem{palaz2015analysis}
D.~Palaz, R.~Collobert \emph{et~al.}, ``Analysis of cnn-based speech
  recognition system using raw speech as input,'' Idiap, Tech. Rep., 2015.

\bibitem{palaz2015convolutional}
D.~Palaz, M.~M. Doss, and R.~Collobert, ``Convolutional neural networks-based
  continuous speech recognition using raw speech signal,'' in \emph{2015 IEEE
  International Conference on Acoustics, Speech and Signal Processing
  (ICASSP)}.\hskip 1em plus 0.5em minus 0.4em\relax IEEE, 2015, pp. 4295--4299.

\bibitem{aldeneh2017using}
Z.~Aldeneh and E.~M. Provost, ``Using regional saliency for speech emotion
  recognition,'' in \emph{2017 IEEE International Conference on Acoustics,
  Speech and Signal Processing (ICASSP)}.\hskip 1em plus 0.5em minus
  0.4em\relax IEEE, 2017, pp. 2741--2745.

\bibitem{sutskever2014sequence}
I.~Sutskever, O.~Vinyals, and Q.~V. Le, ``Sequence to sequence learning with
  neural networks,'' in \emph{Advances in neural information processing
  systems}, 2014, pp. 3104--3112.

\bibitem{cho2014learning}
K.~Cho, B.~Van~Merri{\"e}nboer, C.~Gulcehre, D.~Bahdanau, F.~Bougares,
  H.~Schwenk, and Y.~Bengio, ``Learning phrase representations using {RNN}
  encoder-decoder for statistical machine translation,'' \emph{arXiv preprint
  arXiv:1406.1078}, 2014.

\bibitem{hochreiter1997long}
S.~Hochreiter and J.~Schmidhuber, ``Long short-term memory,'' \emph{Neural
  computation}, vol.~9, no.~8, pp. 1735--1780, 1997.

\bibitem{schuster1997bidirectional}
M.~Schuster and K.~K. Paliwal, ``Bidirectional recurrent neural networks,''
  \emph{IEEE Transactions on Signal Processing}, vol.~45, no.~11, pp.
  2673--2681, 1997.

\bibitem{sainath2019two}
T.~N. Sainath, R.~Pang, D.~Rybach, Y.~He, R.~Prabhavalkar, W.~Li, M.~Visontai,
  Q.~Liang, T.~Strohman, Y.~Wu \emph{et~al.}, ``Two-pass end-to-end speech
  recognition,'' \emph{arXiv preprint arXiv:1908.10992}, 2019.

\bibitem{hinton1994autoencoders}
G.~E. Hinton and R.~S. Zemel, ``Autoencoders, minimum description length and
  helmholtz free energy,'' in \emph{Advances in neural information processing
  systems}, 1994, pp. 3--10.

\bibitem{ng2011sparse}
A.~Ng \emph{et~al.}, ``Sparse autoencoder,'' \emph{CS294A Lecture notes},
  vol.~72, no. 2011, pp. 1--19, 2011.

\bibitem{makhzani2013k}
A.~Makhzani and B.~Frey, ``K-sparse autoencoders,'' \emph{arXiv preprint
  arXiv:1312.5663}, 2013.

\bibitem{deng2013sparse}
J.~Deng, Z.~Zhang, E.~Marchi, and B.~Schuller, ``Sparse autoencoder-based
  feature transfer learning for speech emotion recognition,'' in \emph{2013
  Humaine Association Conference on Affective Computing and Intelligent
  Interaction}.\hskip 1em plus 0.5em minus 0.4em\relax IEEE, 2013, pp.
  511--516.

\bibitem{vincent2008extracting}
P.~Vincent, H.~Larochelle, Y.~Bengio, and P.-A. Manzagol, ``Extracting and
  composing robust features with denoising autoencoders,'' in \emph{Proceedings
  of the 25th international conference on Machine learning}.\hskip 1em plus
  0.5em minus 0.4em\relax ACM, 2008, pp. 1096--1103.

\bibitem{rifai2011contractive}
S.~Rifai, P.~Vincent, X.~Muller, X.~Glorot, and Y.~Bengio, ``Contractive
  auto-encoders: Explicit invariance during feature extraction,'' in
  \emph{Proceedings of the 28th International Conference on International
  Conference on Machine Learning}.\hskip 1em plus 0.5em minus 0.4em\relax
  Omnipress, 2011, pp. 833--840.

\bibitem{hinton2006fast}
G.~E. Hinton, S.~Osindero, and Y.-W. Teh, ``A fast learning algorithm for deep
  belief nets,'' \emph{Neural computation}, vol.~18, no.~7, pp. 1527--1554,
  2006.

\bibitem{ackley1985learning}
D.~H. Ackley, G.~E. Hinton, and T.~J. Sejnowski, ``A learning algorithm for
  boltzmann machines,'' \emph{Cognitive science}, vol.~9, no.~1, pp. 147--169,
  1985.

\bibitem{bengio2014deep}
Y.~Bengio, E.~Laufer, G.~Alain, and J.~Yosinski, ``Deep generative stochastic
  networks trainable by backprop,'' in \emph{International Conference on
  Machine Learning}, 2014, pp. 226--234.

\bibitem{yu2013feature}
D.~Yu, M.~L. Seltzer, J.~Li, J.-T. Huang, and F.~Seide, ``Feature learning in
  deep neural networks-studies on speech recognition tasks,'' \emph{arXiv
  preprint arXiv:1301.3605}, 2013.

\bibitem{li2015constructing}
X.~Li and X.~Wu, ``Constructing long short-term memory based deep recurrent
  neural networks for large vocabulary speech recognition,'' in
  \emph{Acoustics, Speech and Signal Processing (ICASSP), 2015 IEEE
  International Conference on}.\hskip 1em plus 0.5em minus 0.4em\relax IEEE,
  2015, pp. 4520--4524.

\bibitem{radford2015unsupervised}
A.~Radford, L.~Metz, and S.~Chintala, ``Unsupervised representation learning
  with deep convolutional generative adversarial networks,'' \emph{arXiv
  preprint arXiv:1511.06434}, 2015.

\bibitem{donahue2016adversarial}
J.~Donahue, P.~Kr{\"a}henb{\"u}hl, and T.~Darrell, ``Adversarial feature
  learning,'' \emph{arXiv preprint arXiv:1605.09782}, 2016.

\bibitem{chen2016infogan}
X.~Chen, Y.~Duan, R.~Houthooft, J.~Schulman, I.~Sutskever, and P.~Abbeel,
  ``Infogan: Interpretable representation learning by information maximizing
  generative adversarial nets,'' in \emph{Advances in neural information
  processing systems}, 2016, pp. 2172--2180.

\bibitem{higgins2017beta}
I.~Higgins, L.~Matthey, A.~Pal, C.~Burgess, X.~Glorot, M.~Botvinick,
  S.~Mohamed, and A.~Lerchner, ``beta-vae: Learning basic visual concepts with
  a constrained variational framework.'' \emph{ICLR}, vol.~2, no.~5, p.~6,
  2017.

\bibitem{zhao2017infovae}
S.~Zhao, J.~Song, and S.~Ermon, ``Infovae: Information maximizing variational
  autoencoders,'' \emph{arXiv preprint arXiv:1706.02262}, 2017.

\bibitem{gulrajani2016pixelvae}
I.~Gulrajani, K.~Kumar, F.~Ahmed, A.~A. Taiga, F.~Visin, D.~Vazquez, and
  A.~Courville, ``Pixelvae: A latent variable model for natural images,''
  \emph{arXiv preprint arXiv:1611.05013}, 2016.

\bibitem{tschannen2018recent}
M.~Tschannen, O.~Bachem, and M.~Lucic, ``Recent advances in autoencoder-based
  representation learning,'' \emph{arXiv preprint arXiv:1812.05069}, 2018.

\bibitem{van2016conditional}
A.~Van~den Oord, N.~Kalchbrenner, L.~Espeholt, O.~Vinyals, A.~Graves
  \emph{et~al.}, ``Conditional image generation with pixelcnn decoders,'' in
  \emph{Advances in neural information processing systems}, 2016, pp.
  4790--4798.

\bibitem{rethage2018wavenet}
D.~Rethage, J.~Pons, and X.~Serra, ``A wavenet for speech denoising,'' in
  \emph{2018 IEEE International Conference on Acoustics, Speech and Signal
  Processing (ICASSP)}.\hskip 1em plus 0.5em minus 0.4em\relax IEEE, 2018, pp.
  5069--5073.

\bibitem{chorowski2019unsupervised}
J.~Chorowski, R.~J. Weiss, S.~Bengio, and A.~v.~d. Oord, ``Unsupervised speech
  representation learning using wavenet autoencoders,'' \emph{arXiv preprint
  arXiv:1901.08810}, 2019.

\bibitem{lee2009unsupervised}
H.~Lee, P.~Pham, Y.~Largman, and A.~Y. Ng, ``Unsupervised feature learning for
  audio classification using convolutional deep belief networks,'' in
  \emph{Advances in neural information processing systems}, 2009, pp.
  1096--1104.

\bibitem{ali2018speaker}
H.~Ali, S.~N. Tran, E.~Benetos, and A.~S.~d. Garcez, ``Speaker recognition with
  hybrid features from a deep belief network,'' \emph{Neural Computing and
  Applications}, vol.~29, no.~6, pp. 13--19, 2018.

\bibitem{yaman2012bottleneck}
S.~Yaman, J.~Pelecanos, and R.~Sarikaya, ``Bottleneck features for speaker
  recognition,'' in \emph{Odyssey 2012-The Speaker and Language Recognition
  Workshop}, 2012.

\bibitem{cairong2016novel}
Z.~Cairong, Z.~Xinran, Z.~Cheng, and Z.~Li, ``A novel dbn feature fusion model
  for cross-corpus speech emotion recognition,'' \emph{Journal of Electrical
  and Computer Engineering}, vol. 2016, 2016.

\bibitem{dahl2010phone}
G.~Dahl, A.-r. Mohamed, G.~E. Hinton \emph{et~al.}, ``Phone recognition with
  the mean-covariance restricted boltzmann machine,'' in \emph{Advances in
  neural information processing systems}, 2010, pp. 469--477.

\bibitem{muckenhirn2018towards}
H.~Muckenhirn, M.~M. Doss, and S.~Marcell, ``Towards directly modeling raw
  speech signal for speaker verification using cnns,'' in \emph{2018 IEEE
  International Conference on Acoustics, Speech and Signal Processing
  (ICASSP)}.\hskip 1em plus 0.5em minus 0.4em\relax IEEE, 2018, pp. 4884--4888.

\bibitem{tzirakis2018end}
P.~Tzirakis, J.~Zhang, and B.~W. Schuller, ``End-to-end speech emotion
  recognition using deep neural networks,'' in \emph{2018 IEEE International
  Conference on Acoustics, Speech and Signal Processing (ICASSP)}.\hskip 1em
  plus 0.5em minus 0.4em\relax IEEE, 2018, pp. 5089--5093.

\bibitem{sarma2018emotion}
M.~Sarma, P.~Ghahremani, D.~Povey, N.~K. Goel, K.~K. Sarma, and N.~Dehak,
  ``Emotion identification from raw speech signals using dnns.'' in
  \emph{Interspeech}, 2018, pp. 3097--3101.

\bibitem{sainath2015learning}
T.~N. Sainath, R.~J. Weiss, A.~Senior, K.~W. Wilson, and O.~Vinyals, ``Learning
  the speech front-end with raw waveform cldnns,'' in \emph{Sixteenth Annual
  Conference of the International Speech Communication Association}, 2015.

\bibitem{neumann2019improving}
M.~Neumann and N.~T. Vu, ``Improving speech emotion recognition with
  unsupervised representation learning on unlabeled speech,'' in \emph{ICASSP
  2019-2019 IEEE International Conference on Acoustics, Speech and Signal
  Processing (ICASSP)}.\hskip 1em plus 0.5em minus 0.4em\relax IEEE, 2019, pp.
  7390--7394.

\bibitem{hsu2019disentangling}
W.-N. Hsu, Y.~Zhang, R.~J. Weiss, Y.-A. Chung, Y.~Wang, Y.~Wu, and J.~Glass,
  ``Disentangling correlated speaker and noise for speech synthesis via data
  augmentation and adversarial factorization,'' in \emph{ICASSP 2019-2019 IEEE
  International Conference on Acoustics, Speech and Signal Processing
  (ICASSP)}.\hskip 1em plus 0.5em minus 0.4em\relax IEEE, 2019, pp. 5901--5905.

\bibitem{feng2014speech}
X.~Feng, Y.~Zhang, and J.~Glass, ``Speech feature denoising and dereverberation
  via deep autoencoders for noisy reverberant speech recognition,'' in
  \emph{2014 IEEE international conference on acoustics, speech and signal
  processing (ICASSP)}.\hskip 1em plus 0.5em minus 0.4em\relax IEEE, 2014, pp.
  1759--1763.

\bibitem{weninger2014deep}
F.~Weninger, S.~Watanabe, Y.~Tachioka, and B.~Schuller, ``Deep recurrent
  de-noising auto-encoder and blind de-reverberation for reverberated speech
  recognition,'' in \emph{2014 IEEE International Conference on Acoustics,
  Speech and Signal Processing (ICASSP)}.\hskip 1em plus 0.5em minus
  0.4em\relax IEEE, 2014, pp. 4623--4627.

\bibitem{zhao2015music}
M.~Zhao, D.~Wang, Z.~Zhang, and X.~Zhang, ``Music removal by convolutional
  denoising autoencoder in speech recognition,'' in \emph{2015 Asia-Pacific
  Signal and Information Processing Association Annual Summit and Conference
  (APSIPA)}.\hskip 1em plus 0.5em minus 0.4em\relax IEEE, 2015, pp. 338--341.

\bibitem{sailor2016unsupervised}
H.~B. Sailor and H.~A. Patil, ``Unsupervised deep auditory model using stack of
  convolutional rbms for speech recognition.'' in \emph{INTERSPEECH}, 2016, pp.
  3379--3383.

\bibitem{mohamed2009deep}
A.-r. Mohamed, G.~Dahl, and G.~Hinton, ``Deep belief networks for phone
  recognition,'' in \emph{Nips workshop on deep learning for speech recognition
  and related applications}, vol.~1, no.~9.\hskip 1em plus 0.5em minus
  0.4em\relax Vancouver, Canada, 2009, p.~39.

\bibitem{ghosh2016representation}
S.~Ghosh, E.~Laksana, L.-P. Morency, and S.~Scherer, ``Representation learning
  for speech emotion recognition.'' in \emph{Interspeech}, 2016, pp.
  3603--3607.

\bibitem{ghosh2015learning}
------, ``Learning representations of affect from speech,'' \emph{arXiv
  preprint arXiv:1511.04747}, 2015.

\bibitem{huang2015speech}
Z.-w. Huang, W.-t. Xue, and Q.-r. Mao, ``Speech emotion recognition with
  unsupervised feature learning,'' \emph{Frontiers of Information Technology \&
  Electronic Engineering}, vol.~16, no.~5, pp. 358--366, 2015.

\bibitem{xia2014modeling}
R.~Xia, J.~Deng, B.~Schuller, and Y.~Liu, ``Modeling gender information for
  emotion recognition using denoising autoencoder,'' in \emph{2014 IEEE
  International Conference on Acoustics, Speech and Signal Processing
  (ICASSP)}.\hskip 1em plus 0.5em minus 0.4em\relax IEEE, 2014, pp. 990--994.

\bibitem{xia2013using}
R.~Xia and Y.~Liu, ``Using denoising autoencoder for emotion recognition.'' in
  \emph{Interspeech}, 2013, pp. 2886--2889.

\bibitem{chang2017learning}
J.~Chang and S.~Scherer, ``Learning representations of emotional speech with
  deep convolutional generative adversarial networks,'' in \emph{Acoustics,
  Speech and Signal Processing (ICASSP), 2017 IEEE International Conference
  on}.\hskip 1em plus 0.5em minus 0.4em\relax IEEE, 2017, pp. 2746--2750.

\bibitem{yu2017adversarial}
H.~Yu, Z.-H. Tan, Z.~Ma, and J.~Guo, ``Adversarial network bottleneck features
  for noise robust speaker verification,'' \emph{arXiv preprint
  arXiv:1706.03397}, 2017.

\bibitem{donahue2018synthesizing}
C.~Donahue, J.~McAuley, and M.~Puckette, ``Synthesizing audio with generative
  adversarial networks,'' \emph{arXiv preprint arXiv:1802.04208}, 2018.

\bibitem{sahu2018adversarial}
S.~Sahu, R.~Gupta, G.~Sivaraman, W.~AbdAlmageed, and C.~Espy-Wilson,
  ``Adversarial auto-encoders for speech based emotion recognition,''
  \emph{arXiv preprint arXiv:1806.02146}, 2018.

\bibitem{ravanelli2018learning}
M.~Ravanelli and Y.~Bengio, ``Learning speaker representations with mutual
  information,'' \emph{arXiv preprint arXiv:1812.00271}, 2018.

\bibitem{latif2018transfer}
S.~Latif, R.~Rana, S.~Younis, J.~Qadir, and J.~Epps, ``Transfer learning for
  improving speech emotion classification accuracy,'' \emph{arXiv preprint
  arXiv:1801.06353}, 2018.

\bibitem{latif2019unsupervised}
S.~Latif, J.~Qadir, and M.~Bilal, ``Unsupervised adversarial domain adaptation
  for cross-lingual speech emotion recognition,'' \emph{arXiv preprint
  arXiv:1907.06083}, 2019.

\bibitem{latif2018cross}
S.~Latif, A.~Qayyum, M.~Usman, and J.~Qadir, ``Cross lingual speech emotion
  recognition: Urdu vs. western languages,'' in \emph{2018 International
  Conference on Frontiers of Information Technology (FIT)}.\hskip 1em plus
  0.5em minus 0.4em\relax IEEE, 2018, pp. 88--93.

\bibitem{rana2019multi}
R.~Rana, S.~Latif, S.~Khalifa, and R.~Jurdak, ``Multi-task semi-supervised
  adversarial autoencoding for speech emotion,'' \emph{arXiv preprint
  arXiv:1907.06078}, 2019.

\bibitem{zhu2005semi}
X.~J. Zhu, ``Semi-supervised learning literature survey,'' University of
  Wisconsin-Madison Department of Computer Sciences, Tech. Rep., 2005.

\bibitem{Huang}
Z.~Huang, M.~Dong, Q.~Mao, and Y.~Zhan, ``Speech emotion recognition using
  cnn,'' in \emph{Proceedings of the 22Nd ACM International Conference on
  Multimedia}, 2014.

\bibitem{huang2018speech}
J.~Huang, Y.~Li, J.~Tao, Z.~Lian, M.~Niu, and J.~Yi, ``Speech emotion
  recognition using semi-supervised learning with ladder networks,'' in
  \emph{2018 First Asian Conference on Affective Computing and Intelligent
  Interaction (ACII Asia)}.\hskip 1em plus 0.5em minus 0.4em\relax IEEE, 2018,
  pp. 1--5.

\bibitem{parthasarathy2019semi}
S.~Parthasarathy and C.~Busso, ``Semi-supervised speech emotion recognition
  with ladder networks,'' \emph{arXiv preprint arXiv:1905.02921}, 2019.

\bibitem{parthasarathy2018ladder}
------, ``Ladder networks for emotion recognition: Using unsupervised auxiliary
  tasks to improve predictions of emotional attributes,'' \emph{Proc.
  Interspeech 2018}, pp. 3698--3702, 2018.

\bibitem{deng2018semisupervised}
J.~Deng, X.~Xu, Z.~Zhang, S.~Fr{\"u}hholz, and B.~Schuller, ``Semisupervised
  autoencoders for speech emotion recognition,'' \emph{IEEE/ACM Transactions on
  Audio, Speech, and Language Processing}, vol.~26, no.~1, pp. 31--43, 2018.

\bibitem{thomas2013deep}
S.~Thomas, M.~L. Seltzer, K.~Church, and H.~Hermansky, ``Deep neural network
  features and semi-supervised training for low resource speech recognition,''
  in \emph{2013 IEEE international conference on acoustics, speech and signal
  processing}.\hskip 1em plus 0.5em minus 0.4em\relax IEEE, 2013, pp.
  6704--6708.

\bibitem{cui2015multilingual}
J.~Cui, B.~Kingsbury, B.~Ramabhadran, A.~Sethy, K.~Audhkhasi, X.~Cui,
  E.~Kislal, L.~Mangu, M.~Nussbaum-Thom, M.~Picheny \emph{et~al.},
  ``Multilingual representations for low resource speech recognition and
  keyword search,'' in \emph{2015 IEEE Workshop on Automatic Speech Recognition
  and Understanding (ASRU)}.\hskip 1em plus 0.5em minus 0.4em\relax IEEE, 2015,
  pp. 259--266.

\bibitem{karita2018semi}
S.~Karita, S.~Watanabe, T.~Iwata, A.~Ogawa, and M.~Delcroix, ``Semi-supervised
  end-to-end speech recognition.'' in \emph{Interspeech}, 2018, pp. 2--6.

\bibitem{tu2014speech}
Y.~Tu, J.~Du, Y.~Xu, L.~Dai, and C.-H. Lee, ``Speech separation based on
  improved deep neural networks with dual outputs of speech features for both
  target and interfering speakers,'' in \emph{The 9th International Symposium
  on Chinese Spoken Language Processing}.\hskip 1em plus 0.5em minus
  0.4em\relax IEEE, 2014, pp. 250--254.

\bibitem{pal2019study}
M.~Pal, M.~Kumar, R.~Peri, and S.~Narayanan, ``A study of semi-supervised
  speaker diarization system using {GAN} mixture model,'' \emph{arXiv preprint
  arXiv:1910.11416}, 2019.

\bibitem{bengio2012deep}
Y.~Bengio, ``Deep learning of representations for unsupervised and transfer
  learning,'' in \emph{Proceedings of ICML workshop on unsupervised and
  transfer learning}, 2012, pp. 17--36.

\bibitem{thrun2012learning}
S.~Thrun and L.~Pratt, \emph{Learning to learn}.\hskip 1em plus 0.5em minus
  0.4em\relax Springer Science \& Business Media, 2012.

\bibitem{sun2017unsupervised}
S.~Sun, B.~Zhang, L.~Xie, and Y.~Zhang, ``An unsupervised deep domain
  adaptation approach for robust speech recognition,'' \emph{Neurocomputing},
  vol. 257, pp. 79--87, 2017.

\bibitem{swietojanski2016learning}
P.~Swietojanski, J.~Li, and S.~Renals, ``Learning hidden unit contributions for
  unsupervised acoustic model adaptation,'' \emph{IEEE/ACM Transactions on
  Audio, Speech, and Language Processing}, vol.~24, no.~8, pp. 1450--1463,
  2016.

\bibitem{hsu2018extracting}
W.-N. Hsu and J.~Glass, ``Extracting domain invariant features by unsupervised
  learning for robust automatic speech recognition,'' in \emph{2018 IEEE
  International Conference on Acoustics, Speech and Signal Processing
  (ICASSP)}.\hskip 1em plus 0.5em minus 0.4em\relax IEEE, 2018, pp. 5614--5618.

\bibitem{tang2018study}
H.~Tang, W.-N. Hsu, F.~Grondin, and J.~Glass, ``A study of enhancement,
  augmentation, and autoencoder methods for domain adaptation in distant speech
  recognition,'' \emph{arXiv preprint arXiv:1806.04841}, 2018.

\bibitem{hsu2018unsupervised}
W.-N. Hsu, H.~Tang, and J.~Glass, ``Unsupervised adaptation with interpretable
  disentangled representations for distant conversational speech recognition,''
  \emph{arXiv preprint arXiv:1806.04872}, 2018.

\bibitem{fan2016unsupervised}
Y.~Fan, Y.~Qian, F.~K. Soong, and L.~He, ``Unsupervised speaker adaptation for
  dnn-based tts synthesis,'' in \emph{2016 IEEE International Conference on
  Acoustics, Speech and Signal Processing (ICASSP)}.\hskip 1em plus 0.5em minus
  0.4em\relax IEEE, 2016, pp. 5135--5139.

\bibitem{qin2018towards}
C.-X. Qin, D.~Qu, and L.-H. Zhang, ``Towards end-to-end speech recognition with
  transfer learning,'' \emph{EURASIP Journal on Audio, Speech, and Music
  Processing}, vol. 2018, no.~1, p.~18, 2018.

\bibitem{huang2013cross}
J.-T. Huang, J.~Li, D.~Yu, L.~Deng, and Y.~Gong, ``Cross-language knowledge
  transfer using multilingual deep neural network with shared hidden layers,''
  in \emph{2013 IEEE International Conference on Acoustics, Speech and Signal
  Processing}.\hskip 1em plus 0.5em minus 0.4em\relax IEEE, 2013, pp.
  7304--7308.

\bibitem{knill2014language}
K.~M. Knill, M.~J. Gales, A.~Ragni, and S.~P. Rath, ``Language independent and
  unsupervised acoustic models for speech recognition and keyword spotting,''
  2014.

\bibitem{denisov2018unsupervised}
P.~Denisov, N.~T. Vu, and M.~F. Font, ``Unsupervised domain adaptation by
  adversarial learning for robust speech recognition,'' in \emph{Speech
  Communication; 13th ITG-Symposium}.\hskip 1em plus 0.5em minus 0.4em\relax
  VDE, 2018, pp. 1--5.

\bibitem{sun2018domain}
S.~Sun, C.-F. Yeh, M.-Y. Hwang, M.~Ostendorf, and L.~Xie, ``Domain adversarial
  training for accented speech recognition,'' in \emph{2018 IEEE International
  Conference on Acoustics, Speech and Signal Processing (ICASSP)}.\hskip 1em
  plus 0.5em minus 0.4em\relax IEEE, 2018, pp. 4854--4858.

\bibitem{shinohara2016adversarial}
Y.~Shinohara, ``Adversarial multi-task learning of deep neural networks for
  robust speech recognition.'' in \emph{INTERSPEECH}.\hskip 1em plus 0.5em
  minus 0.4em\relax San Francisco, CA, USA, 2016, pp. 2369--2372.

\bibitem{hosseini2018multi}
E.~Hosseini-Asl, Y.~Zhou, C.~Xiong, and R.~Socher, ``A multi-discriminator
  cyclegan for unsupervised non-parallel speech domain adaptation,''
  \emph{arXiv preprint arXiv:1804.00522}, 2018.

\bibitem{meng2018adversarial}
Z.~Meng, J.~Li, Y.~Gong, and B.-H. Juang, ``Adversarial teacher-student
  learning for unsupervised domain adaptation,'' in \emph{2018 IEEE
  International Conference on Acoustics, Speech and Signal Processing
  (ICASSP)}.\hskip 1em plus 0.5em minus 0.4em\relax IEEE, 2018, pp. 5949--5953.

\bibitem{meng2018speaker}
Z.~Meng, J.~Li, Z.~Chen, Y.~Zhao, V.~Mazalov, Y.~Gang, and B.-H. Juang,
  ``Speaker-invariant training via adversarial learning,'' in \emph{2018 IEEE
  International Conference on Acoustics, Speech and Signal Processing
  (ICASSP)}.\hskip 1em plus 0.5em minus 0.4em\relax IEEE, 2018, pp. 5969--5973.

\bibitem{meng2019adversarial}
Z.~Meng, J.~Li, and Y.~Gong, ``Adversarial speaker adaptation,'' in
  \emph{ICASSP 2019-2019 IEEE International Conference on Acoustics, Speech and
  Signal Processing (ICASSP)}.\hskip 1em plus 0.5em minus 0.4em\relax IEEE,
  2019, pp. 5721--5725.

\bibitem{tripathi2018adversarial}
A.~Tripathi, A.~Mohan, S.~Anand, and M.~Singh, ``Adversarial learning of raw
  speech features for domain invariant speech recognition,'' in \emph{2018 IEEE
  International Conference on Acoustics, Speech and Signal Processing
  (ICASSP)}.\hskip 1em plus 0.5em minus 0.4em\relax IEEE, 2018, pp. 5959--5963.

\bibitem{shon2017autoencoder}
S.~Shon, S.~Mun, W.~Kim, and H.~Ko, ``Autoencoder based domain adaptation for
  speaker recognition under insufficient channel information,'' \emph{arXiv
  preprint arXiv:1708.01227}, 2017.

\bibitem{wang2018unsupervised}
Q.~Wang, W.~Rao, S.~Sun, L.~Xie, E.~S. Chng, and H.~Li, ``Unsupervised domain
  adaptation via domain adversarial training for speaker recognition,'' in
  \emph{2018 IEEE International Conference on Acoustics, Speech and Signal
  Processing (ICASSP)}.\hskip 1em plus 0.5em minus 0.4em\relax IEEE, 2018, pp.
  4889--4893.

\bibitem{bhattacharya2019generative}
G.~Bhattacharya, J.~Monteiro, J.~Alam, and P.~Kenny, ``Generative adversarial
  speaker embedding networks for domain robust end-to-end speaker
  verification,'' in \emph{ICASSP 2019-2019 IEEE International Conference on
  Acoustics, Speech and Signal Processing (ICASSP)}.\hskip 1em plus 0.5em minus
  0.4em\relax IEEE, 2019, pp. 6226--6230.

\bibitem{deng2014autoencoder}
J.~Deng, Z.~Zhang, F.~Eyben, and B.~Schuller, ``Autoencoder-based unsupervised
  domain adaptation for speech emotion recognition,'' \emph{IEEE Signal
  Processing Letters}, vol.~21, no.~9, pp. 1068--1072, 2014.

\bibitem{deng2017universum}
J.~Deng, X.~Xu, Z.~Zhang, S.~Fr{\"u}hholz, and B.~Schuller, ``Universum
  autoencoder-based domain adaptation for speech emotion recognition,''
  \emph{IEEE Signal Processing Letters}, vol.~24, no.~4, pp. 500--504, 2017.

\bibitem{zhou2018transferable}
H.~Zhou and K.~Chen, ``Transferable positive/negative speech emotion
  recognition via class-wise adversarial domain adaptation,'' \emph{arXiv
  preprint arXiv:1810.12782}, 2018.

\bibitem{gideon2019barking}
J.~Gideon, M.~G. McInnis, and E.~Mower~Provost, ``Barking up the right tree:
  Improving cross-corpus speech emotion recognition with adversarial
  discriminative domain generalization (addog),'' \emph{arXiv preprint
  arXiv:1903.12094}, 2019.

\bibitem{collobert2008unified}
R.~Collobert and J.~Weston, ``A unified architecture for natural language
  processing: Deep neural networks with multitask learning,'' in
  \emph{Proceedings of the 25th international conference on Machine
  learning}.\hskip 1em plus 0.5em minus 0.4em\relax ACM, 2008, pp. 160--167.

\bibitem{deng2013new}
L.~Deng, G.~Hinton, and B.~Kingsbury, ``New types of deep neural network
  learning for speech recognition and related applications: An overview,'' in
  \emph{2013 IEEE International Conference on Acoustics, Speech and Signal
  Processing}.\hskip 1em plus 0.5em minus 0.4em\relax IEEE, 2013, pp.
  8599--8603.

\bibitem{girshick2015fast}
R.~Girshick, ``Fast r-cnn,'' in \emph{Proceedings of the IEEE international
  conference on computer vision}, 2015, pp. 1440--1448.

\bibitem{caruana1998learning}
R.~Caruana, ``Learning to learn, chapter multitask learning,'' 1998.

\bibitem{stadermann2005multi}
J.~Stadermann, W.~Koska, and G.~Rigoll, ``Multi-task learning strategies for a
  recurrent neural net in a hybrid tied-posteriors acoustic model,'' in
  \emph{Ninth European Conference on Speech Communication and Technology},
  2005.

\bibitem{huang2015rapid}
Z.~Huang, J.~Li, S.~M. Siniscalchi, I.-F. Chen, J.~Wu, and C.-H. Lee, ``Rapid
  adaptation for deep neural networks through multi-task learning,'' in
  \emph{Sixteenth Annual Conference of the International Speech Communication
  Association}, 2015.

\bibitem{price2014speaker}
R.~Price, K.-i. Iso, and K.~Shinoda, ``Speaker adaptation of deep neural
  networks using a hierarchy of output layers,'' in \emph{2014 IEEE Spoken
  Language Technology Workshop (SLT)}.\hskip 1em plus 0.5em minus 0.4em\relax
  IEEE, 2014, pp. 153--158.

\bibitem{lu2004multitask}
Y.~Lu, F.~Lu, S.~Sehgal, S.~Gupta, J.~Du, C.~H. Tham, P.~Green, and V.~Wan,
  ``Multitask learning in connectionist speech recognition,'' in
  \emph{Proceedings of the Australian International Conference on Speech
  Science and Technology}, 2004.

\bibitem{chen2015speech}
Z.~Chen, S.~Watanabe, H.~Erdogan, and J.~R. Hershey, ``Speech enhancement and
  recognition using multi-task learning of long short-term memory recurrent
  neural networks,'' in \emph{Sixteenth Annual Conference of the International
  Speech Communication Association}, 2015.

\bibitem{han2019emobed}
J.~Han, Z.~Zhang, Z.~Ren, and B.~W. Schuller, ``Emobed: Strengthening monomodal
  emotion recognition via training with crossmodal emotion embeddings,''
  \emph{IEEE Transactions on Affective Computing}, 2019.

\bibitem{kim2017towards}
J.~Kim, G.~Englebienne, K.~P. Truong, and V.~Evers, ``Towards speech emotion
  recognition" in the wild" using aggregated corpora and deep multi-task
  learning,'' \emph{arXiv preprint arXiv:1708.03920}, 2017.

\bibitem{parthasarathy2017jointly}
S.~Parthasarathy and C.~Busso, ``Jointly predicting arousal, valence and
  dominance with multi-task learning,'' \emph{INTERSPEECH, Stockholm, Sweden},
  2017.

\bibitem{xia2017multi}
R.~Xia and Y.~Liu, ``A multi-task learning framework for emotion recognition
  using {2D} continuous space,'' \emph{IEEE Transactions on Affective
  Computing}, no.~1, pp. 3--14, 2017.

\bibitem{Lotfian2018}
\BIBentryALTinterwordspacing
R.~Lotfian and C.~Busso, ``Predicting categorical emotions by jointly learning
  primary and secondary emotions through multitask learning,'' in \emph{Proc.
  Interspeech 2018}, 2018, pp. 951--955. [Online]. Available:
  \url{http://dx.doi.org/10.21437/Interspeech.2018-2464}
\BIBentrySTDinterwordspacing

\bibitem{tao2018advanced}
F.~Tao and G.~Liu, ``Advanced {LSTM}: A study about better time dependency
  modeling in emotion recognition,'' in \emph{2018 IEEE International
  Conference on Acoustics, Speech and Signal Processing (ICASSP)}.\hskip 1em
  plus 0.5em minus 0.4em\relax IEEE, 2018, pp. 2906--2910.

\bibitem{zhang2017cross}
B.~Zhang, E.~M. Provost, and G.~Essl, ``Cross-corpus acoustic emotion
  recognition with multi-task learning: Seeking common ground while preserving
  differences,'' \emph{IEEE Transactions on Affective Computing}, no.~1, pp.
  1--1, 2017.

\bibitem{pironkov2016multi}
G.~Pironkov, S.~Dupont, and T.~Dutoit, ``Multi-task learning for speech
  recognition: an overview.'' in \emph{ESANN}, 2016.

\bibitem{raina2007self}
R.~Raina, A.~Battle, H.~Lee, B.~Packer, and A.~Y. Ng, ``Self-taught learning:
  transfer learning from unlabeled data,'' in \emph{Proceedings of the 24th
  international conference on Machine learning}.\hskip 1em plus 0.5em minus
  0.4em\relax ACM, 2007, pp. 759--766.

\bibitem{ons2013self}
B.~Ons, N.~Tessema, J.~Van De~Loo, J.~Gemmeke, G.~De~Pauw, W.~Daelemans, and
  H.~Van~Hamme, ``A self learning vocal interface for speech-impaired users,''
  in \emph{Proceedings of the Fourth Workshop on Speech and Language Processing
  for Assistive Technologies}, 2013, pp. 73--81.

\bibitem{feng2018autoencoder}
S.~Feng and M.~F. Duarte, ``Autoencoder based sample selection for self-taught
  learning,'' \emph{arXiv preprint arXiv:1808.01574}, 2018.

\bibitem{kober2013reinforcement}
J.~Kober, J.~A. Bagnell, and J.~Peters, ``Reinforcement learning in robotics: A
  survey,'' \emph{The International Journal of Robotics Research}, vol.~32,
  no.~11, pp. 1238--1274, 2013.

\bibitem{arulkumaran2017deep}
K.~Arulkumaran, M.~P. Deisenroth, M.~Brundage, and A.~A. Bharath, ``Deep
  reinforcement learning: A brief survey,'' \emph{IEEE Signal Processing
  Magazine}, vol.~34, no.~6, pp. 26--38, 2017.

\bibitem{gelada2019deepmdp}
C.~Gelada, S.~Kumar, J.~Buckman, O.~Nachum, and M.~G. Bellemare, ``Deepmdp:
  Learning continuous latent space models for representation learning,''
  \emph{arXiv preprint arXiv:1906.02736}, 2019.

\bibitem{zhang2018learning}
T.~Zhang, M.~Huang, and L.~Zhao, ``Learning structured representation for text
  classification via reinforcement learning,'' in \emph{Thirty-Second AAAI
  Conference on Artificial Intelligence}, 2018.

\bibitem{cuayahuitl2006reinforcement}
H.~Cuay{\'a}huitl, S.~Renals, O.~Lemon, and H.~Shimodaira, ``Reinforcement
  learning of dialogue strategies with hierarchical abstract machines,'' in
  \emph{2006 IEEE Spoken Language Technology Workshop}.\hskip 1em plus 0.5em
  minus 0.4em\relax IEEE, 2006, pp. 182--185.

\bibitem{li2016deep}
J.~Li, W.~Monroe, A.~Ritter, M.~Galley, J.~Gao, and D.~Jurafsky, ``Deep
  reinforcement learning for dialogue generation,'' \emph{arXiv preprint
  arXiv:1606.01541}, 2016.

\bibitem{lee1990corrective}
K.-F. Lee and S.~Mahajan, ``Corrective and reinforcement learning for
  speaker-independent continuous speech recognition,'' \emph{Computer Speech \&
  Language}, vol.~4, no.~3, pp. 231--245, 1990.

\bibitem{lakomkin2018emorl}
E.~Lakomkin, M.~A. Zamani, C.~Weber, S.~Magg, and S.~Wermter, ``Emorl:
  continuous acoustic emotion classification using deep reinforcement
  learning,'' in \emph{2018 IEEE International Conference on Robotics and
  Automation (ICRA)}.\hskip 1em plus 0.5em minus 0.4em\relax IEEE, 2018, pp.
  1--6.

\bibitem{zhang2017active}
Y.~Zhang, M.~Lease, and B.~C. Wallace, ``Active discriminative text
  representation learning,'' in \emph{Thirty-First AAAI Conference on
  Artificial Intelligence}, 2017.

\bibitem{hakkani2004unsupervised}
D.~Hakkani-Tur, G.~Tur, M.~Rahim, and G.~Riccardi, ``Unsupervised and active
  learning in automatic speech recognition for call classification,'' in
  \emph{2004 IEEE International Conference on Acoustics, Speech, and Signal
  Processing}, vol.~1.\hskip 1em plus 0.5em minus 0.4em\relax IEEE, 2004, pp.
  I--429.

\bibitem{riccardi2005active}
G.~Riccardi and D.~Hakkani-Tur, ``Active learning: Theory and applications to
  automatic speech recognition,'' \emph{IEEE transactions on speech and audio
  processing}, vol.~13, no.~4, pp. 504--511, 2005.

\bibitem{huang2016active}
J.~Huang, R.~Child, V.~Rao, H.~Liu, S.~Satheesh, and A.~Coates, ``Active
  learning for speech recognition: the power of gradients,'' \emph{arXiv
  preprint arXiv:1612.03226}, 2016.

\bibitem{zhang2015cooperative}
Z.~Zhang, E.~Coutinho, J.~Deng, and B.~Schuller, ``Cooperative learning and its
  application to emotion recognition from speech,'' \emph{IEEE/ACM Transactions
  on Audio, Speech and Language Processing (TASLP)}, vol.~23, no.~1, pp.
  115--126, 2015.

\bibitem{dong2003human}
M.~Dong and Z.~Sun, ``On human machine cooperative learning control,'' in
  \emph{Proceedings of the 2003 IEEE International Symposium on Intelligent
  Control}.\hskip 1em plus 0.5em minus 0.4em\relax IEEE, 2003, pp. 81--86.

\bibitem{schuller2015speech}
B.~W. Schuller, ``Speech analysis in the big data era,'' in \emph{International
  Conference on Text, Speech, and Dialogue}.\hskip 1em plus 0.5em minus
  0.4em\relax Springer, 2015, pp. 3--11.

\bibitem{wagner2018applying}
J.~Wagner, T.~Baur, Y.~Zhang, M.~F. Valstar, B.~Schuller, and E.~Andr{\'e},
  ``Applying cooperative machine learning to speed up the annotation of social
  signals in large multi-modal corpora,'' \emph{arXiv preprint
  arXiv:1802.02565}, 2018.

\bibitem{zhang2018leveraging}
Z.~Zhang, J.~Han, J.~Deng, X.~Xu, F.~Ringeval, and B.~Schuller, ``Leveraging
  unlabeled data for emotion recognition with enhanced collaborative
  semi-supervised learning,'' \emph{IEEE Access}, vol.~6, pp. 22\,196--22\,209,
  2018.

\bibitem{tu2015speech}
Y.-H. Tu, J.~Du, L.-R. Dai, and C.-H. Lee, ``Speech separation based on
  signal-noise-dependent deep neural networks for robust speech recognition,''
  in \emph{2015 IEEE International Conference on Acoustics, Speech and Signal
  Processing (ICASSP)}.\hskip 1em plus 0.5em minus 0.4em\relax IEEE, 2015, pp.
  61--65.

\bibitem{narayanan2014investigation}
A.~Narayanan and D.~Wang, ``Investigation of speech separation as a front-end
  for noise robust speech recognition,'' \emph{IEEE/ACM Transactions on Audio,
  Speech, and Language Processing}, vol.~22, no.~4, pp. 826--835, 2014.

\bibitem{liu2014graph}
Y.~Liu and K.~Kirchhoff, ``Graph-based semi-supervised acoustic modeling in
  dnn-based speech recognition,'' in \emph{2014 IEEE Spoken Language Technology
  Workshop (SLT)}.\hskip 1em plus 0.5em minus 0.4em\relax IEEE, 2014, pp.
  177--182.

\bibitem{mohamed2010investigation}
A.-r. Mohamed, D.~Yu, and L.~Deng, ``Investigation of full-sequence training of
  deep belief networks for speech recognition,'' in \emph{Eleventh Annual
  Conference of the International Speech Communication Association}, 2010.

\bibitem{mohamed2011deep}
A.-r. Mohamed, T.~N. Sainath, G.~E. Dahl, B.~Ramabhadran, G.~E. Hinton, M.~A.
  Picheny \emph{et~al.}, ``Deep belief networks using discriminative features
  for phone recognition.'' in \emph{ICASSP}, 2011, pp. 5060--5063.

\bibitem{gholamipoor2014feature}
M.~Gholamipoor and B.~Nasersharif, ``Feature mapping using deep belief networks
  for robust speech recognition,'' \emph{The Modares Journal of Electrical
  Engineering}, vol.~14, no.~3, pp. 24--30, 2014.

\bibitem{huang2013audio}
J.~Huang and B.~Kingsbury, ``Audio-visual deep learning for noise robust speech
  recognition,'' in \emph{2013 IEEE International Conference on Acoustics,
  Speech and Signal Processing}.\hskip 1em plus 0.5em minus 0.4em\relax IEEE,
  2013, pp. 7596--7599.

\bibitem{hu2016dbn}
Y.-J. Hu and Z.-H. Ling, ``Dbn-based spectral feature representation for
  statistical parametric speech synthesis,'' \emph{IEEE Signal Processing
  Letters}, vol.~23, no.~3, pp. 321--325, 2016.

\bibitem{yu2010roles}
D.~Yu, L.~Deng, and G.~Dahl, ``Roles of pre-training and fine-tuning in
  context-dependent dbn-hmms for real-world speech recognition,'' in
  \emph{Proc. NIPS Workshop on Deep Learning and Unsupervised Feature
  Learning}, 2010.

\bibitem{yu2011improved}
D.~Yu and M.~L. Seltzer, ``Improved bottleneck features using pretrained deep
  neural networks,'' in \emph{Twelfth annual conference of the international
  speech communication association}, 2011.

\bibitem{wu2015deep}
Z.~Wu, C.~Valentini-Botinhao, O.~Watts, and S.~King, ``Deep neural networks
  employing multi-task learning and stacked bottleneck features for speech
  synthesis,'' in \emph{2015 IEEE international conference on acoustics, speech
  and signal processing (ICASSP)}.\hskip 1em plus 0.5em minus 0.4em\relax IEEE,
  2015, pp. 4460--4464.

\bibitem{yu2012exploiting}
D.~Yu, F.~Seide, G.~Li, and L.~Deng, ``Exploiting sparseness in deep neural
  networks for large vocabulary speech recognition,'' in \emph{2012 IEEE
  International conference on acoustics, speech and signal processing
  (ICASSP)}.\hskip 1em plus 0.5em minus 0.4em\relax IEEE, 2012, pp. 4409--4412.

\bibitem{dighe2016exploiting}
P.~Dighe, G.~Luyet, A.~Asaei, and H.~Bourlard, ``Exploiting low-dimensional
  structures to enhance {DNN} based acoustic modeling in speech recognition,''
  in \emph{2016 IEEE International Conference on Acoustics, Speech and Signal
  Processing (ICASSP)}.\hskip 1em plus 0.5em minus 0.4em\relax IEEE, 2016, pp.
  5690--5694.

\bibitem{abdel2014convolutional}
O.~Abdel-Hamid, A.-r. Mohamed, H.~Jiang, L.~Deng, G.~Penn, and D.~Yu,
  ``Convolutional neural networks for speech recognition,'' \emph{IEEE/ACM
  Transactions on audio, speech, and language processing}, vol.~22, no.~10, pp.
  1533--1545, 2014.

\bibitem{mitra2015time}
V.~Mitra and H.~Franco, ``Time-frequency convolutional networks for robust
  speech recognition,'' in \emph{2015 IEEE Workshop on Automatic Speech
  Recognition and Understanding (ASRU)}.\hskip 1em plus 0.5em minus 0.4em\relax
  IEEE, 2015, pp. 317--323.

\bibitem{serdyuk2016invariant}
D.~Serdyuk, K.~Audhkhasi, P.~Brakel, B.~Ramabhadran, S.~Thomas, and Y.~Bengio,
  ``Invariant representations for noisy speech recognition,'' \emph{arXiv
  preprint arXiv:1612.01928}, 2016.

\bibitem{campbell2014using}
W.~M. Campbell, ``Using deep belief networks for vector-based speaker
  recognition,'' in \emph{Fifteenth Annual Conference of the International
  Speech Communication Association}, 2014.

\bibitem{ghahabi2014vector}
O.~Ghahabi and J.~Hernando, ``I-vector modeling with deep belief networks for
  multi-session speaker recognition,'' \emph{network}, vol.~20, p.~13, 2014.

\bibitem{vasilakakis2013speaker}
V.~Vasilakakis, S.~Cumani, P.~Laface, and P.~Torino, ``Speaker recognition by
  means of deep belief networks,'' \emph{Proc. Biometric Technologies in
  Forensic Science}, pp. 52--57, 2013.

\bibitem{yamada2013improvement}
T.~Yamada, L.~Wang, and A.~Kai, ``Improvement of distant-talking speaker
  identification using bottleneck features of dnn.'' in \emph{Interspeech},
  2013, pp. 3661--3664.

\bibitem{heigold2016end}
G.~Heigold, I.~Moreno, S.~Bengio, and N.~Shazeer, ``End-to-end text-dependent
  speaker verification,'' in \emph{2016 IEEE International Conference on
  Acoustics, Speech and Signal Processing (ICASSP)}.\hskip 1em plus 0.5em minus
  0.4em\relax IEEE, 2016, pp. 5115--5119.

\bibitem{heo2017joint}
H.-s. Heo, J.-w. Jung, I.-h. Yang, S.-h. Yoon, and H.-j. Yu, ``Joint training
  of expanded end-to-end {DNN} for text-dependent speaker verification.'' in
  \emph{INTERSPEECH}, 2017, pp. 1532--1536.

\bibitem{huang2015investigation}
H.~Huang and K.~C. Sim, ``An investigation of augmenting speaker
  representations to improve speaker normalisation for dnn-based speech
  recognition,'' in \emph{2015 IEEE International Conference on Acoustics,
  Speech and Signal Processing (ICASSP)}.\hskip 1em plus 0.5em minus
  0.4em\relax IEEE, 2015, pp. 4610--4613.

\bibitem{snyder2018x}
D.~Snyder, D.~Garcia-Romero, G.~Sell, D.~Povey, and S.~Khudanpur, ``X-vectors:
  Robust {DNN} embeddings for speaker recognition,'' in \emph{2018 IEEE
  International Conference on Acoustics, Speech and Signal Processing
  (ICASSP)}.\hskip 1em plus 0.5em minus 0.4em\relax IEEE, 2018, pp. 5329--5333.

\bibitem{icsik2015s}
Y.~Z. I{\c{s}}ik, H.~Erdogan, and R.~Sarikaya, ``S-vector: A discriminative
  representation derived from i-vector for speaker verification,'' in
  \emph{2015 23rd European Signal Processing Conference (EUSIPCO)}.\hskip 1em
  plus 0.5em minus 0.4em\relax IEEE, 2015, pp. 2097--2101.

\bibitem{zhang2016end}
S.-X. Zhang, Z.~Chen, Y.~Zhao, J.~Li, and Y.~Gong, ``End-to-end attention based
  text-dependent speaker verification,'' in \emph{2016 IEEE Spoken Language
  Technology Workshop (SLT)}.\hskip 1em plus 0.5em minus 0.4em\relax IEEE,
  2016, pp. 171--178.

\bibitem{zhang2017end}
C.~Zhang and K.~Koishida, ``End-to-end text-independent speaker verification
  with triplet loss on short utterances.'' in \emph{Interspeech}, 2017, pp.
  1487--1491.

\bibitem{mclaren2014application}
M.~McLaren, Y.~Lei, N.~Scheffer, and L.~Ferrer, ``Application of convolutional
  neural networks to speaker recognition in noisy conditions,'' in
  \emph{Fifteenth Annual Conference of the International Speech Communication
  Association}, 2014.

\bibitem{lukic2016speaker}
Y.~Lukic, C.~Vogt, O.~D{\"u}rr, and T.~Stadelmann, ``Speaker identification and
  clustering using convolutional neural networks,'' in \emph{2016 IEEE 26th
  international workshop on machine learning for signal processing
  (MLSP)}.\hskip 1em plus 0.5em minus 0.4em\relax IEEE, 2016, pp. 1--6.

\bibitem{ranjan2017improved}
S.~Ranjan and J.~H. Hansen, ``Improved gender independent speaker recognition
  using convolutional neural network based bottleneck features.'' in
  \emph{INTERSPEECH}, 2017, pp. 1009--1013.

\bibitem{wang2017does}
S.~Wang, Y.~Qian, and K.~Yu, ``What does the speaker embedding encode?'' in
  \emph{Interspeech}, 2017, pp. 1497--1501.

\bibitem{shon2018frame}
S.~Shon, H.~Tang, and J.~Glass, ``Frame-level speaker embeddings for
  text-independent speaker recognition and analysis of end-to-end model,'' in
  \emph{2018 IEEE Spoken Language Technology Workshop (SLT)}.\hskip 1em plus
  0.5em minus 0.4em\relax IEEE, 2018, pp. 1007--1013.

\bibitem{marchi2018generalised}
E.~Marchi, S.~Shum, K.~Hwang, S.~Kajarekar, S.~Sigtia, H.~Richards, R.~Haynes,
  Y.~Kim, and J.~Bridle, ``Generalised discriminative transform via curriculum
  learning for speaker recognition,'' in \emph{2018 IEEE International
  Conference on Acoustics, Speech and Signal Processing (ICASSP)}.\hskip 1em
  plus 0.5em minus 0.4em\relax IEEE, 2018, pp. 5324--5328.

\bibitem{mclaren2015advances}
M.~McLaren, Y.~Lei, and L.~Ferrer, ``Advances in deep neural network approaches
  to speaker recognition,'' in \emph{2015 IEEE international conference on
  acoustics, speech and signal processing (ICASSP)}.\hskip 1em plus 0.5em minus
  0.4em\relax IEEE, 2015, pp. 4814--4818.

\bibitem{li2017deep}
C.~Li, X.~Ma, B.~Jiang, X.~Li, X.~Zhang, X.~Liu, Y.~Cao, A.~Kannan, and Z.~Zhu,
  ``Deep speaker: an end-to-end neural speaker embedding system,'' \emph{arXiv
  preprint arXiv:1705.02304}, 2017.

\bibitem{michelsanti2017conditional}
D.~Michelsanti and Z.-H. Tan, ``Conditional generative adversarial networks for
  speech enhancement and noise-robust speaker verification,'' \emph{arXiv
  preprint arXiv:1709.01703}, 2017.

\bibitem{lorenzo2018investigating}
J.~Lorenzo-Trueba, G.~E. Henter, S.~Takaki, J.~Yamagishi, Y.~Morino, and
  Y.~Ochiai, ``Investigating different representations for modeling and
  controlling multiple emotions in dnn-based speech synthesis,'' \emph{Speech
  Communication}, vol.~99, pp. 135--143, 2018.

\bibitem{wang2017learning}
Z.-Q. Wang and I.~Tashev, ``Learning utterance-level representations for speech
  emotion and age/gender recognition using deep neural networks,'' in
  \emph{2017 IEEE international conference on acoustics, speech and signal
  processing (ICASSP)}.\hskip 1em plus 0.5em minus 0.4em\relax IEEE, 2017, pp.
  5150--5154.

\bibitem{stolar2014optimized}
M.~N. Stolar, M.~Lech, and I.~S. Burnett, ``Optimized multi-channel deep neural
  network with {2D} graphical representation of acoustic speech features for
  emotion recognition,'' in \emph{2014 8th International Conference on Signal
  Processing and Communication Systems (ICSPCS)}.\hskip 1em plus 0.5em minus
  0.4em\relax IEEE, 2014, pp. 1--6.

\bibitem{lee2015high}
J.~Lee and I.~Tashev, ``High-level feature representation using recurrent
  neural network for speech emotion recognition,'' in \emph{Sixteenth Annual
  Conference of the International Speech Communication Association}, 2015.

\bibitem{huang2016attention}
C.-W. Huang and S.~S. Narayanan, ``Attention assisted discovery of
  sub-utterance structure in speech emotion recognition.'' in
  \emph{INTERSPEECH}, 2016, pp. 1387--1391.

\bibitem{kim2017learning}
J.~Kim, K.~P. Truong, G.~Englebienne, and V.~Evers, ``Learning spectro-temporal
  features with {3D CNNs} for speech emotion recognition,'' in \emph{2017
  Seventh International Conference on Affective Computing and Intelligent
  Interaction (ACII)}.\hskip 1em plus 0.5em minus 0.4em\relax IEEE, 2017, pp.
  383--388.

\bibitem{zheng2015experimental}
W.~Zheng, J.~Yu, and Y.~Zou, ``An experimental study of speech emotion
  recognition based on deep convolutional neural networks,'' in \emph{2015
  international conference on affective computing and intelligent interaction
  (ACII)}.\hskip 1em plus 0.5em minus 0.4em\relax IEEE, 2015, pp. 827--831.

\bibitem{mao2014learning}
Q.~Mao, M.~Dong, Z.~Huang, and Y.~Zhan, ``Learning salient features for speech
  emotion recognition using convolutional neural networks,'' \emph{IEEE
  transactions on multimedia}, vol.~16, no.~8, pp. 2203--2213, 2014.

\bibitem{li2018attention}
P.~Li, Y.~Song, I.~V. McLoughlin, W.~Guo, and L.~Dai, ``An attention pooling
  based representation learning method for speech emotion recognition.'' in
  \emph{Interspeech}, 2018, pp. 3087--3091.

\bibitem{zhao2018exploring}
Z.~Zhao, Y.~Zheng, Z.~Zhang, H.~Wang, Y.~Zhao, and C.~Li, ``Exploring
  spatio-temporal representations by integrating attention-based
  bidirectional-lstm-rnns and fcns for speech emotion recognition,'' 2018.

\bibitem{luo2018investigation}
D.~Luo, Y.~Zou, and D.~Huang, ``Investigation on joint representation learning
  for robust feature extraction in speech emotion recognition.'' in
  \emph{Interspeech}, 2018, pp. 152--156.

\bibitem{lim2016speech}
W.~Lim, D.~Jang, and T.~Lee, ``Speech emotion recognition using convolutional
  and recurrent neural networks,'' in \emph{2016 Asia-Pacific Signal and
  Information Processing Association Annual Summit and Conference
  (APSIPA)}.\hskip 1em plus 0.5em minus 0.4em\relax IEEE, 2016, pp. 1--4.

\bibitem{palaz2013estimating}
D.~Palaz, R.~Collobert, and M.~M. Doss, ``Estimating phoneme class conditional
  probabilities from raw speech signal using convolutional neural networks,''
  \emph{arXiv preprint arXiv:1304.1018}, 2013.

\bibitem{kabil2018learning}
S.~H. Kabil, H.~Muckenhirn, and M.~Magimai-Doss, ``On learning to identify
  genders from raw speech signal using cnns.'' in \emph{Interspeech}, 2018, pp.
  287--291.

\bibitem{golik2015convolutional}
P.~Golik, Z.~T{\"u}ske, R.~Schl{\"u}ter, and H.~Ney, ``Convolutional neural
  networks for acoustic modeling of raw time signal in lvcsr,'' in
  \emph{Sixteenth annual conference of the international speech communication
  association}, 2015.

\bibitem{dai2017very}
W.~Dai, C.~Dai, S.~Qu, J.~Li, and S.~Das, ``Very deep convolutional neural
  networks for raw waveforms,'' in \emph{2017 IEEE International Conference on
  Acoustics, Speech and Signal Processing (ICASSP)}.\hskip 1em plus 0.5em minus
  0.4em\relax IEEE, 2017, pp. 421--425.

\bibitem{liu2018deep}
X.~Liu, ``Deep convolutional and {LSTM} neural networks for acoustic modelling
  in automatic speech recognition,'' 2018.

\bibitem{zeghidour2018learning}
N.~Zeghidour, N.~Usunier, I.~Kokkinos, T.~Schaiz, G.~Synnaeve, and E.~Dupoux,
  ``Learning filterbanks from raw speech for phone recognition,'' in \emph{2018
  IEEE International Conference on Acoustics, Speech and Signal Processing
  (ICASSP)}.\hskip 1em plus 0.5em minus 0.4em\relax IEEE, 2018, pp. 5509--5513.

\bibitem{zazo2016feature}
R.~Zazo~Candil, T.~N. Sainath, G.~Simko, and C.~Parada, ``Feature learning with
  raw-waveform cldnns for voice activity detection,'' 2016.

\bibitem{ravanelli2018speaker}
M.~Ravanelli and Y.~Bengio, ``Speaker recognition from raw waveform with
  sincnet,'' in \emph{2018 IEEE Spoken Language Technology Workshop
  (SLT)}.\hskip 1em plus 0.5em minus 0.4em\relax IEEE, 2018, pp. 1021--1028.

\bibitem{jung2018avoiding}
J.-W. Jung, H.-S. Heo, I.-H. Yang, H.-J. Shim, and H.-J. Yu, ``Avoiding speaker
  overfitting in end-to-end dnns using raw waveform for text-independent
  speaker verification,'' \emph{extraction}, vol.~8, no.~12, pp. 23--24, 2018.

\bibitem{bengiospeaker}
M.~Ravanelli and Y.~Bengio, ``Speaker recognition from raw waveform with
  {SincNet},'' in \emph{2018 IEEE Spoken Language Technology Workshop
  (SLT)}.\hskip 1em plus 0.5em minus 0.4em\relax IEEE, 2018, pp. 1021--1028.

\bibitem{jung2018complete}
J.-W. Jung, H.-S. Heo, I.-H. Yang, H.-J. Shim, and H.-J. Yu, ``A complete
  end-to-end speaker verification system using deep neural networks: From raw
  signals to verification result,'' in \emph{2018 IEEE International Conference
  on Acoustics, Speech and Signal Processing (ICASSP)}.\hskip 1em plus 0.5em
  minus 0.4em\relax IEEE, 2018, pp. 5349--5353.

\bibitem{sarma2019improving}
M.~Sarma, P.~Ghahremani, D.~Povey, N.~K. Goel, K.~K. Sarma, and N.~Dehak,
  ``Improving emotion identification using phone posteriors in raw speech
  waveform based dnn,'' \emph{Proc. Interspeech 2019}, pp. 3925--3929, 2019.

\bibitem{lu2013combining}
H.~Lu, S.~King, and O.~Watts, ``Combining a vector space representation of
  linguistic context with a deep neural network for text-to-speech synthesis,''
  in \emph{Eighth ISCA Workshop on Speech Synthesis}, 2013.

\bibitem{hau2011exploring}
D.~Hau and K.~Chen, ``Exploring hierarchical speech representations with a deep
  convolutional neural network,'' \emph{UKCI 2011 Accepted Papers}, p.~37,
  2011.

\bibitem{chung2019unsupervised}
Y.-A. Chung, W.-N. Hsu, H.~Tang, and J.~Glass, ``An unsupervised autoregressive
  model for speech representation learning,'' \emph{arXiv preprint
  arXiv:1904.03240}, 2019.

\bibitem{chung2016audio}
Y.-A. Chung, C.-C. Wu, C.-H. Shen, H.-Y. Lee, and L.-S. Lee, ``Audio word2vec:
  Unsupervised learning of audio segment representations using
  sequence-to-sequence autoencoder,'' \emph{arXiv preprint arXiv:1603.00982},
  2016.

\bibitem{ishii2013reverberant}
T.~Ishii, H.~Komiyama, T.~Shinozaki, Y.~Horiuchi, and S.~Kuroiwa, ``Reverberant
  speech recognition based on denoising autoencoder.'' in \emph{Interspeech},
  2013, pp. 3512--3516.

\bibitem{xia2013speech}
B.~Xia and C.~Bao, ``Speech enhancement with weighted denoising auto-encoder.''
  in \emph{INTERSPEECH}, 2013, pp. 3444--3448.

\bibitem{zhang2015deep}
Z.~Zhang, L.~Wang, A.~Kai, T.~Yamada, W.~Li, and M.~Iwahashi, ``Deep neural
  network-based bottleneck feature and denoising autoencoder-based
  dereverberation for distant-talking speaker identification,'' \emph{EURASIP
  Journal on Audio, Speech, and Music Processing}, vol. 2015, no.~1, p.~12,
  2015.

\bibitem{van2017neural}
A.~van~den Oord, O.~Vinyals \emph{et~al.}, ``Neural discrete representation
  learning,'' in \emph{Advances in Neural Information Processing Systems},
  2017, pp. 6306--6315.

\bibitem{hsu2018hierarchical}
W.-N. Hsu, Y.~Zhang, R.~J. Weiss, H.~Zen, Y.~Wu, Y.~Wang, Y.~Cao, Y.~Jia,
  Z.~Chen, J.~Shen \emph{et~al.}, ``Hierarchical generative modeling for
  controllable speech synthesis,'' \emph{arXiv preprint arXiv:1810.07217},
  2018.

\bibitem{pal2019speaker}
M.~Pal, M.~Kumar, R.~Peri, T.~J. Park, S.~H. Kim, C.~Lord, S.~Bishop, and
  S.~Narayanan, ``Speaker diarization using latent space clustering in
  generative adversarial network,'' \emph{arXiv preprint arXiv:1910.11398},
  2019.

\bibitem{cibau2013speech}
N.~E. Cibau, E.~M. Albornoz, and H.~L. Rufiner, ``Speech emotion recognition
  using a deep autoencoder,'' \emph{Anales de la XV Reunion de Procesamiento de
  la Informacion y Control}, vol.~16, pp. 934--939, 2013.

\bibitem{eskimez2018unsupervised}
S.~E. Eskimez, Z.~Duan, and W.~Heinzelman, ``Unsupervised learning approach to
  feature analysis for automatic speech emotion recognition,'' in \emph{2018
  IEEE International Conference on Acoustics, Speech and Signal Processing
  (ICASSP)}.\hskip 1em plus 0.5em minus 0.4em\relax IEEE, 2018, pp. 5099--5103.

\bibitem{sahu2019modeling}
S.~Sahu, R.~Gupta, and C.~Espy-Wilson, ``Modeling feature representations for
  affective speech using generative adversarial networks,'' \emph{arXiv
  preprint arXiv:1911.00030}, 2019.

\bibitem{chung2019semi}
Y.-A. Chung, Y.~Wang, W.-N. Hsu, Y.~Zhang, and R.~Skerry-Ryan,
  ``Semi-supervised training for improving data efficiency in end-to-end speech
  synthesis,'' in \emph{ICASSP 2019-2019 IEEE International Conference on
  Acoustics, Speech and Signal Processing (ICASSP)}.\hskip 1em plus 0.5em minus
  0.4em\relax IEEE, 2019, pp. 6940--6944.

\bibitem{zhao2018wasserstein}
Y.~Zhao, S.~Takaki, H.-T. Luong, J.~Yamagishi, D.~Saito, and N.~Minematsu,
  ``Wasserstein {GAN} and waveform loss-based acoustic model training for
  multi-speaker text-to-speech synthesis systems using a wavenet vocoder,''
  \emph{IEEE Access}, vol.~6, pp. 60\,478--60\,488, 2018.

\bibitem{huang2016unified}
Z.~Huang, S.~M. Siniscalchi, and C.-H. Lee, ``A unified approach to transfer
  learning of deep neural networks with applications to speaker adaptation in
  automatic speech recognition,'' \emph{Neurocomputing}, vol. 218, pp.
  448--459, 2016.

\bibitem{swietojanski2012unsupervised}
P.~Swietojanski, A.~Ghoshal, and S.~Renals, ``Unsupervised cross-lingual
  knowledge transfer in {DNN}-based {LVCSR},'' in \emph{2012 IEEE Spoken
  Language Technology Workshop (SLT)}.\hskip 1em plus 0.5em minus 0.4em\relax
  IEEE, 2012, pp. 246--251.

\bibitem{latif2018crosscorpus}
S.~Latif, R.~Rana, S.~Younis, J.~Qadir, and J.~Epps, ``Cross corpus speech
  emotion classification-an effective transfer learning technique,''
  \emph{arXiv preprint arXiv:1801.06353}, 2018.

\bibitem{neumann2018cross}
M.~Neumann \emph{et~al.}, ``Cross-lingual and multilingual speech emotion
  recognition on english and french,'' in \emph{2018 IEEE International
  Conference on Acoustics, Speech and Signal Processing (ICASSP)}.\hskip 1em
  plus 0.5em minus 0.4em\relax IEEE, 2018, pp. 5769--5773.

\bibitem{abdelwahab2018domain}
M.~Abdelwahab and C.~Busso, ``Domain adversarial for acoustic emotion
  recognition,'' \emph{IEEE/ACM Transactions on Audio, Speech, and Language
  Processing}, vol.~26, no.~12, pp. 2423--2435, 2018.

\bibitem{tu2019towards}
M.~Tu, Y.~Tang, J.~Huang, X.~He, and B.~Zhou, ``Towards adversarial learning of
  speaker-invariant representation for speech emotion recognition,''
  \emph{arXiv preprint arXiv:1903.09606}, 2019.

\bibitem{deng2014introducing}
J.~Deng, R.~Xia, Z.~Zhang, Y.~Liu, and B.~Schuller, ``Introducing
  shared-hidden-layer autoencoders for transfer learning and their application
  in acoustic emotion recognition,'' in \emph{2014 IEEE International
  Conference on Acoustics, Speech and Signal Processing (ICASSP)}.\hskip 1em
  plus 0.5em minus 0.4em\relax IEEE, 2014, pp. 4818--4822.

\bibitem{deng2017recognizing}
J.~Deng, S.~Fr{\"u}hholz, Z.~Zhang, and B.~Schuller, ``Recognizing emotions
  from whispered speech based on acoustic feature transfer learning,''
  \emph{IEEE Access}, vol.~5, pp. 5235--5246, 2017.

\bibitem{xu2017multi}
Y.~Xu, J.~Du, Z.~Huang, L.-R. Dai, and C.-H. Lee, ``Multi-objective learning
  and mask-based post-processing for deep neural network based speech
  enhancement,'' \emph{arXiv preprint arXiv:1703.07172}, 2017.

\bibitem{seltzer2013multi}
M.~L. Seltzer and J.~Droppo, ``Multi-task learning in deep neural networks for
  improved phoneme recognition,'' in \emph{2013 IEEE International Conference
  on Acoustics, Speech and Signal Processing}.\hskip 1em plus 0.5em minus
  0.4em\relax IEEE, 2013, pp. 6965--6969.

\bibitem{tan2016speaker}
T.~Tan, Y.~Qian, D.~Yu, S.~Kundu, L.~Lu, K.~C. Sim, X.~Xiao, and Y.~Zhang,
  ``Speaker-aware training of {LSTM-RNNs} for acoustic modelling,'' in
  \emph{2016 IEEE International Conference on Acoustics, Speech and Signal
  Processing (ICASSP)}.\hskip 1em plus 0.5em minus 0.4em\relax IEEE, 2016, pp.
  5280--5284.

\bibitem{li2015modeling}
X.~Li and X.~Wu, ``Modeling speaker variability using long short-term memory
  networks for speech recognition,'' in \emph{Sixteenth Annual Conference of
  the International Speech Communication Association}, 2015.

\bibitem{tang2017collaborative}
Z.~Tang, L.~Li, D.~Wang, R.~Vipperla, Z.~Tang, L.~Li, D.~Wang, and R.~Vipperla,
  ``Collaborative joint training with multitask recurrent model for speech and
  speaker recognition,'' \emph{IEEE/ACM Transactions on Audio, Speech and
  Language Processing (TASLP)}, vol.~25, no.~3, pp. 493--504, 2017.

\bibitem{qian2016noise}
Y.~Qian, J.~Tao, D.~Suendermann-Oeft, K.~Evanini, A.~V. Ivanov, and
  V.~Ramanarayanan, ``Noise and metadata sensitive bottleneck features for
  improving speaker recognition with non-native speech input.'' in
  \emph{INTERSPEECH}, 2016, pp. 3648--3652.

\bibitem{chen2015multi}
N.~Chen, Y.~Qian, and K.~Yu, ``Multi-task learning for text-dependent speaker
  verification,'' in \emph{Sixteenth annual conference of the international
  speech communication association}, 2015.

\bibitem{yadav2018learning}
S.~Yadav and A.~Rai, ``Learning discriminative features for speaker
  identification and verification.'' in \emph{Interspeech}, 2018, pp.
  2237--2241.

\bibitem{tang2016multi}
Z.~Tang, L.~Li, and D.~Wang, ``Multi-task recurrent model for speech and
  speaker recognition,'' in \emph{2016 Asia-Pacific Signal and Information
  Processing Association Annual Summit and Conference (APSIPA)}.\hskip 1em plus
  0.5em minus 0.4em\relax IEEE, 2016, pp. 1--4.

\bibitem{xia2015multi}
R.~Xia and Y.~Liu, ``A multi-task learning framework for emotion recognition
  using {2D} continuous space,'' \emph{IEEE Transactions on Affective
  Computing}, vol.~8, no.~1, pp. 3--14, 2015.

\bibitem{zhang2017multi}
Y.~Zhang, Y.~Liu, F.~Weninger, and B.~Schuller, ``Multi-task deep neural
  network with shared hidden layers: Breaking down the wall between emotion
  representations,'' in \emph{2017 IEEE international conference on acoustics,
  speech and signal processing (ICASSP)}.\hskip 1em plus 0.5em minus
  0.4em\relax IEEE, 2017, pp. 4990--4994.

\bibitem{ma2018speech}
F.~Ma, W.~Gu, W.~Zhang, S.~Ni, S.-L. Huang, and L.~Zhang, ``Speech emotion
  recognition via attention-based {DNN} from multi-task learning,'' in
  \emph{Proceedings of the 16th ACM Conference on Embedded Networked Sensor
  Systems}.\hskip 1em plus 0.5em minus 0.4em\relax ACM, 2018, pp. 363--364.

\bibitem{zhang2019attention}
Z.~Zhang, B.~Wu, and B.~Schuller, ``Attention-augmented end-to-end multi-task
  learning for emotion prediction from speech,'' in \emph{ICASSP 2019-2019 IEEE
  International Conference on Acoustics, Speech and Signal Processing
  (ICASSP)}.\hskip 1em plus 0.5em minus 0.4em\relax IEEE, 2019, pp. 6705--6709.

\bibitem{eyben2012multitask}
F.~Eyben, M.~W{\"o}llmer, and B.~Schuller, ``A multitask approach to continuous
  five-dimensional affect sensing in natural speech,'' \emph{ACM Transactions
  on Interactive Intelligent Systems (TiiS)}, vol.~2, no.~1, p.~6, 2012.

\bibitem{le2017discretized}
D.~Le, Z.~Aldeneh, and E.~M. Provost, ``Discretized continuous speech emotion
  recognition with multi-task deep recurrent neural network,''
  \emph{Interspeech, 2017 (to apear)}, 2017.

\bibitem{li2019speaker}
H.~Li, M.~Tu, J.~Huang, S.~Narayanan, and P.~Georgiou, ``Speaker-invariant
  affective representation learning via adversarial training,'' \emph{arXiv
  preprint arXiv:1911.01533}, 2019.

\bibitem{srivastava2015training}
R.~K. Srivastava, K.~Greff, and J.~Schmidhuber, ``Training very deep
  networks,'' in \emph{Advances in neural information processing systems},
  2015, pp. 2377--2385.

\bibitem{szegedy2015going}
C.~Szegedy, W.~Liu, Y.~Jia, P.~Sermanet, S.~Reed, D.~Anguelov, D.~Erhan,
  V.~Vanhoucke, and A.~Rabinovich, ``Going deeper with convolutions,'' in
  \emph{Proceedings of the IEEE conference on computer vision and pattern
  recognition}, 2015, pp. 1--9.

\bibitem{saxe2013exact}
A.~M. Saxe, J.~L. McClelland, and S.~Ganguli, ``Exact solutions to the
  nonlinear dynamics of learning in deep linear neural networks,'' \emph{arXiv
  preprint arXiv:1312.6120}, 2013.

\bibitem{he2015delving}
K.~He, X.~Zhang, S.~Ren, and J.~Sun, ``Delving deep into rectifiers: Surpassing
  human-level performance on imagenet classification,'' in \emph{Proceedings of
  the IEEE international conference on computer vision}, 2015, pp. 1026--1034.

\bibitem{jia2017improving}
K.~Jia, D.~Tao, S.~Gao, and X.~Xu, ``Improving training of deep neural networks
  via singular value bounding,'' in \emph{Proceedings of the IEEE Conference on
  Computer Vision and Pattern Recognition}, 2017, pp. 4344--4352.

\bibitem{salimans2016weight}
T.~Salimans and D.~P. Kingma, ``Weight normalization: A simple
  reparameterization to accelerate training of deep neural networks,'' in
  \emph{Advances in Neural Information Processing Systems}, 2016, pp. 901--909.

\bibitem{ioffe2015batch}
S.~Ioffe and C.~Szegedy, ``Batch normalization: Accelerating deep network
  training by reducing internal covariate shift,'' \emph{arXiv preprint
  arXiv:1502.03167}, 2015.

\bibitem{li2017unsupervised}
H.~Li, B.~Baucom, and P.~Georgiou, ``Unsupervised latent behavior manifold
  learning from acoustic features: Audio2behavior,'' in \emph{2017 IEEE
  International Conference on Acoustics, Speech and Signal Processing
  (ICASSP)}.\hskip 1em plus 0.5em minus 0.4em\relax IEEE, 2017, pp. 5620--5624.

\bibitem{gong2018towards}
Y.~Gong and C.~Poellabauer, ``Towards learning fine-grained disentangled
  representations from speech,'' \emph{arXiv preprint arXiv:1808.02939}, 2018.

\bibitem{arjovsky2017wasserstein}
M.~Arjovsky, S.~Chintala, and L.~Bottou, ``Wasserstein gan,'' \emph{arXiv
  preprint arXiv:1701.07875}, 2017.

\bibitem{arjovsky2017towards}
M.~Arjovsky and L.~Bottou, ``Towards principled methods for training generative
  adversarial networks. arxiv,'' 2017.

\bibitem{roth2017stabilizing}
K.~Roth, A.~Lucchi, S.~Nowozin, and T.~Hofmann, ``Stabilizing training of
  generative adversarial networks through regularization,'' in \emph{Advances
  in neural information processing systems}, 2017, pp. 2018--2028.

\bibitem{asakawa2007automatic}
S.~Asakawa, N.~Minematsu, and K.~Hirose, ``Automatic recognition of connected
  vowels only using speaker-invariant representation of speech dynamics,'' in
  \emph{Eighth Annual Conference of the International Speech Communication
  Association}, 2007.

\bibitem{goodfellow2014explaining}
I.~J. Goodfellow, J.~Shlens, and C.~Szegedy, ``Explaining and harnessing
  adversarial examples,'' \emph{arXiv preprint arXiv:1412.6572}, 2014.

\bibitem{papernot2016limitations}
N.~Papernot, P.~McDaniel, S.~Jha, M.~Fredrikson, Z.~B. Celik, and A.~Swami,
  ``The limitations of deep learning in adversarial settings,'' in \emph{2016
  IEEE European Symposium on Security and Privacy (EuroS\&P)}.\hskip 1em plus
  0.5em minus 0.4em\relax IEEE, 2016, pp. 372--387.

\bibitem{moosavi2016deepfool}
S.-M. Moosavi-Dezfooli, A.~Fawzi, and P.~Frossard, ``Deepfool: a simple and
  accurate method to fool deep neural networks,'' in \emph{Proceedings of the
  IEEE conference on computer vision and pattern recognition}, 2016, pp.
  2574--2582.

\bibitem{carlini2018audio}
N.~Carlini and D.~Wagner, ``Audio adversarial examples: Targeted attacks on
  speech-to-text,'' in \emph{2018 IEEE Security and Privacy Workshops
  (SPW)}.\hskip 1em plus 0.5em minus 0.4em\relax IEEE, 2018, pp. 1--7.

\bibitem{hannun2014deep}
A.~Hannun, C.~Case, J.~Casper, B.~Catanzaro, G.~Diamos, E.~Elsen, R.~Prenger,
  S.~Satheesh, S.~Sengupta, A.~Coates \emph{et~al.}, ``Deep speech: Scaling up
  end-to-end speech recognition,'' \emph{arXiv preprint arXiv:1412.5567}, 2014.

\bibitem{alzantot2018did}
M.~Alzantot, B.~Balaji, and M.~Srivastava, ``Did you hear that? adversarial
  examples against automatic speech recognition,'' \emph{arXiv preprint
  arXiv:1801.00554}, 2018.

\bibitem{schonherr2018adversarial}
L.~Sch{\"o}nherr, K.~Kohls, S.~Zeiler, T.~Holz, and D.~Kolossa, ``Adversarial
  attacks against automatic speech recognition systems via psychoacoustic
  hiding,'' \emph{arXiv preprint arXiv:1808.05665}, 2018.

\bibitem{akhtar2018threat}
N.~Akhtar and A.~Mian, ``Threat of adversarial attacks on deep learning in
  computer vision: A survey,'' \emph{IEEE Access}, vol.~6, pp.
  14\,410--14\,430, 2018.

\bibitem{yin2015noisy}
S.~Yin, C.~Liu, Z.~Zhang, Y.~Lin, D.~Wang, J.~Tejedor, T.~F. Zheng, and Y.~Li,
  ``Noisy training for deep neural networks in speech recognition,''
  \emph{EURASIP Journal on Audio, Speech, and Music Processing}, vol. 2015,
  no.~1, p.~2, 2015.

\bibitem{bu2014incomplete}
F.~Bu, Z.~Chen, Q.~Zhang, and X.~Wang, ``Incomplete big data clustering
  algorithm using feature selection and partial distance,'' in \emph{2014 5th
  International Conference on Digital Home}.\hskip 1em plus 0.5em minus
  0.4em\relax IEEE, 2014, pp. 263--266.

\bibitem{wang2016non}
R.~Wang and D.~Tao, ``Non-local auto-encoder with collaborative stabilization
  for image restoration,'' \emph{IEEE Transactions on Image Processing},
  vol.~25, no.~5, pp. 2117--2129, 2016.

\bibitem{mcfee2015librosa}
B.~McFee, C.~Raffel, D.~Liang, D.~P. Ellis, M.~McVicar, E.~Battenberg, and
  O.~Nieto, ``librosa: Audio and music signal analysis in python,'' in
  \emph{Proceedings of the 14th python in science conference}, vol.~8, 2015.

\bibitem{giannakopoulos2015pyaudioanalysis}
T.~Giannakopoulos, ``pyaudioanalysis: An open-source python library for audio
  signal analysis,'' \emph{PloS one}, vol.~10, no.~12, p. e0144610, 2015.

\bibitem{eyben2010opensmile}
F.~Eyben, M.~W{\"o}llmer, and B.~Schuller, ``{OpenSMILE -- the Munich versatile
  and fast open-source audio feature extractor},'' in \emph{Proceedings of the
  18th ACM international conference on Multimedia}.\hskip 1em plus 0.5em minus
  0.4em\relax ACM, 2010, pp. 1459--1462.

\bibitem{lamere2003cmu}
P.~Lamere, P.~Kwok, E.~Gouvea, B.~Raj, R.~Singh, W.~Walker, M.~Warmuth, and
  P.~Wolf, ``The cmu sphinx-4 speech recognition system,'' in \emph{IEEE Intl.
  Conf. on Acoustics, Speech and Signal Processing (ICASSP 2003), Hong Kong},
  vol.~1, 2003, pp. 2--5.

\bibitem{povey2011kaldi}
D.~Povey, A.~Ghoshal, G.~Boulianne, L.~Burget, O.~Glembek, N.~Goel,
  M.~Hannemann, P.~Motlicek, Y.~Qian, P.~Schwarz \emph{et~al.}, ``The kaldi
  speech recognition toolkit,'' in \emph{IEEE 2011 workshop on automatic speech
  recognition and understanding}, no. CONF.\hskip 1em plus 0.5em minus
  0.4em\relax IEEE Signal Processing Society, 2011.

\bibitem{lee2001julius}
A.~Lee, T.~Kawahara, and K.~Shikano, ``Julius---an open source real-time large
  vocabulary recognition engine,'' 2001.

\bibitem{watanabe2018espnet}
S.~Watanabe, T.~Hori, S.~Karita, T.~Hayashi, J.~Nishitoba, Y.~Unno, N.~E.~Y.
  Soplin, J.~Heymann, M.~Wiesner, N.~Chen \emph{et~al.}, ``Espnet: End-to-end
  speech processing toolkit,'' \emph{arXiv preprint arXiv:1804.00015}, 2018.

\bibitem{woodland1994large}
P.~C. Woodland, J.~J. Odell, V.~Valtchev, and S.~J. Young, ``Large vocabulary
  continuous speech recognition using htk.'' in \emph{ICASSP (2)}, 1994, pp.
  125--128.

\bibitem{larcher2013alize}
A.~Larcher, J.-F. Bonastre, B.~Fauve, K.~Lee, C.~L{\'e}vy, H.~Li, J.~Mason, and
  J.-Y. Parfait, ``Alize 3.0-open source toolkit for state-of-the-art speaker
  recognition,'' 2013.

\bibitem{eyben2009openear}
F.~Eyben, M.~W{\"o}llmer, and B.~Schuller, ``Openear—introducing the munich
  open-source emotion and affect recognition toolkit,'' in \emph{2009 3rd
  international conference on affective computing and intelligent interaction
  and workshops}.\hskip 1em plus 0.5em minus 0.4em\relax IEEE, 2009, pp. 1--6.

\bibitem{jouppi2017datacenter}
N.~P. Jouppi, C.~Young, N.~Patil, D.~Patterson, G.~Agrawal, R.~Bajwa, S.~Bates,
  S.~Bhatia, N.~Boden, A.~Borchers \emph{et~al.}, ``In-datacenter performance
  analysis of a tensor processing unit,'' in \emph{2017 ACM/IEEE 44th Annual
  International Symposium on Computer Architecture (ISCA)}.\hskip 1em plus
  0.5em minus 0.4em\relax IEEE, 2017, pp. 1--12.

\bibitem{arute2019quantum}
F.~Arute, K.~Arya, R.~Babbush, D.~Bacon, J.~C. Bardin, R.~Barends, R.~Biswas,
  S.~Boixo, F.~G. Brandao, D.~A. Buell \emph{et~al.}, ``Quantum supremacy using
  a programmable superconducting processor,'' \emph{Nature}, vol. 574, no.
  7779, pp. 505--510, 2019.

\bibitem{engel2019gansynth}
J.~Engel, K.~K. Agrawal, S.~Chen, I.~Gulrajani, C.~Donahue, and A.~Roberts,
  ``Gansynth: Adversarial neural audio synthesis,'' \emph{arXiv preprint
  arXiv:1902.08710}, 2019.

\bibitem{kumar2019melgan}
K.~Kumar, R.~Kumar, T.~de~Boissiere, L.~Gestin, W.~Z. Teoh, J.~Sotelo,
  A.~de~Brebisson, Y.~Bengio, and A.~Courville, ``Melgan: Generative
  adversarial networks for conditional waveform synthesis,'' \emph{arXiv
  preprint arXiv:1910.06711}, 2019.

\bibitem{chesney2018deep}
R.~Chesney and D.~K. Citron, ``Deep fakes: a looming challenge for privacy,
  democracy, and national security,'' 2018.

\bibitem{zhu2017unpaired}
J.-Y. Zhu, T.~Park, P.~Isola, and A.~A. Efros, ``Unpaired image-to-image
  translation using cycle-consistent adversarial networks,'' in
  \emph{Proceedings of the IEEE international conference on computer vision},
  2017, pp. 2223--2232.

\bibitem{nidadavolu2019low}
P.~S. Nidadavolu, S.~Kataria, J.~Villalba, and N.~Dehak, ``Low-resource domain
  adaptation for speaker recognition using cycle-gans,'' \emph{arXiv preprint
  arXiv:1910.11909}, 2019.

\bibitem{bao2019cyclegan}
F.~Bao, M.~Neumann, and N.~T. Vu, ``Cyclegan-based emotion style transfer as
  data augmentation for speech emotion recognition,'' \emph{Manuscript
  submitted for publication}, pp. 35--37, 2019.

\bibitem{thomas2018disentangling}
V.~Thomas, E.~Bengio, W.~Fedus, J.~Pondard, P.~Beaudoin, H.~Larochelle,
  J.~Pineau, D.~Precup, and Y.~Bengio, ``Disentangling the independently
  controllable factors of variation by interacting with the world,''
  \emph{arXiv preprint arXiv:1802.09484}, 2018.

\bibitem{pathak2013privacy}
M.~A. Pathak, B.~Raj, S.~D. Rane, and P.~Smaragdis, ``Privacy-preserving speech
  processing: cryptographic and string-matching frameworks show promise,''
  \emph{IEEE signal processing magazine}, vol.~30, no.~2, pp. 62--74, 2013.

\bibitem{srivastava2019privacy}
B.~M.~L. Srivastava, A.~Bellet, M.~Tommasi, and E.~Vincent,
  ``Privacy-preserving adversarial representation learning in asr: Reality or
  illusion?'' \emph{Proc. INTERPSPEECH}, pp. 3700--3704, 2019.

\bibitem{jaiswal2019privacy}
M.~Jaiswal and E.~M. Provost, ``Privacy enhanced multimodal neural
  representations for emotion recognition,'' \emph{arXiv preprint
  arXiv:1910.13212}, 2019.

\bibitem{shokri2015privacy}
R.~Shokri and V.~Shmatikov, ``Privacy-preserving deep learning,'' in
  \emph{Proceedings of the 22nd ACM SIGSAC conference on computer and
  communications security}.\hskip 1em plus 0.5em minus 0.4em\relax ACM, 2015,
  pp. 1310--1321.

\end{thebibliography}


\end{document}